\documentclass[preprintnumbers,superscriptaddress,nofootinbib,aps,prd,floatfix]{revtex4}

\usepackage{datetime}
\usepackage{color}

\usepackage{footmisc,multirow}
\usepackage{subfigure}
\usepackage{amsmath,slashed}

\usepackage{graphicx}
\renewcommand\[{\left[}
\renewcommand\]{\right]}
\newcommand{\Lagr}{\mathcal{L}}

\def\missET {\slashed{E}_T}
\def\misspT {\slashed{p}_T}

\def\beq{\begin{equation}}
\def\eeq{\end{equation}}
\def\[{\begin{equation}}
\def\]{\end{equation}}

\newcommand{\be}{\begin{eqnarray*}}
\newcommand{\ee}{\end{eqnarray*}}

\newcommand{\bee}{\begin{eqnarray}}
\newcommand{\eee}{\end{eqnarray}}
\newcommand{\beeq}{\begin{equation}}
\newcommand{\eeeq}{\end{equation}}

\renewcommand\[{\left[}
\renewcommand\]{\right]}

\preprint{IPPP/15/19} \preprint{DCPT/15/38}

\begin{document}

\title{Spectroscopy of Scalar Mediators to Dark Matter at the LHC and at 100 TeV}

\begin{abstract}
\noindent We investigate simplified models of dark matter with scalar mediators at hadron colliders using the final state topology with 2 jets 
and missing energy. These models can arise in a wide variety of BSM scenarios including the possibility of the mediator mixing with the Higgs.
Our aim is first to determine the projected reach of the LHC and the future circular hadron collider for excluding such models, 
and we also compare these to the relic density and direct detection constraints.
We use the kinematic distributions to extract information on mediator masses at colliders.
At the 13 TeV LHC we can probe mediator masses up to 750 GeV, and at a 100 TeV collider the reach is increased to 2.5 TeV mediators. 
We also explain how individual models with different values of mediator masses can be differentiated from each other.
\end{abstract}

\author{Valentin V Khoze} \email{valya.khoze@durham.ac.uk}
\affiliation{Institute for Particle Physics Phenomenology, Department
  of Physics,\\Durham University, DH1 3LE, United Kingdom\\[0.1cm]}

\author{Gunnar Ro} \email{g.o.i.ro@dur.ac.uk}
\affiliation{Institute for Particle Physics Phenomenology, Department
  of Physics,\\Durham University, DH1 3LE, United Kingdom\\[0.1cm]}

\author{Michael Spannowsky} \email{michael.spannowsky@durham.ac.uk}
\affiliation{Institute for Particle Physics Phenomenology, Department
  of Physics,\\Durham University, DH1 3LE, United Kingdom\\[0.1cm]}

\maketitle


\section{Introduction}
\label{sec:intro}

There is overwhelming cosmological evidence for the existence of dark matter (DM) with a density about five times larger than that of ordinary matter. 
But despite a lot of ongoing effort in direct and indirect detection, and collider experiments we still do not know the fundamental nature and composition of DM. 
None of the particles in the Standard Model can provide a dark matter candidate while many beyond the Standard Model (BSM) scenarios do. A common feature of many BSM extensions are the prediction of Weakly Interacting Massive Particles (WIMPs): if a particle has weak scale mass and interactions, the abundance after standard thermal freeze-out would be close to the observed abundance of dark matter. Specific realisations of DM candidates in different BSM models  can have a rich and interesting phenomenology. However, assuming dark matter can be produced at colliders, a generic feature is the existence of signatures with missing energy, as stable dark matter particles leave the detector unobserved.

\medskip
To discover missing energy signals at colliders, one recoils the invisibly decaying particle against reconstructable objects. 
In the simplest and most direct case, this is a visible mono-object, and such searches for mono-jets and mono-photons have been carried out
at Run 1 of the LHC \cite{Khachatryan:2014rra,Khachatryan:2014rwa,Aad:2015zva}. These studies have so far not discovered any evidence 
for an excess of missing energy events and can in parts of the parameter space be as or more constraining than limits from direct and indirect detection \cite{Birkedal:2004xn,Goodman:2010yf,Goodman:2010ku,Fox:2012ee,Abdallah:2014hon,Malik:2014ggr}. 
It is thus important to formulate and extend the searches of dark matter at the LHC in Run 2 and beyond. 

\medskip

Dark matter can be produced at colliders 
via an exchange of a mediator particle which connects the colliding SM partons to the dark sector. A viable and simple approach
to characterise and interpret dark matter searches at colliders relies on using simplified models with
four basic types of mediators: vectors, axial-vectors, scalars and pseudo-scalars (see white papers \cite{Malik:2014ggr,Abdallah:2014hon}
for early reviews and references). The mediator is a dynamical degree of freedom in this approach, and this is the correct description 
for dark matter searches at the LHC as the energy transfer in the collision can typically exceed mediator masses.
Following the Higgs discovery, there is a renewed interest in the role of scalar degrees of freedom and the possibilities provided by 
extended Higgs sectors in searches for new physics. Of particular interest to dark matter searches are the models with scalar and
pseudo-scalar mediators whose reach at the LHC was studied recently in \cite{Buckley:2014fba,Harris:2014hga,Chala:2015ama}.
It was found that the LHC at 14 TeV will provide a complementary coverage to the low-energy experiments in dark matter searches, and
can be the only experiment to probe dark sectors if the invisible particles produced are not stable at cosmological scales.
These studies have been performed using the so-called mono-jet topology \cite{Khachatryan:2014rra,Khachatryan:2014rwa,Aad:2015zva}.  

\medskip

In this paper we will study simplified models with scalar mediators in the 2-jets plus missing energy topology to determine their 
collider limits and the discovery potential  by analysing the kinematics of the final state jets. 
For scalar mediators mono-jet searches are predominantly relying on gluon fusion production \cite{Buckley:2014fba,Harris:2014hga}.
Now the presence of a second jet allows for a more non-trivial kinematic distribution to characterise the final states. Hence the VBF cuts can be imposed which suppress the gluon fusion 
production channel for scalars. This makes the weak boson fusion processes dominant instead and allows to capture mediators with suppressed couplings to fermions. Thus, 
the kinematic information in the 2 jets+MET final state should allow to probe better the mediator masses and also to give a handle on their interactions with electroweak gauge bosons.
In a slightly different context, the idea of exploring the two jets kinematics to learn more about the SM--DM interactions 
has also been implemented in \cite{Haisch:2013fla}.

We consider a simplified model with a scalar mediator whose SM couplings proportional to the SM Higgs. We start by defining the models 
in Section {\bf \ref{sec:model}} before commenting on non-collider dark matter phenomenology in Section {\bf \ref{DMpheno}}. Then, we move on to consider the phenomenology at the LHC in Section {\bf \ref{sec:LHC}} and at a future 100 TeV collider in Section {\bf \ref{sec:100TeV}}.

\section{Models}
\label{sec:model}

In unitary gauge the Standard Model (SM) contains just a single scalar-field degree of freedom, the neutral scalar Higgs $h$.
At tree level $h$ interacts with the massive vector bosons, $W^{\pm}$ and $Z^0$, and all the SM fermions. The  linear interactions in $h$
of the Higgs boson with 
other SM particles can be written in the form,
\beq
\Lagr_h^{SM}\, \supset \, \left(\frac{2M_W^2}{v} \, W^{+}_\mu W^{-\,\mu}\,+\,
\frac{M_Z^2}{v} \, Z_\mu Z^{\mu}  \, -\, \sum_f \frac{m_f}{v}\, \bar{f}f\right)\, h\,.
\label{eq:Lh}
\eeq
We want to extend the SM by introducing a scalar mediator particle $\phi$ which couples to the SM degrees of freedom as well as 
to fermionic dark matter particles $\chi$ via,
\beq
\Lagr_\phi\, \supset \,  -g_{\rm DM}\, \bar{\chi}\chi\, \phi\,.
\label{eq:Lphi}
\eeq
For the purpose of this article the spin of the dark matter particle is not relevant, i.e. the dark matter particle could be instead a vector or scalar particle.
There are two types of settings where the additional scalar $\phi$ can appear in interactions with the Standard Model.
First, it can be an additional Higgs doublet, for example coming from a two Higgs doublet model, or more generally any scalar field transforming 
non-trivially under the $SU(2)_L$ of the SM. Alternatively, the $\phi$ scalar mediator can be a singlet under the Standard Model. 
In the latter case, it interacts with the SM degrees of freedom only via the mixing with the SM Higgs $h$. In this case, the interactions 
of $\phi$ with the SM are subject to experimental constraints on the mixing angle $\sin^2 \theta \lesssim 0.15$
(see \cite{Robens:2015gla,Falkowski:2015iwa}) arising from experimental bounds on the SM Higgs to invisibles decays
and other Higgs data.

First in section {\bf \ref{sec:mixing}} we consider the more constrained singlet-mixing case, and next in section {\bf \ref{sec:kappa}} we define
the less-constrained generic Higgs-like scenario.
The upshot is that both of these cases will be described by the same simplified model of Eq.~\eqref{eq:Lag}
with the scaling parameter $\kappa$ being either unconstrained $\kappa \sim 1$ or small $\kappa \lesssim 0.15$.

\subsection{The singlet mixing model}
\label{sec:mixing}

Here $\phi$ is a Standard-Model singlet neutral scalar. 
The visible SM sector and the `invisible' $\chi$ sector are coupled to each other only via the mixing between the two neutral scalars, $\phi$ and $h$,
as in the Higgs portal model. The states of definite
masses, $h_1$ and $h_2$, are
\beq
h\,=\,  h_1 \cos \theta \,+\,  h_2 \sin \theta \,, \qquad
\phi\,=\, -h_1 \sin \theta \,+\, h_2 \cos \theta\,,
\label{eq:h1h2}
\eeq
where $\theta$ is the mixing angle.
Combining Eqs.~\eqref{eq:Lh}-\eqref{eq:h1h2} we obtain a Simplified Model for invisible Higgs decays involving two Higgs-like
neutral scalars $h_1$ and $h_2$:

\begin{eqnarray}
\Lagr_{h_1,h_2} = \left(\frac{2M_W^2}{v} \, W^{+}_\mu W^{-\,\mu}\,+\,
\frac{M_Z^2}{v} \, Z_\mu Z^{\mu}  \, -\, \sum_f \frac{m_f}{v}\, \bar{f}f\right) \Big( h_1 \cos \theta \,+\,  h_2 \sin \theta  \Big)\nonumber\\
- g_{\rm DM}\, \bar{\chi}\chi \Big( h_2 \cos \theta  \,-\, h_1 \sin \theta\Big) 
\,-\, \frac{1}{2}m_{h_1}^2 h_1^2 \,-\, \frac{1}{2}m_{h_2}^2 h_2^2 \,-\, m_{\chi} \bar{\chi}\chi \,.
\label{eq:smh1h2}
\end{eqnarray}
The first scalar mass eigenstate $h_1$ plays the role of the observed SM Higgs boson and we also assume that the mediator $h_2$ 
is always  heavier than the SM Higgs,
\beq
m_{h_2}  \,>\, m_{h_1} =125 \, {\rm GeV}\,.
\eeq
With this Lagrangian we can produce $h_2$ as in the SM via both gluon fusion and vector boson fusion mechanisms,
with the corresponding SM cross-sections rescaled by $\sin^2 \theta$. Similarly the 125 GeV Higgs scalar $h_1$ production rates are
rescaled relative to the SM by the factor of $\cos^2\theta$ which is $ \simeq 1$ for sufficiently small values of the mixing angle. 

If both mediators can be produced on-shell, in either channel the cross-section for $\bar{\chi}\chi$+2 jet production in the narrow width approximation 
can be written as,
\beq
\sigma_{\rm DM}^{(i)}\,=\, \sigma_{h_i}\, {\rm Br}_{h_i\rightarrow \bar{\chi}\chi},
\label{eq:csni}
\eeq
where $\sigma_{h_i}$ is the production cross-section for $h_i$ + 2 jets and ${\rm Br}_{h_i\rightarrow \bar{\chi}\chi}$ are the branching ratios,
\begin{eqnarray}
\label{eq:csni1}
\sigma_{h_1} &=& \sigma_{\rm SM}\,\cos^2\theta\, ,\, \quad
{\rm Br}_{h_1\rightarrow \bar{\chi}\chi}= \frac{\sin^2\theta\, \Gamma_{\phi\rightarrow\bar{\chi}\chi}}{\sin^2\theta \Gamma_{\phi\rightarrow\bar{\chi}\chi}
+\cos^2\theta \, \Gamma_{h\rightarrow {\rm SM}}} = \sin^2\theta\,\frac{\Gamma_{\phi\rightarrow\bar{\chi}\chi}(m_{h_1})}{\Gamma^{\rm tot}_{h_1}}
\,,
\\
\label{eq:csni2}
 \sigma_{h_2} &=& \sigma_{\rm SM}\, \sin^2\theta\,, \,\quad
{\rm Br}_{h_2\rightarrow \bar{\chi}\chi}= \frac{\cos^2\theta\,\Gamma_{\phi\rightarrow\bar{\chi}\chi}}{\cos^2\theta\, \Gamma_{\phi\rightarrow\bar{\chi}\chi}+\sin^2\theta\, \Gamma_{h\rightarrow {\rm SM}}}= \cos^2\theta\,\frac{\Gamma_{\phi\rightarrow\bar{\chi}\chi}(m_{h_2})}{\Gamma^{\rm tot}_{h_2}}\,,
\end{eqnarray}
where 
\beq
\Gamma_{\phi\rightarrow \bar{\chi}\chi}=\frac{g_{\rm DM}^2m_{h_i}}{8\pi}\left(1-\frac{4m_{\chi}^2}{m^2_{\phi}}\right)^{\frac{3}{2}}.
\eeq

\begin{figure*}[h]
 \begin{center}
   \hspace{-0.6cm}
   \parbox{0.7\textwidth}{
     \includegraphics[width=0.5\textwidth]{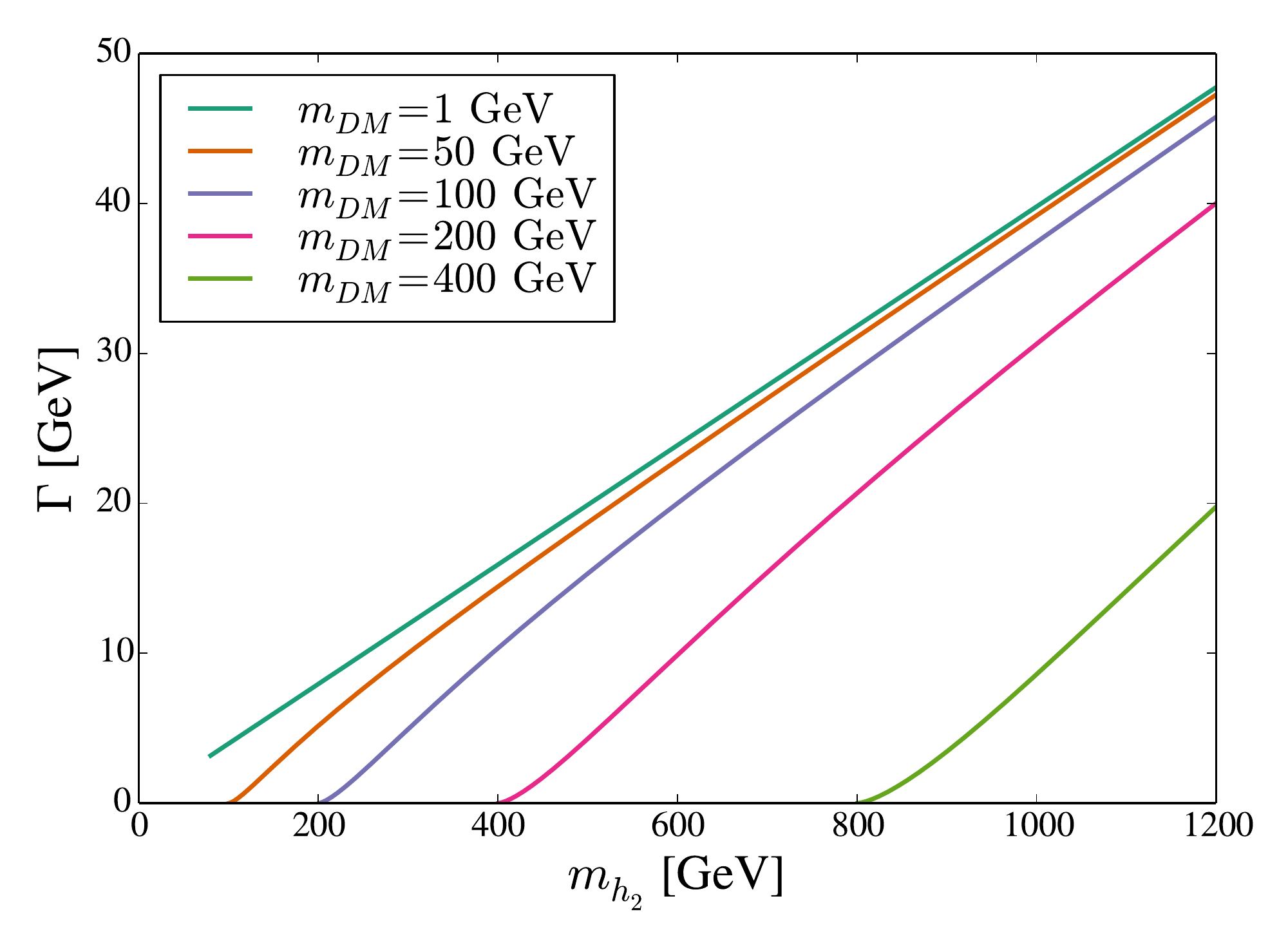}
   }
   \parbox{0.6\textwidth}{ \vspace{0.2cm}  \caption {The decay width of $h_2$ into $\bar{\chi}\chi$ with $g_{\rm DM}=1$. }
     \label{fig:Gammah2}
     }
 \end{center}
\end{figure*}

For dark matter mass below the kinematic threshold of both mediators, $2 m_\chi < m_{h_1} < m_{h_2}$ both mediators can be on-shell
and, in principle, both channels for the dark matter production are open, but the lighter Higgs will dominate, as can be seen from 
\eqref{eq:csni}-\eqref{eq:csni2},
\beq
\sigma_{\rm DM}^{(1)}/\sigma_{\rm DM}^{(2)} \,\propto\,\frac{\Gamma^{\rm tot}_{h_2}}{\Gamma^{\rm tot}_{h_1}}\, \gg 1\,.
\label{eq:ratio}
\eeq
The SM Higgs has a very narrow width of 0.0068 GeV and due to the limits on the Higgs to invisible branching ratio we know that this width,
$\Gamma^{\rm tot}_{h_1}$, can not
increase by more than 35\%. The reason the Higgs width is so small is due to the fact that all the fermions are coupled to the Higgs via Yukawa couplings 
so we cannot have light fermion with large coupling. The total decay width of the second scalar, $\Gamma^{\rm tot}_{h_2}$, on the other hand 
can easily be large as can be inferred from Fig.~\ref{fig:Gammah2}. Even for $g_{\rm DM}=0.1$ the total width of $h_2$ will be an order of magnitude larger than the $h_1$ Higgs width. 
Hence, for light dark matter, when both channels are open, only the $h_1$ Higgs mediator is relevant.

For heavier dark matter,
$m_{h_1} < 2 m_\chi  < m_{h_2},$
only the $h_2$ channel is open and it is efficiently described by the simplified model,
\beq
\Lagr = \, \sin \theta\,\left(\frac{2M_W^2}{v} \, W^{+}_\mu W^{-\,\mu}\,+\,
\frac{M_Z^2}{v} \, Z_\mu Z^{\mu}  \, -\, \sum_f \frac{m_f}{v}\, \bar{f}f\right)  h_2 
\,-\, g_{\rm DM}\, \bar{\chi}\chi h_2
\,-\, \frac{1}{2}m_{h_2}^2 h_2^2 \,-\, m_{\chi} \bar{\chi}\chi~.
\label{eq:Lagh2}
\eeq
Finally, if dark matter masses are higher than $m_{h_2}$, it cannot be produced via an on-shell mediator exchange, and the resulting
rate of its production is too small to be observed.

Current limits on $\sin^2\theta$ mainly come from two sources, the Higgs signal strengths and the electroweak precision tests, for recent papers see
\cite{Englert:2011yb,Englert:2013gz,Lopez-Val:2014jva,Robens:2015gla,Falkowski:2015iwa}. 
Limits from Higgs signal strength measurements constrain $\cos^2\theta$ directly \cite{Falkowski:2015iwa}. This leads to a bound $\sin\theta<0.44$,
independent of the mass of $h_2$. The electroweak precision tests, mainly the $W$ boson mass give a mass-dependent constraint on $\sin\theta$ 
shown in figure 3 in Ref.~\cite{Falkowski:2015iwa}. In the mass range around 1 TeV the limit becomes $\sin\theta<0.3$. 
We also note that the limits coming from a non-observation of the second SM-Higgs-like state are not directly applicable for $h_2$ in our case 
due to its large branching ratio to invisibles.

\medskip
We will only consider these limits in the context of the singlet-mixing simplified model with the $\kappa$-parameter $\kappa=\sin^2\theta$. 
In the simplified model framework we do not know what other particle content there is and additional degrees of freedom could still modify 
both the SM Higgs signal and the loop corrections to the W-mass. 

Recent discussion of theory models for dark matter based on mass mixing between the scalar singlet mediator and the
SM Higgs can be found in Refs.~\cite{Hambye:2013dgv,Carone:2013wla,Khoze:2014xha,Khoze:2014woa,Altmannshofer:2014vra,Heikinheimo:2013fta}.

\subsection{Generic Higgs-like scalar mediator model}
\label{sec:kappa}

More generally, scalar mediators to dark sector can also arise from an independent additional Higgs doublet or Higgs multiplet, for example in 
the two-Higgs-doublet models.
We choose the simplified model for a generic scalar mediator by assuming that it has the same interactions with the SM vector bosons and fermions as the SM Higgs, but
scaled by an overall scaling factor $\kappa$ which is a free parameter of the simplified model,
\beq
\Lagr = \, \sqrt{\kappa}\,\left(\frac{2M_W^2}{v} \, W^{+}_\mu W^{-\,\mu}\,+\,
\frac{M_Z^2}{v} \, Z_\mu Z^{\mu}  \, -\, \sum_f \frac{m_f}{v}\, \bar{f}f\right)  \phi 
\,-\, g_{\rm DM}\, \bar{\chi}\chi \phi
\,-\, \frac{1}{2}M_{\rm med}^2 \phi^2 \,-\, m_{\chi} \bar{\chi}\chi \,.
\label{eq:Lag}
\eeq
In general, the scalar mediator can couple with a different strength to the SM vector bosons and to SM fermions, thus introducing additional
parameters into the simplified model \eqref{eq:Lag}. For clarity and simplicity we will use the minimal model  \eqref{eq:Lag}
with a single scaling factor. Here $\kappa=1$ corresponds to the normal SM Higgs couplings. 
In general we consider values of $\kappa \lesssim 1$ since it is difficult from a model-building perspective to increase the coupling to gauge bosons with additional Higgs singlets or doublets. 
The simplified model for the more constrained singlet mixing case is described by the same Lagrangian with $\kappa = \sin^2\theta \lesssim 0.15$.
In this simplified model framework we do not introduce a direct coupling between the SM Higgs and $\chi\bar{\chi}$ as this interaction can be easily captured with 
giving $\phi$ the same mass as the Higgs.

\section{Comments on the Relic Density and Direct Detection Constraints}
\label{DMpheno}

Simplified models for dark matter are introduced to capture the main aspects of dark matter collider phenomenology, without being complete models. It is therefore customary not to impose constraints from relic density or direct detection stringently. Still, the model introduced above in Eq.~\eqref{eq:Lag} is a valid model that could have cosmologically viable dark matter. Therefore, to give an indication of constraints for models of this type, we calculate the relic density and direct detection constraints 
assuming that 
\begin{itemize}
\item{(i)} the dark sector fermions $\bar\chi$ $\chi$ which enter the simplified model definition \eqref{eq:Lag} is the {\it cosmologically} stable dark matter and not merely one of the dark sector degrees of freedom which are long-lived on a collider scale;
\item{(ii)} the dark matter particles annihilate predominantly via the mediator interaction specified in \eqref{eq:Lag}, and there are
no other DM annihilation channels beyond the simplified model \eqref{eq:Lag} or that they are highly suppressed.
\end{itemize} 
We stress that if either of these additional assumptions are not satisfied, the relic density- and direct detection-related constraints 
discussed in this section will not apply. These are strong assumptions, that can easily be evaded in many well motivated DM models. 

We will now require that the dark matter does not overclose the Universe, and that the direct detection cross section is sufficiently small to not
having been observed so far. 

We calculate the relic density and direct detection limits using the MadDM~\cite{Backovic:2013dpa,Backovic:2015cra} with the simplified 
model \eqref{eq:Lag}. The computed relic density is compared to the observed relic density from the Plank Satellite \cite{Ade:2013zuv} of 
$\Omega h^2 = 0.1199 \pm 0.0027$, and the direct detection cross sections are compared to the limits from 
the LUX experiment \cite{Akerib:2013tjd}.

\begin{figure}
\begin{center}
\begin{tabular}{cccc}
\hspace{-0.5cm}
\includegraphics[width=0.5\textwidth]{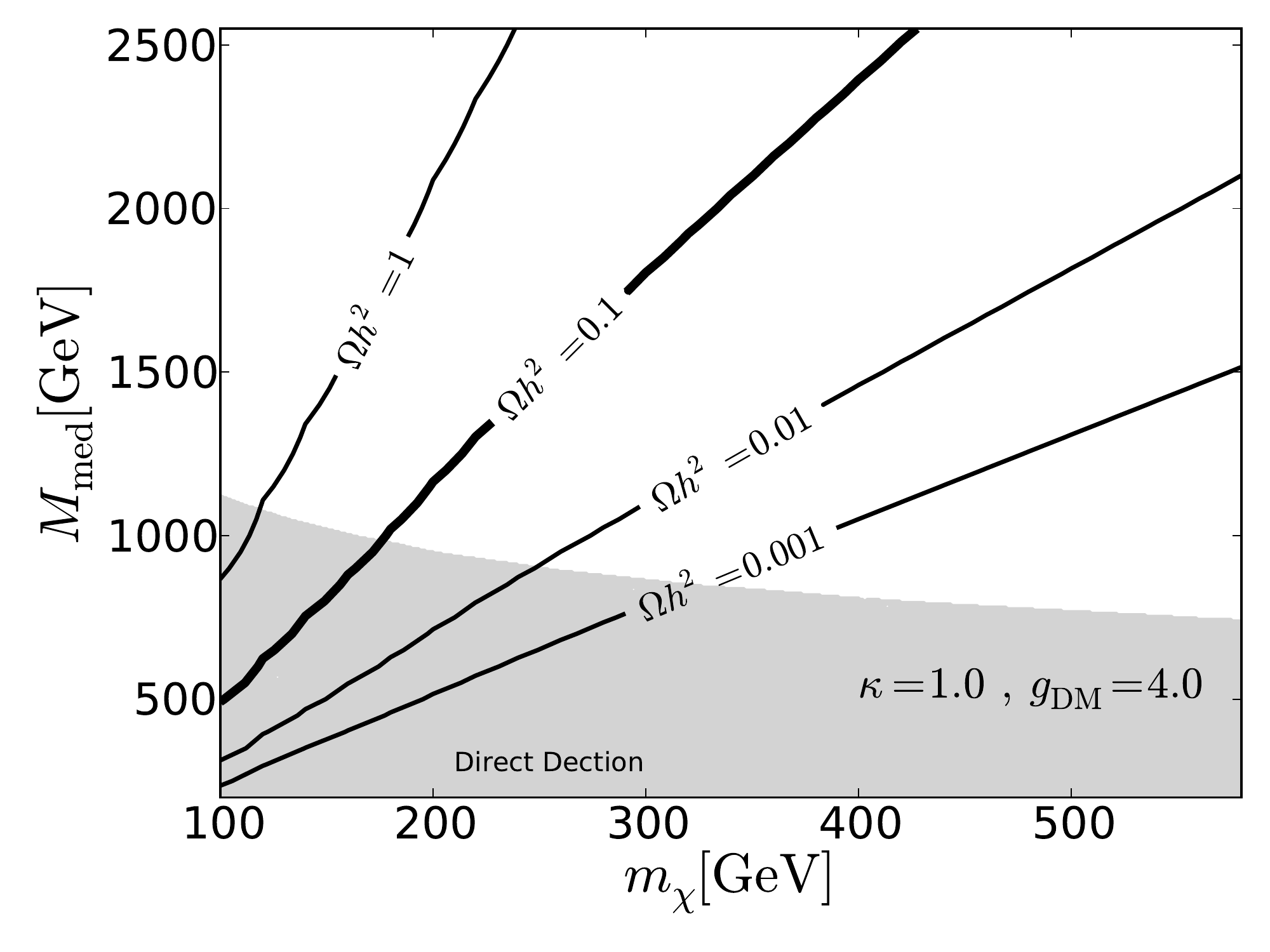}
&
\hspace{.1cm}
\includegraphics[width=0.5\textwidth]{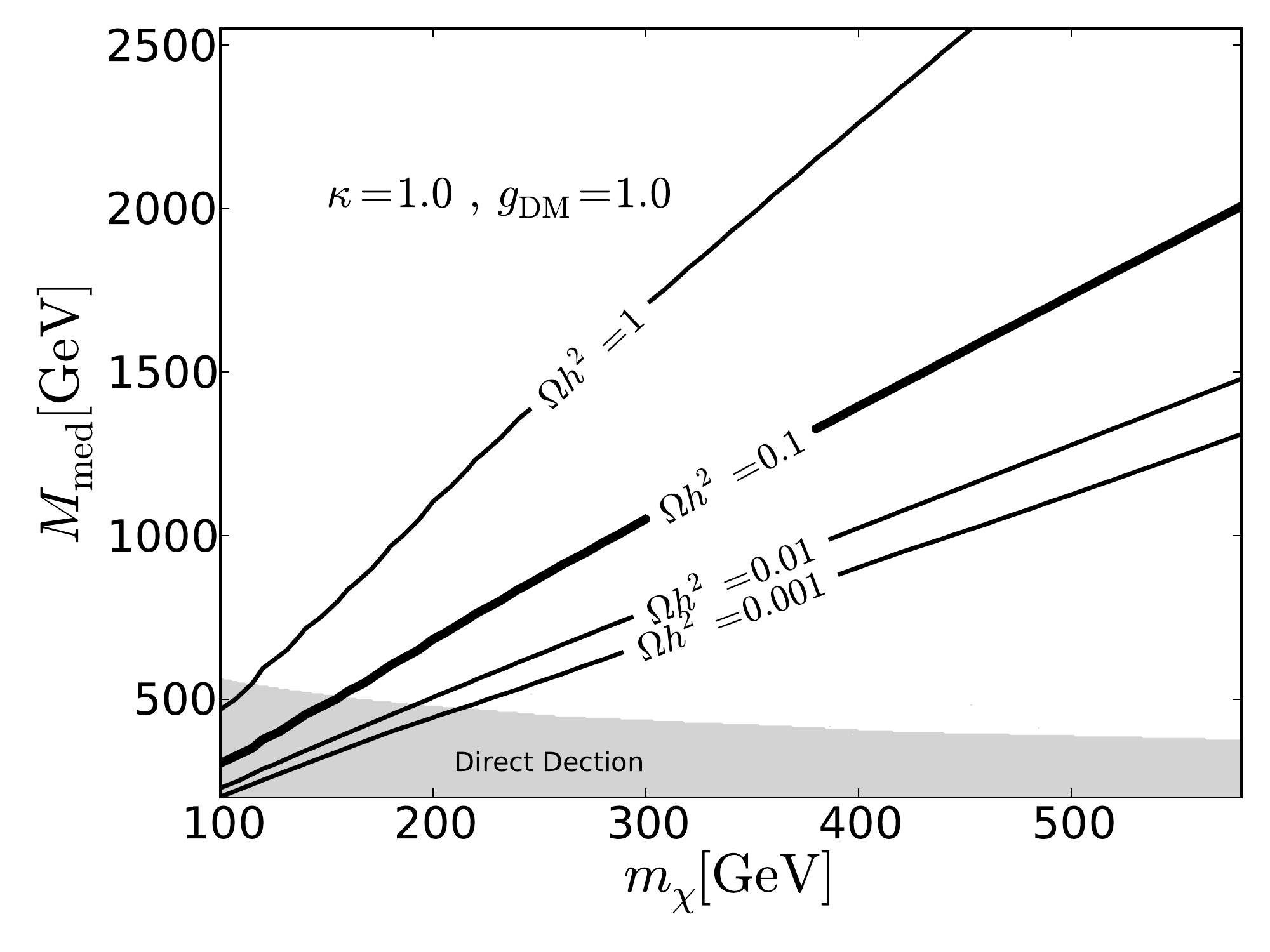}
\\
\hspace{-0.5cm}
\includegraphics[width=0.5\textwidth]{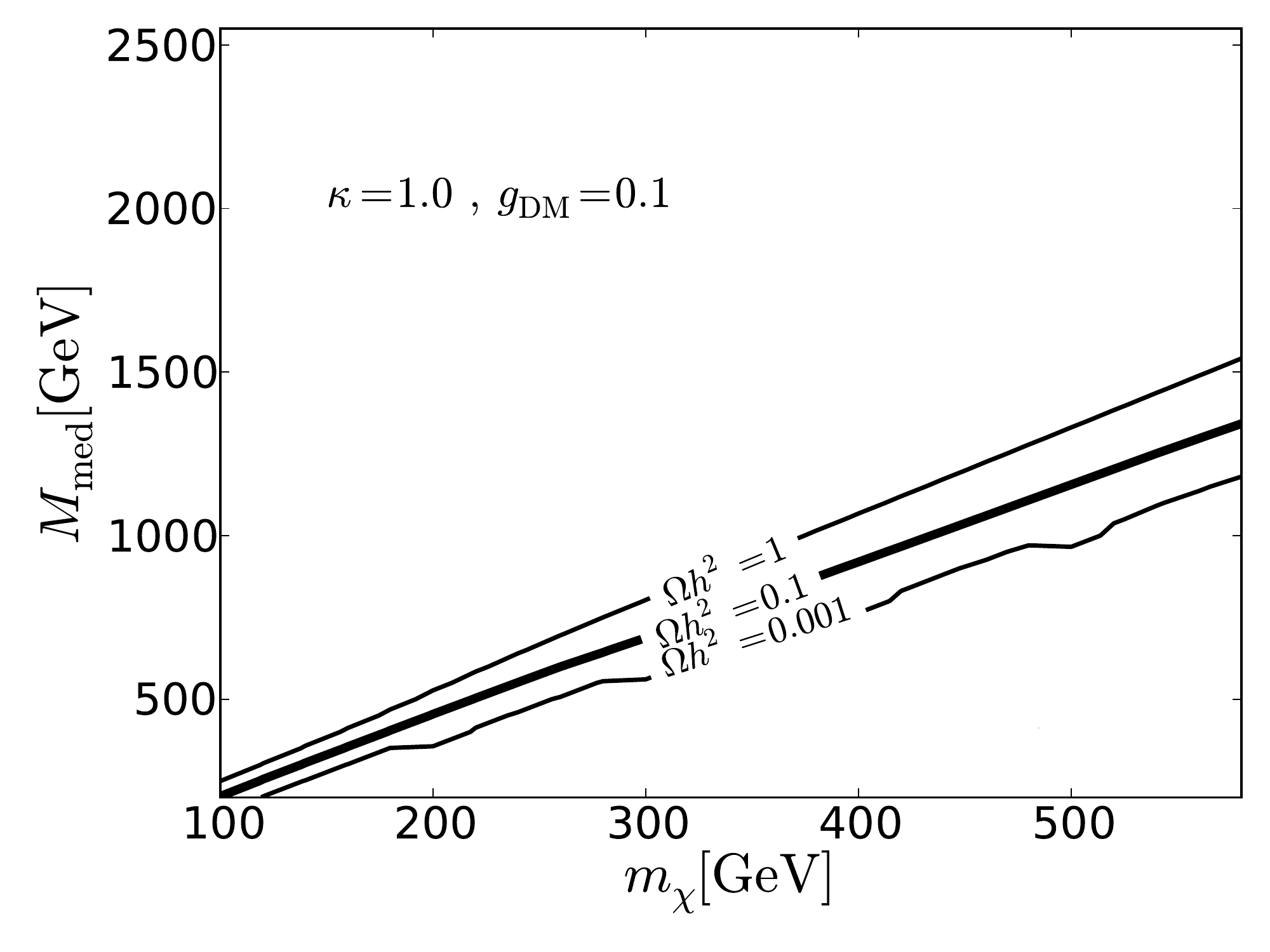}
&
\hspace{.1cm}
\includegraphics[width=0.5\textwidth]{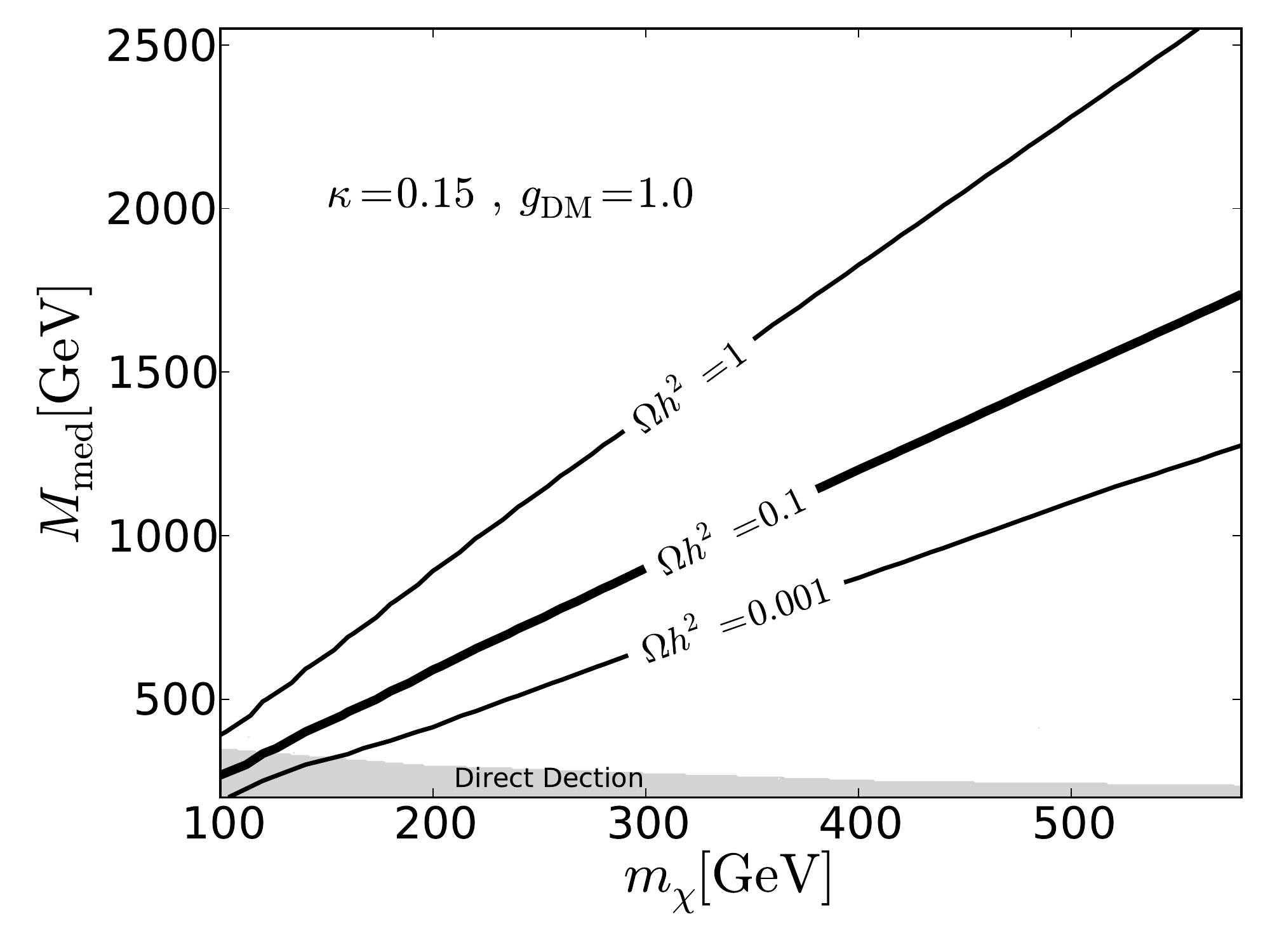}
\\
\end{tabular}
\end{center}
\vskip-0.4cm
\caption{
Dark matter relic density and direct detction constraints for our simplified model for dark matter for various values of $g_{\rm DM}$ 
and $\kappa$. The lines give relic density contours and the grey region shows the area exclude by direct detection constraints.
}
 \label{fig:DMpheno}
\end{figure}

Figure \ref{fig:DMpheno} shows the contours of the computed relic density and the direct detection exclusions on the
mediator mass -- dark matter mass plane and for various values of $g_{\rm DM}$ and 
$\kappa$. For the direct detection constraint we have assumed that the DM density interacting with the detector is given by the canonical value, even if the DM in our model only is a sub-component of the total DM density in this region of parameter space. Therefore, the direct detection limits on our model are weaker than what is shown in the figure in the region of parameter space where the calculated DM density is smaller than the observed value. 

For the collider phenomenology at the LHC and at future colliders, we will be interested in heavy mediators with the dark matter mass and the dark matter coupling largely unconstrained as long as the scalar mediators have a large branching ratio to dark matter. From Figure \ref{fig:DMpheno} we can see that all these models easily avoid direct detection constraints, and as long as the dark matter mass is quite heavy we can have mediator masses up to 2500 GeV, without over-closing the Universe. For $g_{\rm DM}=4$ and a heavy mediator 
$M_{\rm med} \simeq 2.5$ TeV we need $m_\chi \gtrsim 400$ GeV to have viable dark matter 
(another way to put it is that only the DM which is more than 20 times lighter than the mediator is constrained here). 
For smaller couplings the minimal DM mass increases  accordingly (as can be seen from the second and third plots in Fig.~\ref{fig:DMpheno})
to not overclose the universe for the heaviest mediators, but this is not a problem for the models we will consider in the rest of the paper. 
We will therefore now turn to collider phenomenology where we will study models that, if we interpret them as complete models, can provide a viable dark matter candidate. 

 We conclude that the relic density and direct detection considerations can provide useful constraints on our simplified model 
 under certain assumptions, in addition to the collider searches.
This provides an important complementarity to the collider phenomenology we will now study. 
If the LHC or future colliders can resolve and probe the mediator mass-scale and a signal with missing energy is discovered, 
one of the main open questions will be if the signal results in the production of cosmological dark matter and what is its particle identity.

\section{Collider limits on scalar mediators with 2 jets and missing transverse energy at the LHC}
\label{sec:LHC}

For deriving the collider limits on the models with scalar mediators to dark matter sectors and for the ability to distinguish between models
with different mass scales, we will use the search strategy based on final states with missing transverse energy plus two jets.
There are four main kinematic quantities associated with the $\missET$ plus 2 jets signatures:
the missing transverse momentum $\misspT$, the jets invariant mass $M_{jj}$, the azimuthal angle between the tagging jets $\Delta \phi_{jj}$ and the jets
pseudo-rapidity difference $\Delta \eta=\eta_{j1}-\eta_{j2}$. In terms of these we impose the VBF cuts \cite{Eboli:2000ze,Bernaciak:2014pna},
\beq
\misspT>100 \,\text{Gev}\,, \quad M_{jj}>1200\, \text{GeV}\,, \quad \Delta \phi_{jj}<1\,, \quad  \Delta \eta > 4.5 \,,
\quad p_{T,j} >40\, \text{GeV}\,, 
\label{eq:cut1}
\eeq
to separate the signal and background. Here $p_{T,j}$ is the transverse momentum of each jet defined by using the anti-kt jet algorithm with $R=0.4$. We reconstruct jets using Fastjet \cite{fastjet}.
After imposing these cuts, the main production channel of the scalar mediator is largely reduced
to weak vector boson fusion (WBF), leaving only a small contribution from the gluon fusion (GGF) channel ({\it cf.} Table~1).
In spite of the relative smallness of the GGF process after cuts \eqref{eq:cut1}, one should not be tempted to approximate them by the Higgs-gluon effective vertex.
The inclusion of finite top quark mass effects in the top-loop in the GGF production is known to be important (in the context of DM searches at the LHC
this was emphasised in \cite{Haisch:2012kf}), especially for heavier scalar mediators
where the heavy top mass approximation breaks down.
We therefore simulate both the WBF and GGF contributions to the signal with VBFNLO \cite{Baglio:2014uba,Arnold:2011wj,Arnold:2008rz} which includes the 
full top-loop dependence to GGF.

The background is simulated at leading order using MadGraph\cite{Alwall:2011uj}. Both signal and background are then showered with Herwig++ 
\cite{Bellm:2013lba}.
The main backgrounds are Z + 2 jets with the Z decaying to neutrinos and W$^\pm$ + 2 jets where the W decays to a neutrino and a missing lepton. We count the lepton as missed if it has $|\eta_l| > 2.5$ or $p_T<10$ GeV. We have also checked that the $\bar{t}t$ background is negligible after the cuts.
The projected LHC exclusion limits for these final states have been studied previously in \cite{Eboli:2000ze,Bai:2011wz,Ghosh:2012ep,Bernaciak:2014pna} in the context of an invisible
 branching ratio for the SM Higgs.

\subsection{Width effect on differential distributions}
\label{sec:width}
\begin{figure}
\begin{center}
\begin{tabular}{cccc}
\hspace{-0.5cm}
\includegraphics[width=0.5\textwidth]{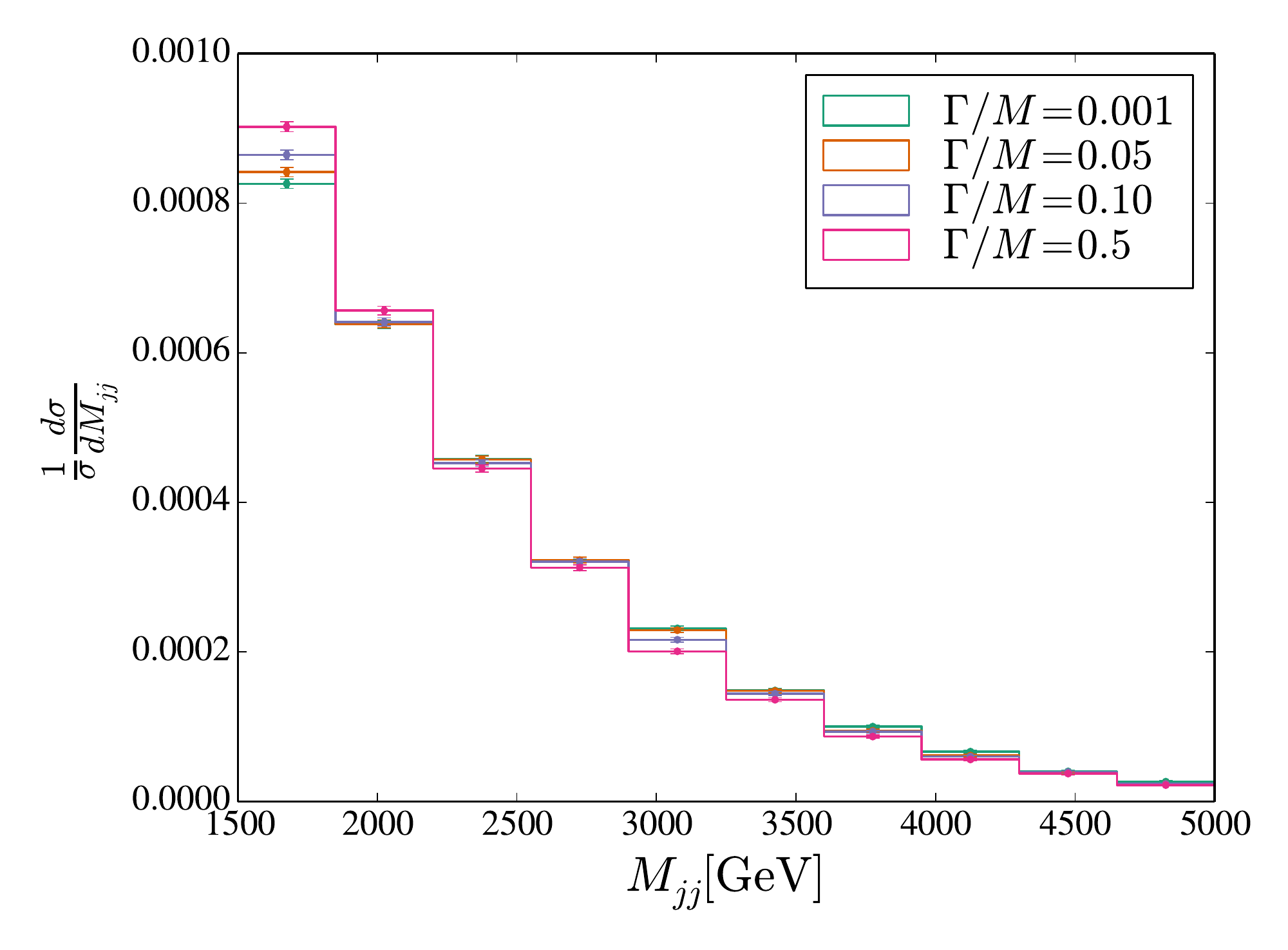}
&
\hspace{.1cm}
\includegraphics[width=0.5\textwidth]{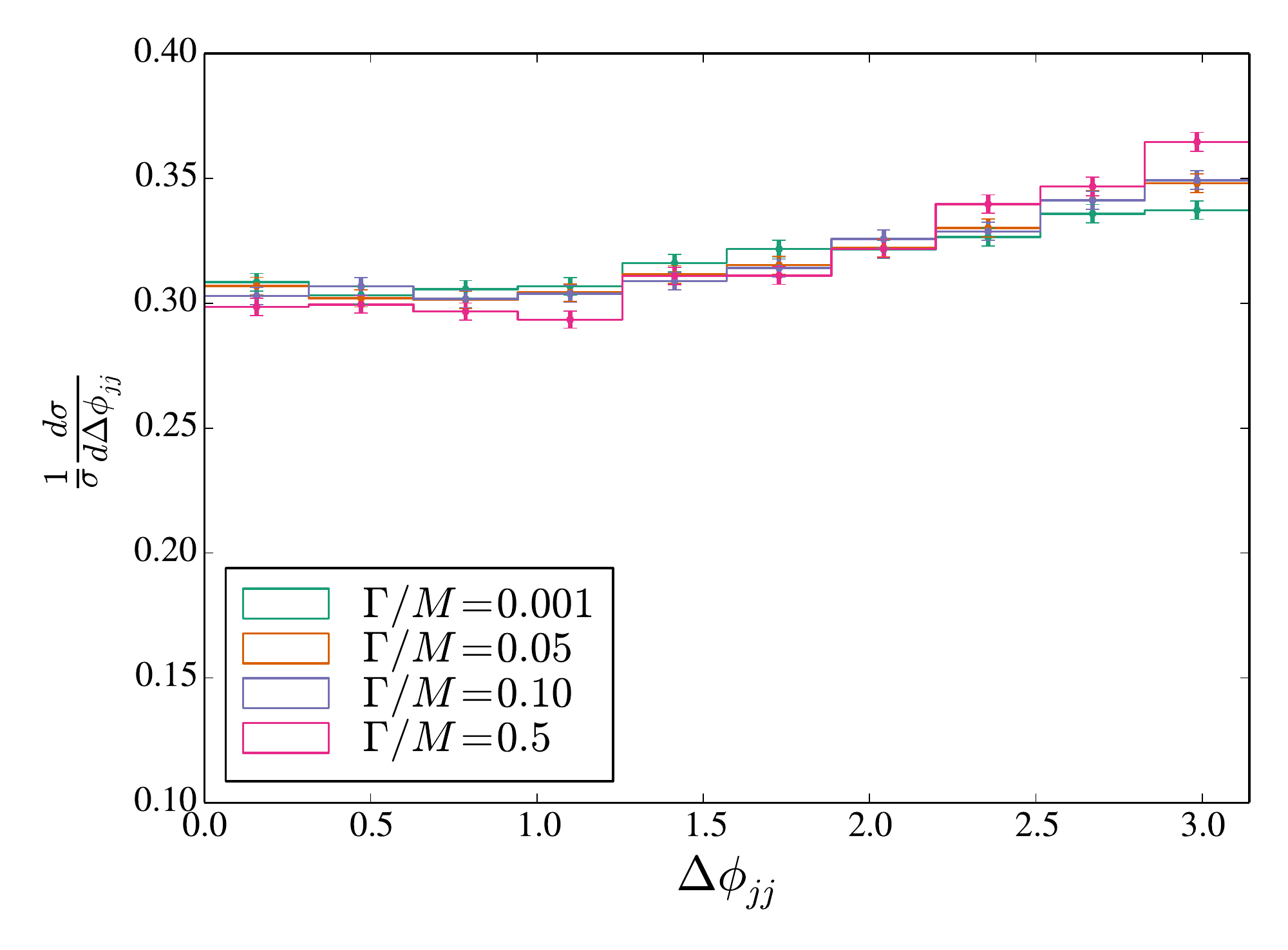}
\end{tabular}
\end{center}
\vskip-0.4cm
\caption{
Kinematic distributions for different values of the mediator width at $\sqrt{s} = 13$ TeV when $M_{\rm med}=800$ GeV. }
 \label{fig:width}
\end{figure}

In Figure~\ref{fig:width} we can see the effect of varying the width of the mediator on the differential distributions of $M_{jj}$ and $\Delta \phi_{jj}$ for a mediator with $M_{\rm med}=800$ GeV. A smaller width leads to a slightly broader $M_{jj}$ tail and flatter $\Delta \phi_{jj}$ distribution. For reasonably small total widths this effect is not very large. We will therefore use the narrow width approximation, where we produce the mediator on-shell with subsequent decay to $\bar{\chi}\chi$ with a branching ratio determined by the coupling constants and dark matter mass, when we simulate the signal.

 \subsection{The LHC exclusion limits reach}

Our first goal is to establish the projected LHC exclusions for models with scalar mediators based on the 2 jets and $\missET$ final states. 
We aim to evaluate the upper limit on the mediator mass for the model to be within the LHC reach.

The left panel in Table~1 
shows the cross-sections for the signal at the LHC at the 13 TeV center-of-mass energy, assuming a 100\% branching 
ratio of the scalar mediators to dark matter and $\kappa=1$. The cross-sections for SM backgrounds are shown in the table on the right.
Using these one can calculate the simple projected exclusion limits for these models from a standard cut-and-count procedure.

\medskip
\begin{center}
\begin{tabular}{cc}
\begin{tabular}{ l | c | c | c}
  \hline                       
  $M_{\rm med}$ & VBF & GGF &Total  \\
  125 GeV &89&17&107  \\
  250 GeV &61&13&74  \\
  500 GeV &26&10&36  \\
  750 GeV &12&3&15  \\
  1000 GeV &6&0.7&6.7  \\
  1500 GeV &2&0.08&2.01  \\
  \hline 
\end{tabular}
\hspace{.5cm}
&
\hspace{.5cm}
\begin{tabular}{ l |  l}
  \hline                       
  Background & Cross-section(fb)  \\
   Zjj&128  \\
   W$^+$jj&116  \\
   W$^-$jj&40  \\
  \hline 
\end{tabular}
\end{tabular}
\end{center}
\centerline{Table 1. Cross-sections (fb) at partonic level after VBF cuts in \eqref{eq:cut1} at 13 TeV. }

\medskip

 \begin{figure}
\begin{center}
\begin{tabular}{cccc}
\hspace{-0.5cm}
\includegraphics[width=0.5\textwidth]{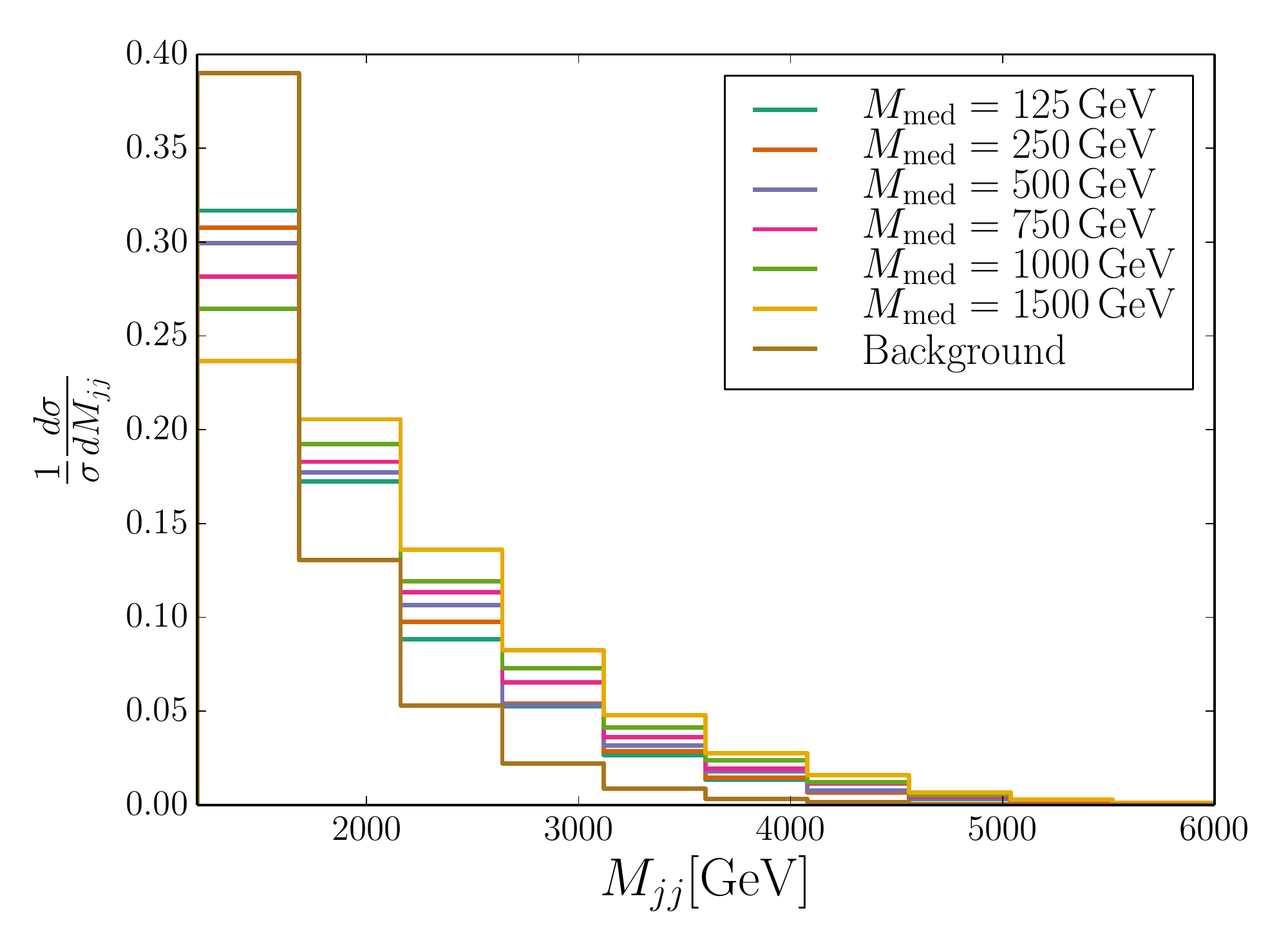}
&
\hspace{.1cm}
\includegraphics[width=0.5\textwidth]{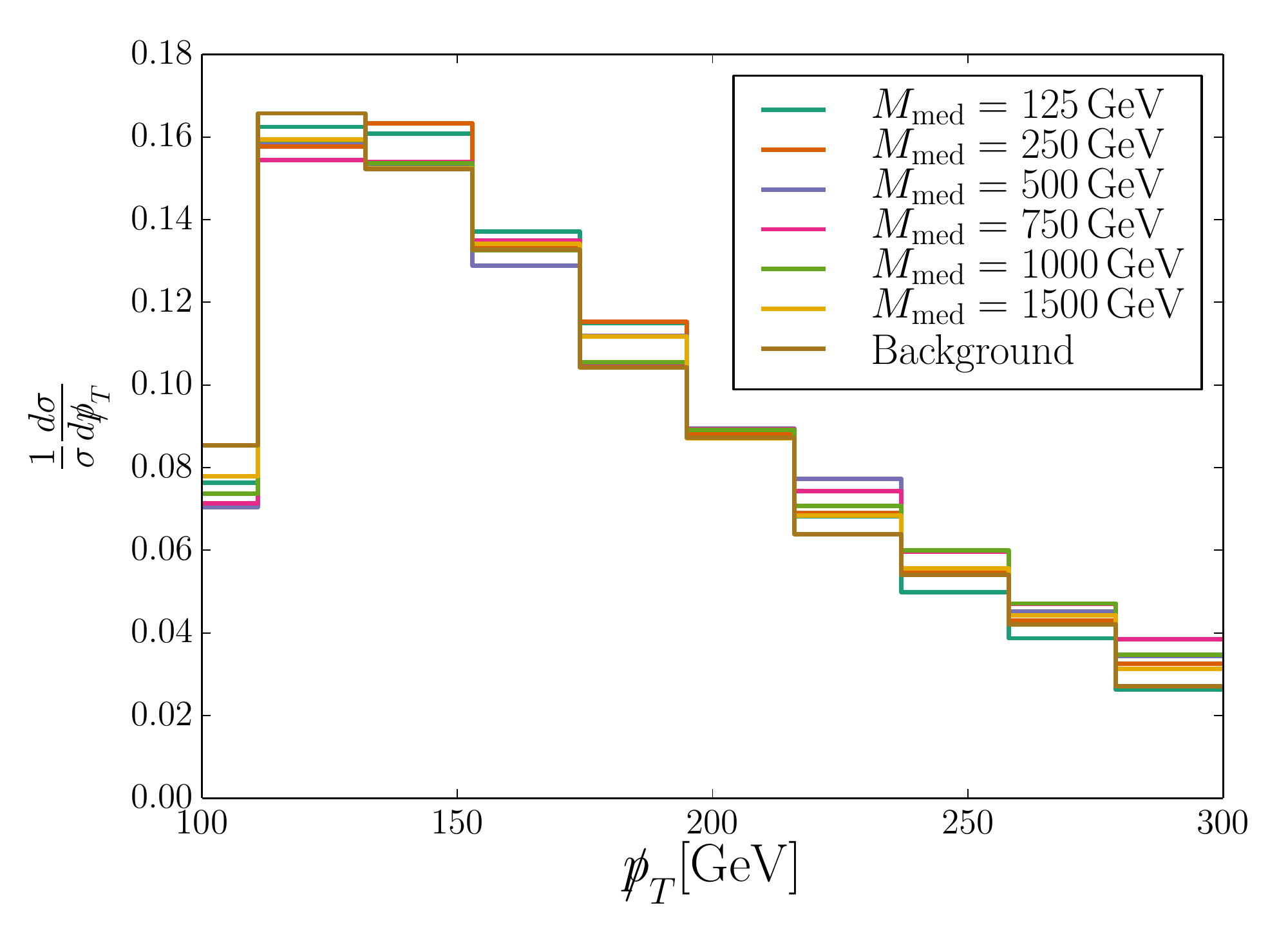}
\\
\hspace{-0.5cm}
\includegraphics[width=0.5\textwidth]{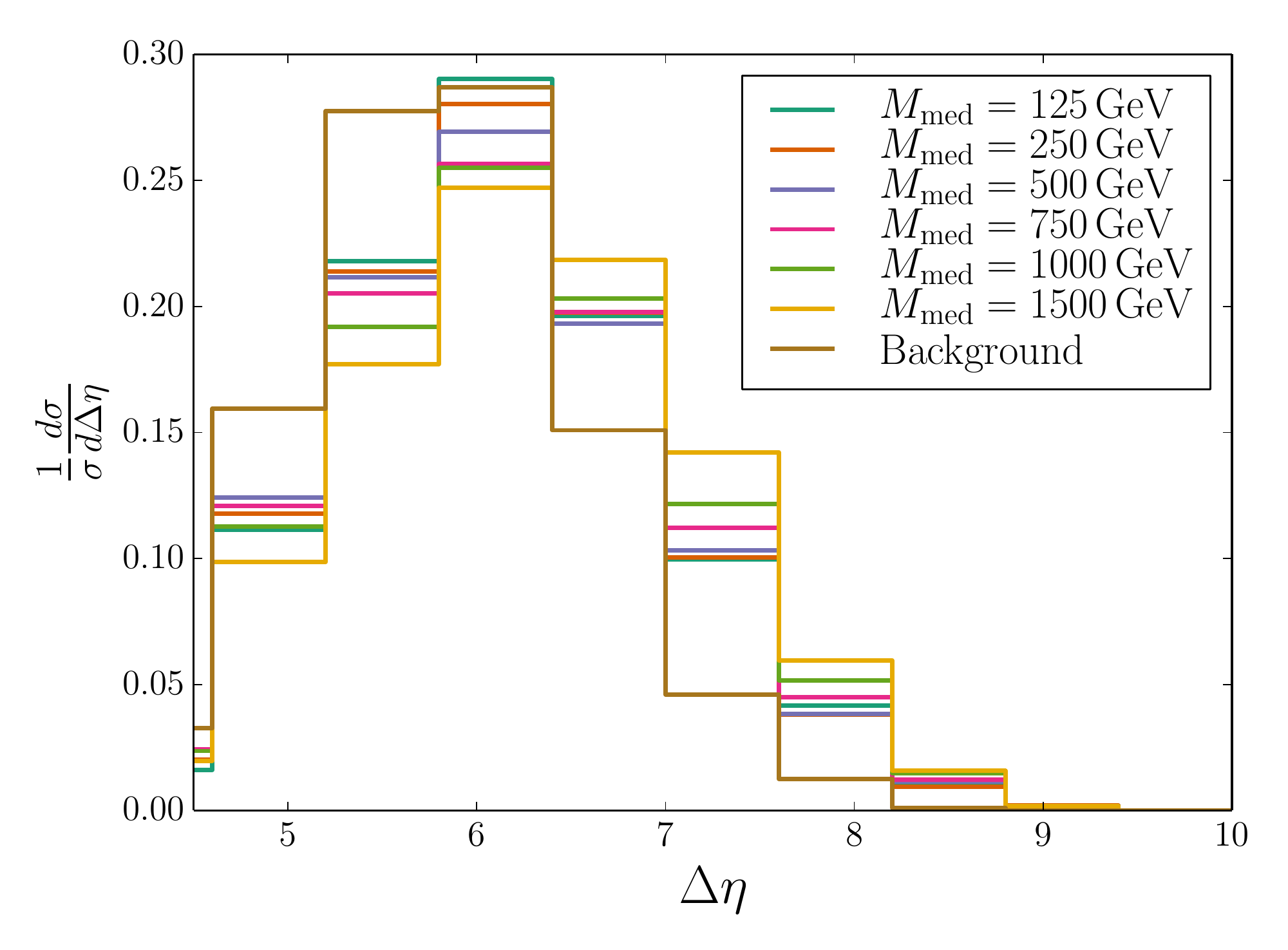}
&
\hspace{.1cm}
\includegraphics[width=0.5\textwidth]{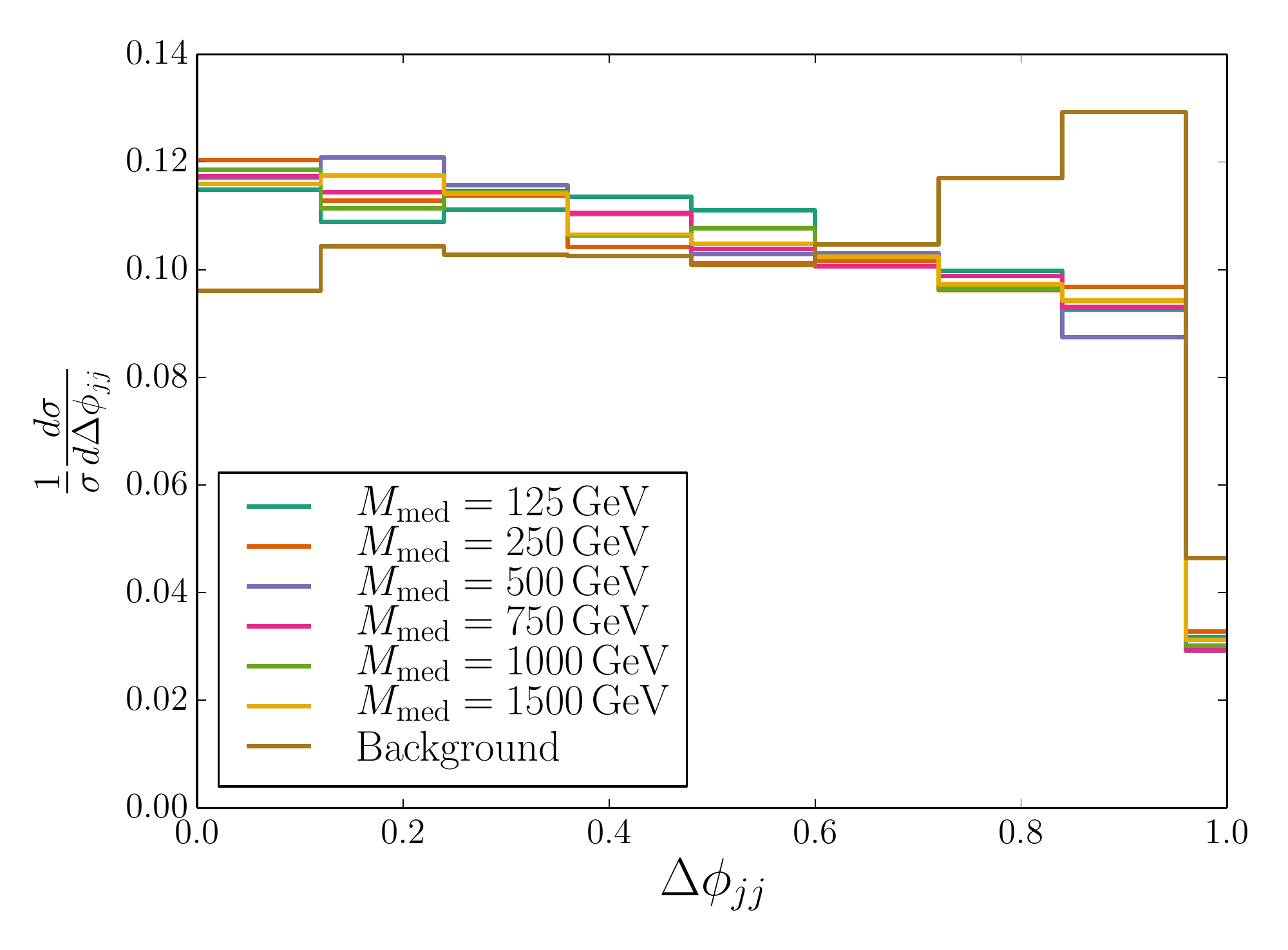}
\\
\end{tabular}
\end{center}
\vskip-0.4cm
\caption{
Kinematic distributions for different values of the mediator mass for the signal, and for the background at the LHC. $M_{jj}$ distributions
are shown on the top left panel, $\misspT$ is on top right right, $\Delta \eta$ and $\phi_{jj}$ distributions are on the bottom left and right panels respectively.
}
 \label{fig:Mjj}
\end{figure}
 \begin{figure}
\begin{center}
\begin{tabular}{cc}
\hspace{-0.5cm}
\includegraphics[width=0.5\textwidth]{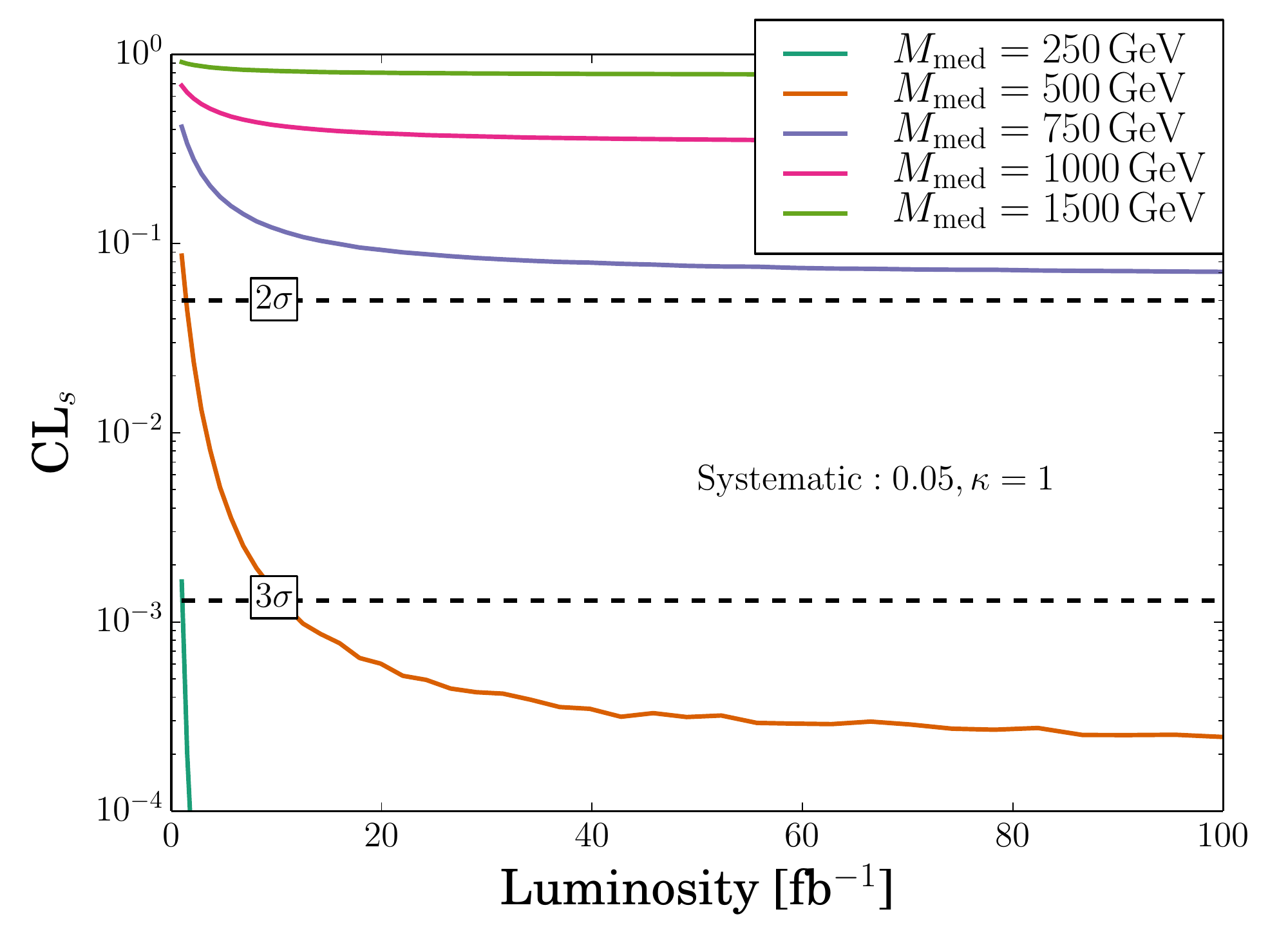}
&
\hspace{.1cm}
\includegraphics[width=0.5\textwidth]{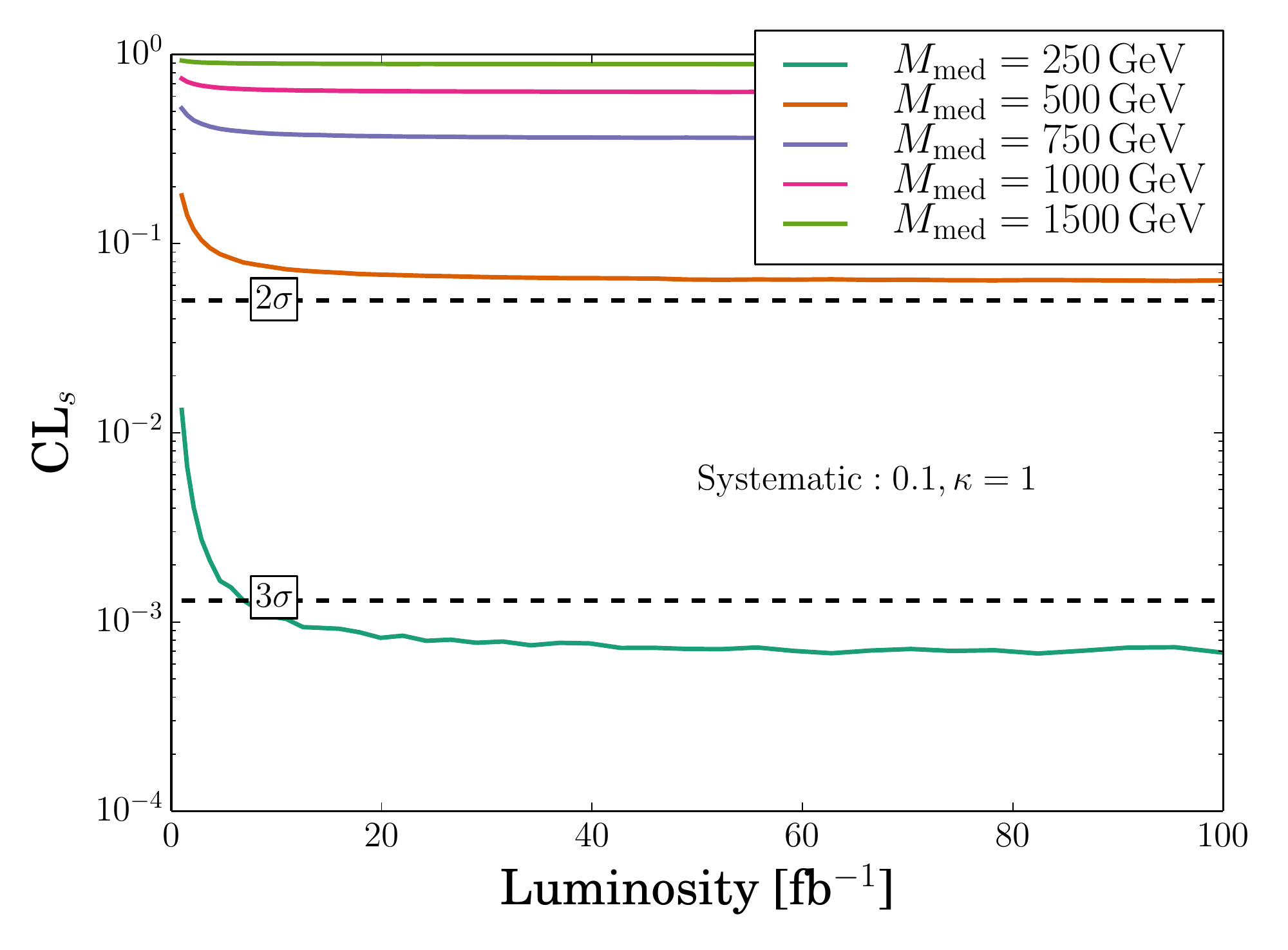}
\\
\end{tabular}
\end{center}
\vskip-0.4cm
\caption{We characterise the LHC reach for models with different values of $M_{\rm med}$ by computing
confidence levels for excluding signals from the SM backgrounds. We consider models with $\kappa=1$
and on the left panel use a systematic uncertainty of 5\%,  the panel on the right 
corresponds to 10\% systematic uncertainty. }
 \label{fig:reach1}
 \end{figure}
 \begin{figure}
\begin{center}
\begin{tabular}{cc}
\hspace{-0.5cm}
\includegraphics[width=0.5\textwidth]{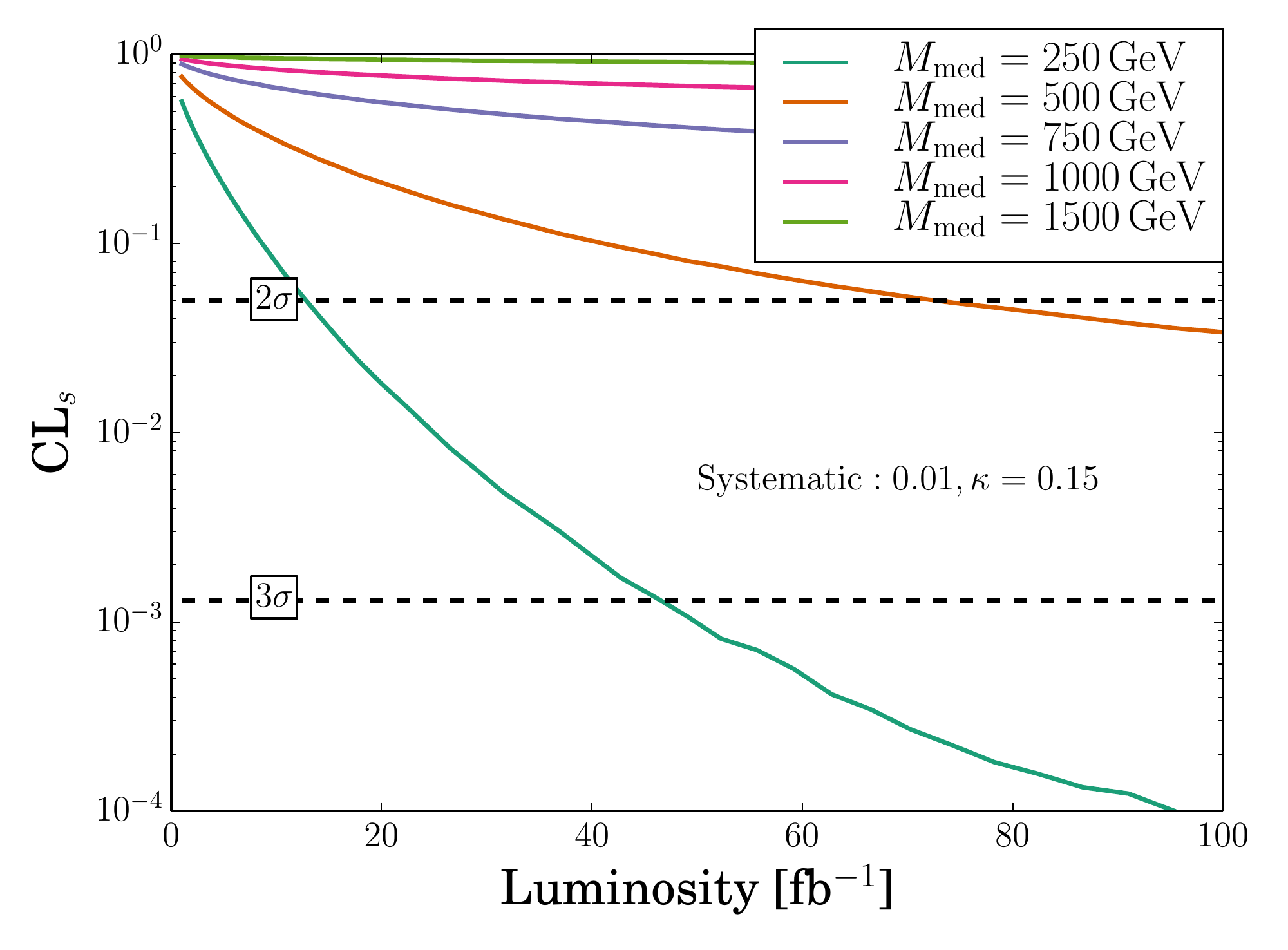}
&
\hspace{.1cm}
\includegraphics[width=0.5\textwidth]{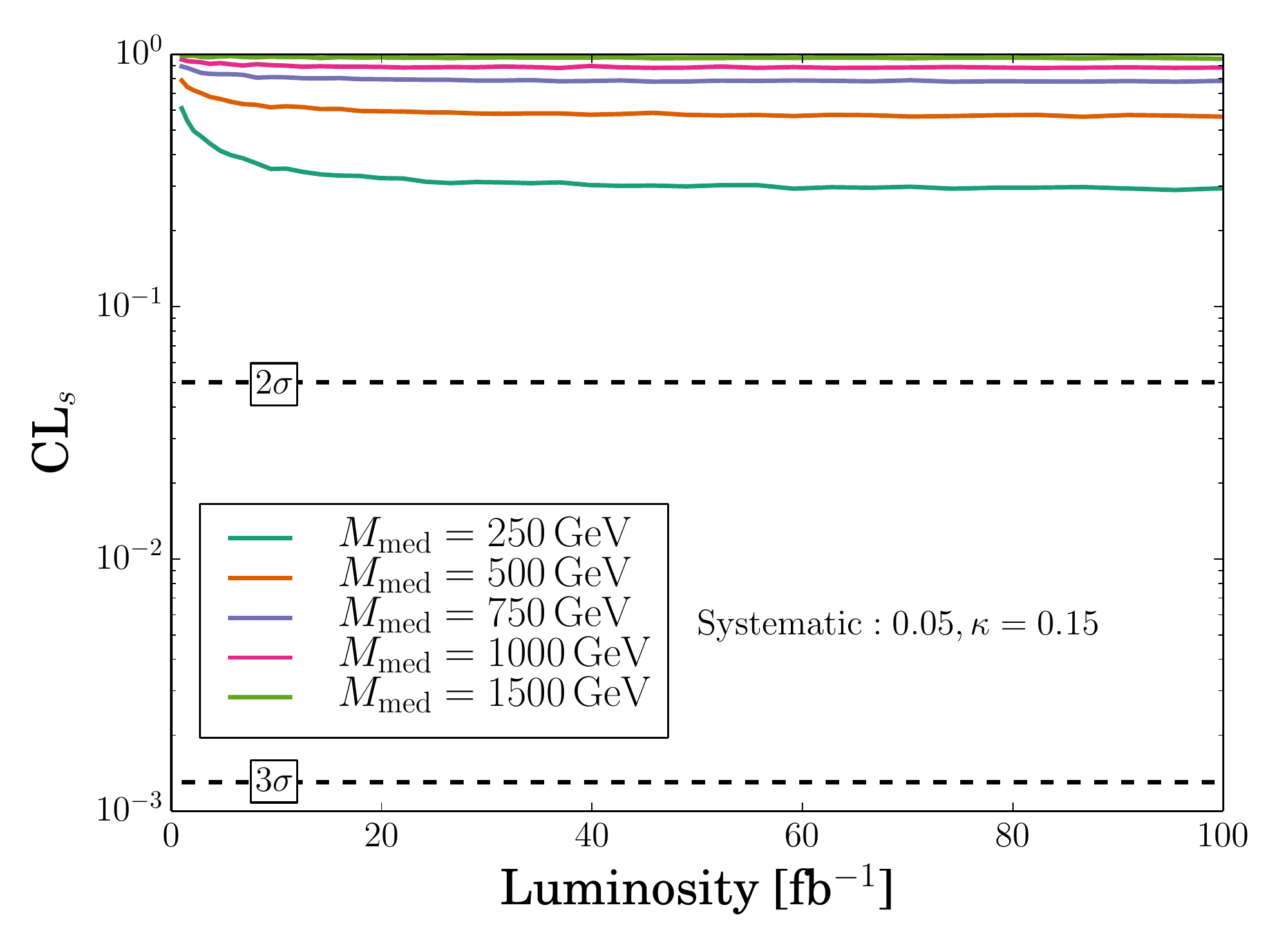}
\\
\end{tabular}
\end{center}
\vskip-0.4cm
\caption{
The LHC reach for different $M_{\rm med}$ models with $\kappa=0.15$ in terms of confidence levels to exclude signal from SM background. 
On the left panel we use a systematic uncertainty of 1\%, and the panel on the right 
corresponds to 5\% systematic uncertainty. }
 \label{fig:reach015}
 \end{figure}
 \begin{figure}
\begin{center}
\begin{tabular}{cc}
\hspace{-0.5cm}
\includegraphics[width=0.5\textwidth]{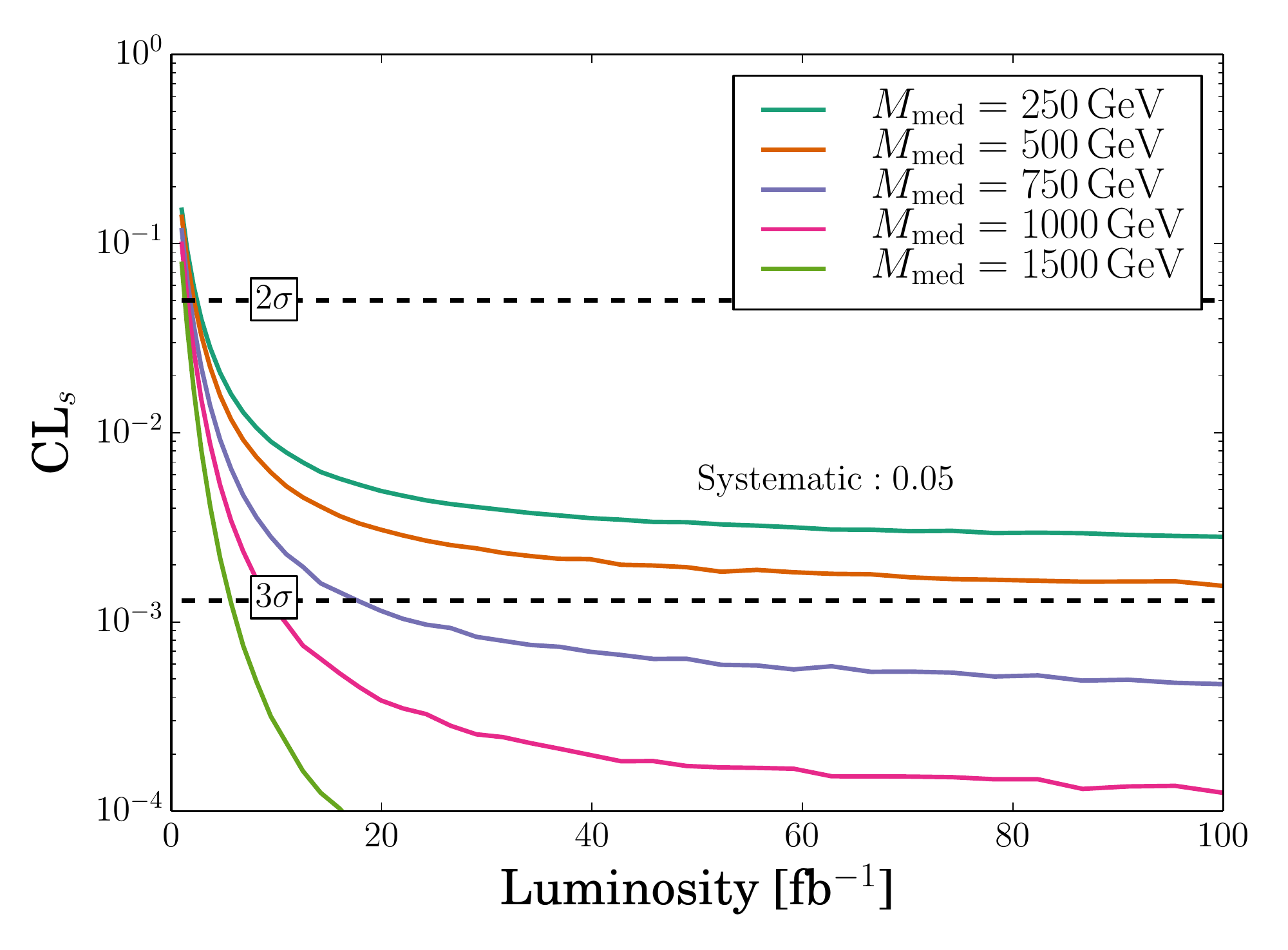}
&
\hspace{.1cm}
\includegraphics[width=0.5\textwidth]{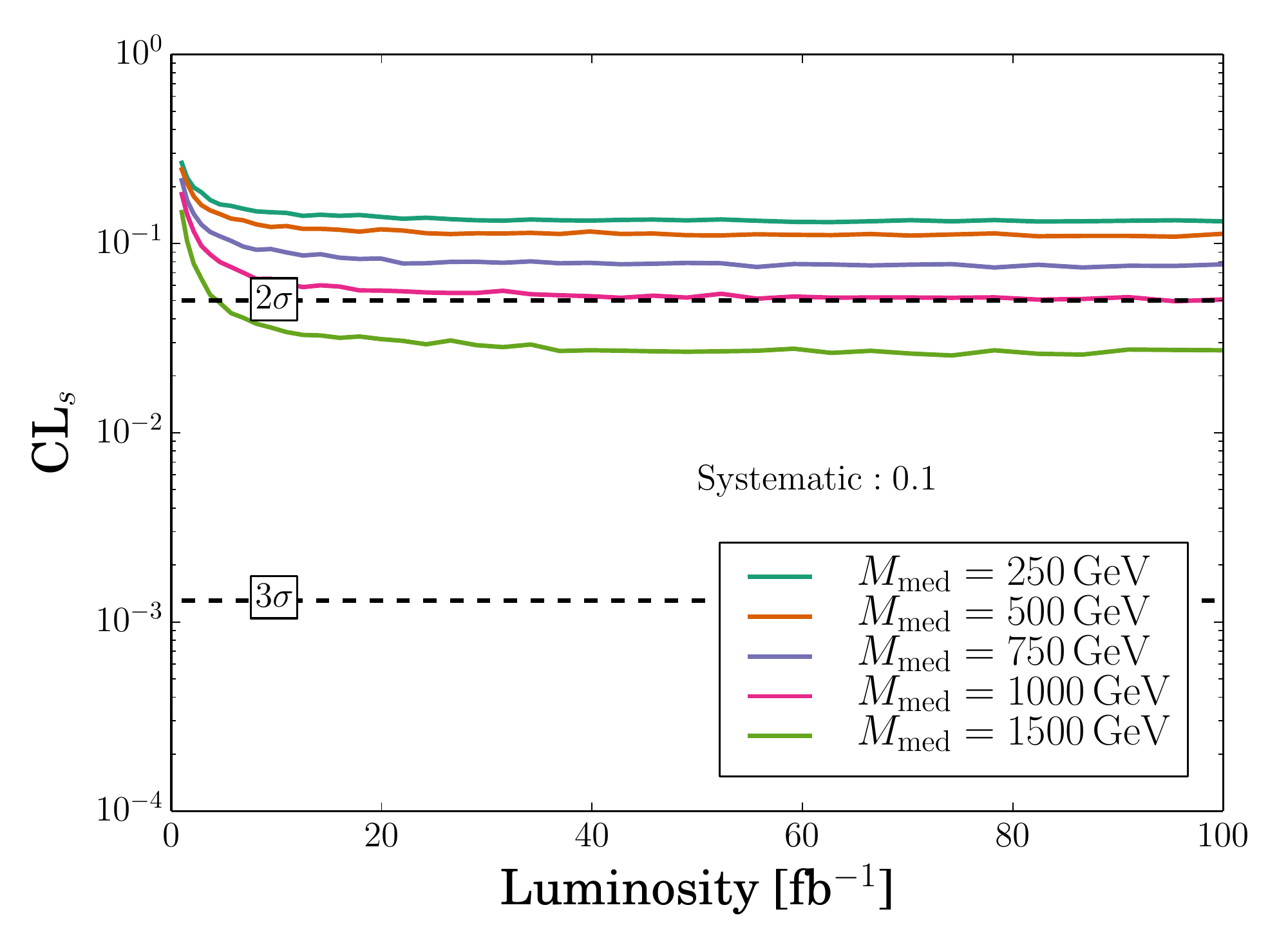}
\\
\end{tabular}
\end{center}
\vskip-0.4cm
\caption{
The LHC reach for different $M_{\rm med}$ models. We normalise the cross-sections for all models to the SM Higgs cross-section 
with ${\rm Br}_{\rm inv} =30$\% and a systematic uncertainty of 5\% (left panel) and 10\% (right panel).
}
 \label{fig:reach_equal}
 \end{figure}

For our analysis we will use the differential cross-sections to perform a binned log-likelihood analysis \cite{Junk:1999kv} 
to compute confidence levels (CLs) for experimental exclusions
 \cite{Read:2002hq}.
In the four plots of  Fig.~\ref{fig:Mjj} we show the normalised differential distributions for signal and background as functions of the four kinematic variables
$M_{jj},\, \misspT,\,\Delta \eta, \,\Delta \phi_{jj}$.
These kinematic distributions are plotted for different values of the mediator mass ranging from $M_{\rm med}=125$ GeV to 1500 GeV\footnote{Compared to differential distributions at $e^+e^-$ colliders \cite{Andersen:2013rda,Chacko:2013lna}, at the LHC differences between the models are less pronounced and more difficult to exploit.}.

The differences in their shapes for models with different values of $M_{\rm med}$ can be used for differentiating between 
them.
The  binned log-likelihood technique for computing confidence levels 
is based on regarding each bin in a histogram for the measured variable as an independent search channel to be combined with all others. Systematic
uncertainty is taken into account by running many pseudo Monte Carlo
experiments where the normalisation of the background histogram is
varied randomly. The significance is then given by the fraction of these
experiments that has a smaller likelihood ratio than that for the
expected background distribution. We use the $M_{jj}$ distribution(the
$\Delta \eta$ distribution gives similar results) with ten bins to both
determine the signal exclusion limits and later to distinguish between
different signal models.

In Figures~\ref{fig:reach1}-\ref{fig:reach_equal}  we show the LHC reach for excluding scalar mediator models for different values of mediator masses.
Figure~\ref{fig:reach1} applies to generic models with $\kappa=1$ and assumes a 5\% and a 10\% level of systematic uncertainty.
Figure~\ref{fig:reach015} shows the LHC exclusion contours in the context of the mediator-Higgs mixing models, here we set $\kappa=0.15$
and assumes a 1\% and a 5\% systematic uncertainty.
Plots in Fig.~\ref{fig:reach_equal} show the LHC exclusion limits without fixing the $\kappa$ parameter to a specific value.
Here one allows $\kappa$ to float such that for each model the computed cross-section is set equal to a cross-section that corresponds to a 30\% 
invisible branching ratio for the 125 GeV Higgs.

The conclusions we draw is  that for generic scalar mediator models with $\kappa \simeq 1$, with the 13 TeV LHC we can probe models
with mediator masses  up to 
$M_{\rm med}\approx 750$ GeV (assuming a 5\% level of systematic uncertainty). For the models with small $\kappa$, in particular the models associated with the Higgs--singlet-mediator mixing where
$\kappa=\sin^2\theta \lesssim 0.15$,  we can probe up to $M_{\rm med}\approx 500$ GeV (with an optimistic 1\% systematic uncertainty). 
The decrease in cross-section at small values of $\kappa$ 
not surprisingly makes it very hard to reach to the higher mediator masses in the Higgs portal-type mixing model realisations at the LHC.
While we only focus on CP-even scalar mediators, the same distributions can be used to access the scale of CP-odd scalar mediators. 
$\Delta \phi_{jj}$  can be a powerful observable to discriminate CP-even from CP-odd mediators \cite{Plehn:2001nj,Hankele:2006ja,Dolan:2014upa}.  

 \begin{figure}[h!]
\begin{center}
\begin{tabular}{cc}
\hspace{-0.5cm}
\includegraphics[width=0.5\textwidth]{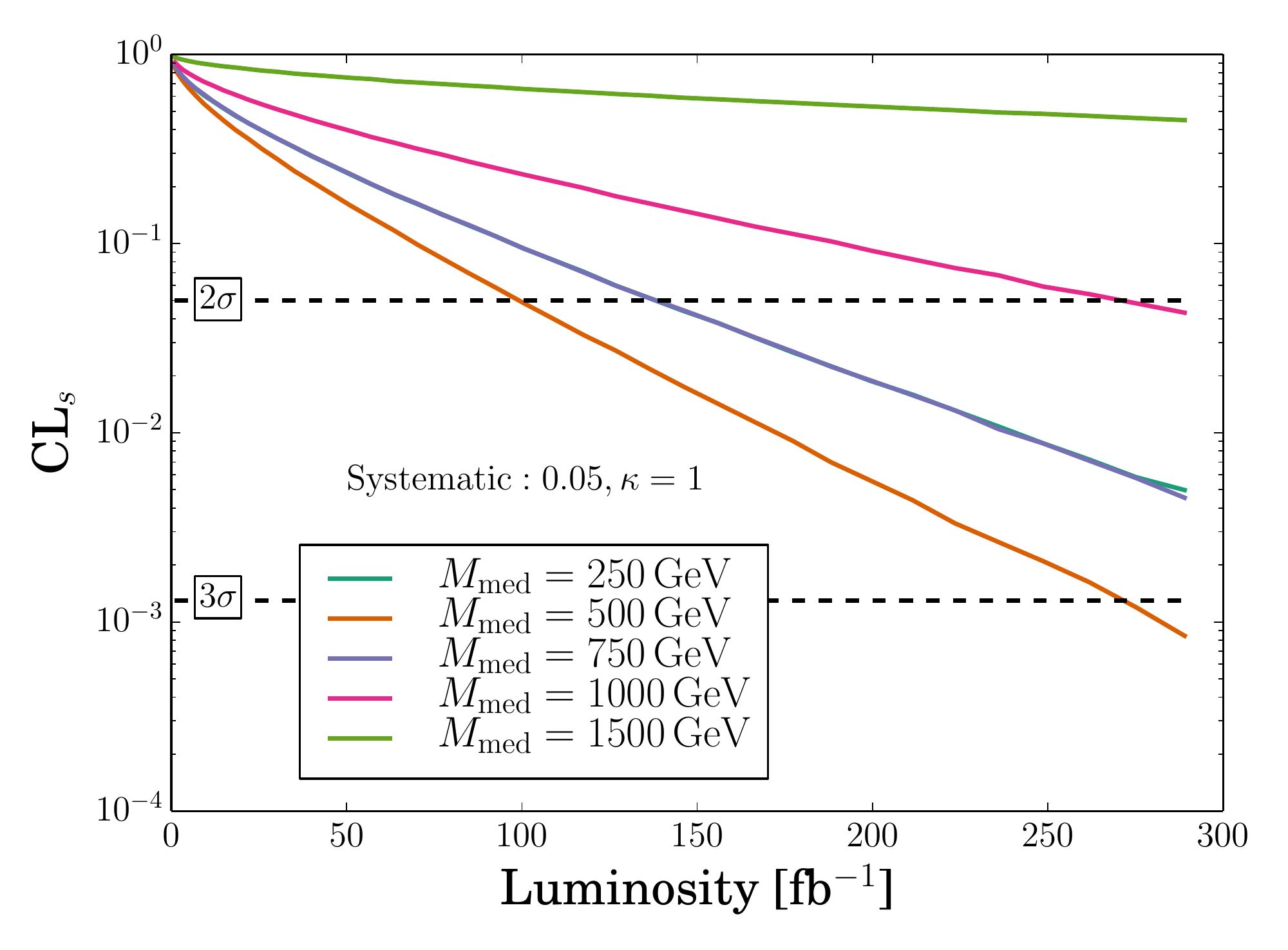}
&
\hspace{.1cm}
\includegraphics[width=0.5\textwidth]{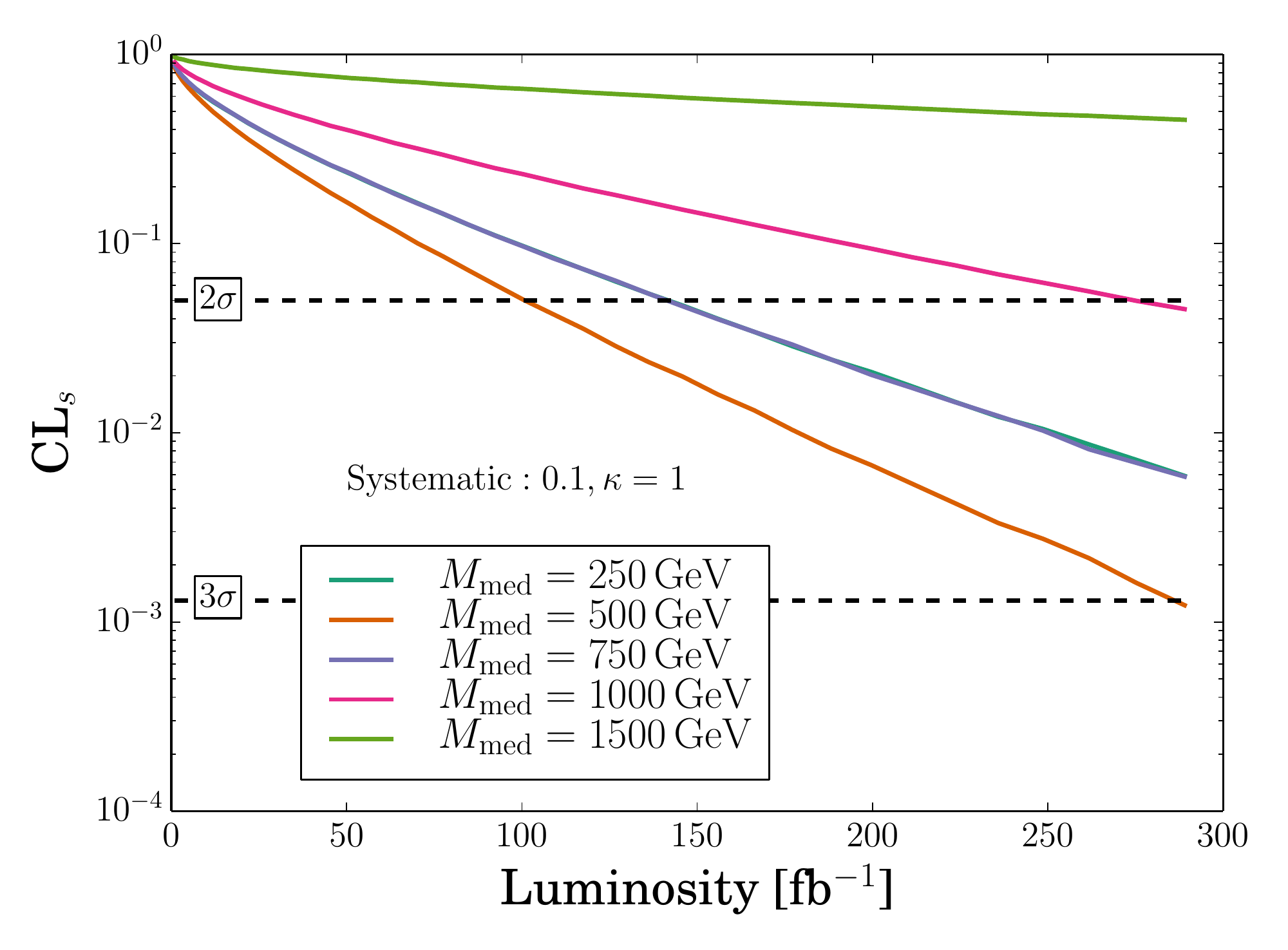}
\\
\end{tabular}
\end{center}
\vskip-0.4cm
\caption{
Differentiating the models at $\kappa=1$ at the LHC. For each value of $M_{\rm med}$ between $250$ and $1500$ GeV the models are compared to the reference model with the 125 GeV mediator.  We assume a systematic uncertainty of 5\% (left panel) and 10\% (right panel).
}
 \label{fig:lhc_dif1}
 \end{figure}
 \begin{figure}[h!]
\begin{center}
\begin{tabular}{cc}
\hspace{-0.5cm}
\includegraphics[width=0.5\textwidth]{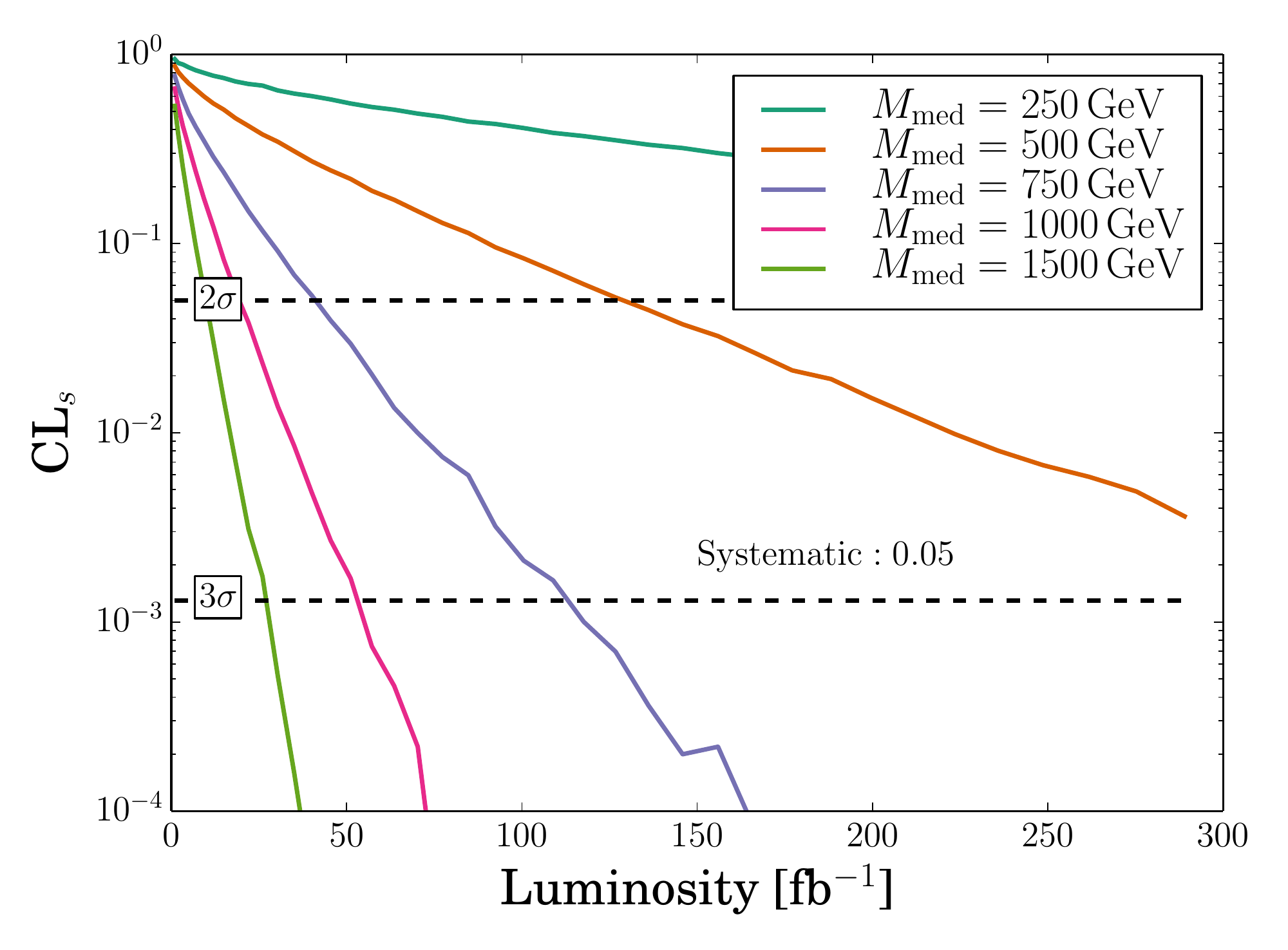}
&
\hspace{.1cm}
\includegraphics[width=0.5\textwidth]{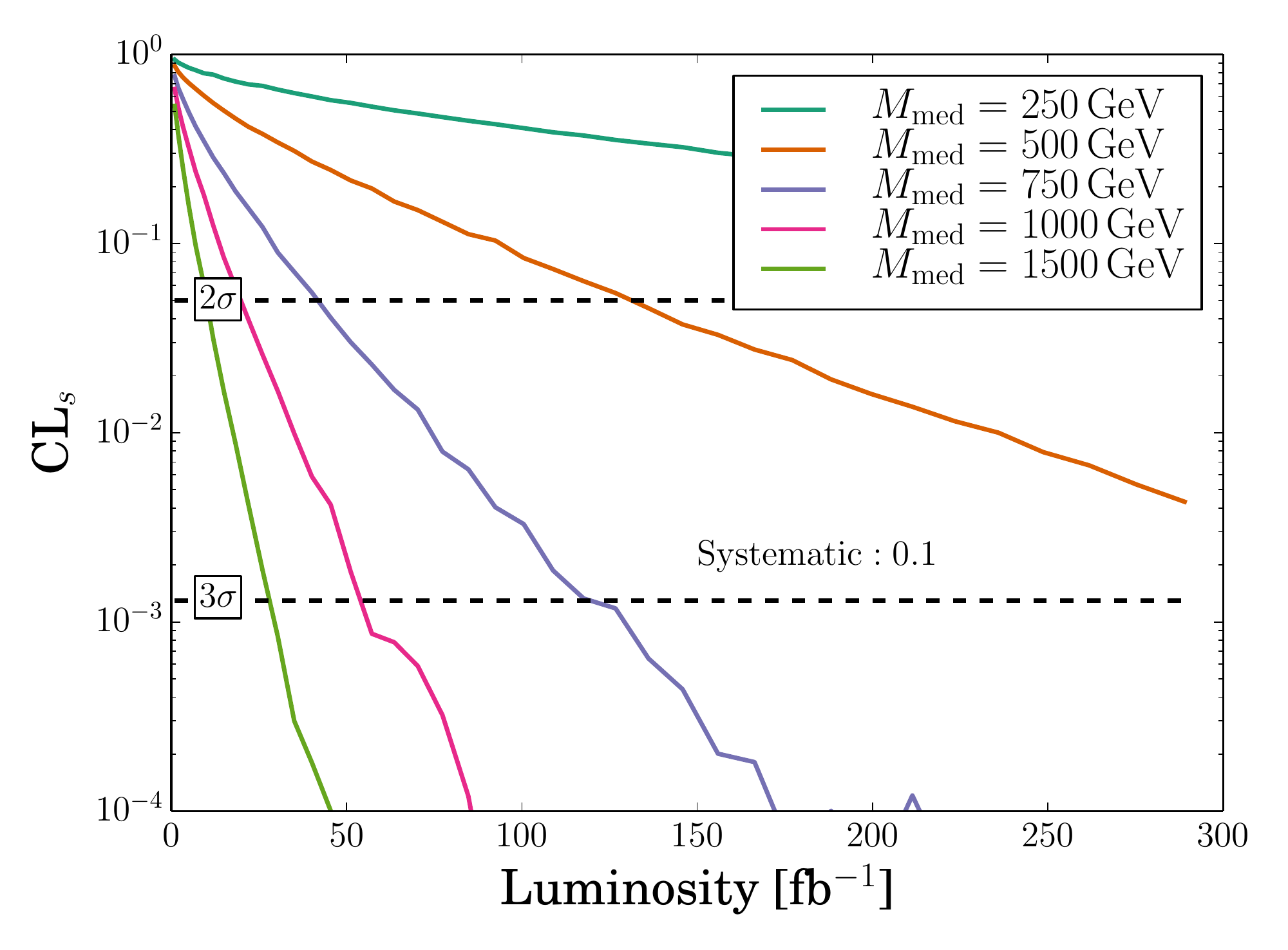}
\\
\end{tabular}
\end{center}
\vskip-0.4cm
\caption{
Differentiating the models with the floating $\kappa$ parameter defined as in the caption of Fig.~\ref{fig:reach_equal}.
Models are compared pairwise to the 125 GeV reference model.
}
 \label{fig:lhc_dif2}
 \end{figure}

\subsection{Distinguishing between models with different mediator masses}

For the models which are within the LHC reach, i.e. with $M_{\rm med}$ below the upper bounds set be the exclusion contours in
Figs.~\ref{fig:reach1}-\ref{fig:reach_equal}, the next step is to be able to distinguish between different models.

This is achieved by comparing the shapes of the kinematic distributions plotted in Fig.~\ref{fig:Mjj} for different mediator masses.
As we increase $M_{\rm med}$  the visible jets will recoil against a heavier object and we can use the kinematic variables to distinguish the models. 
We will use the following procedure. 
Before we even start comparing  different models, we would need an excess of signal events over the SM background in the data after the VBF cuts. 
The cross-section of this signal can be used to infer an upper limit for the mass of the mediator as a function of $\kappa$. The question then becomes 
if we can distinguish between the different models that can achieve the measured cross-section. We will again use a binned log-likelihood method 
and we will be comparing models pairwise. For each cross-section we select the two extreme models: first one with the maximal mass,
and the second (reference model) -- with the 125 GeV mediator.

 \begin{figure}[t]
\begin{center}
\begin{tabular}{cc}
\hspace{-0.5cm}
\includegraphics[width=0.5\textwidth]{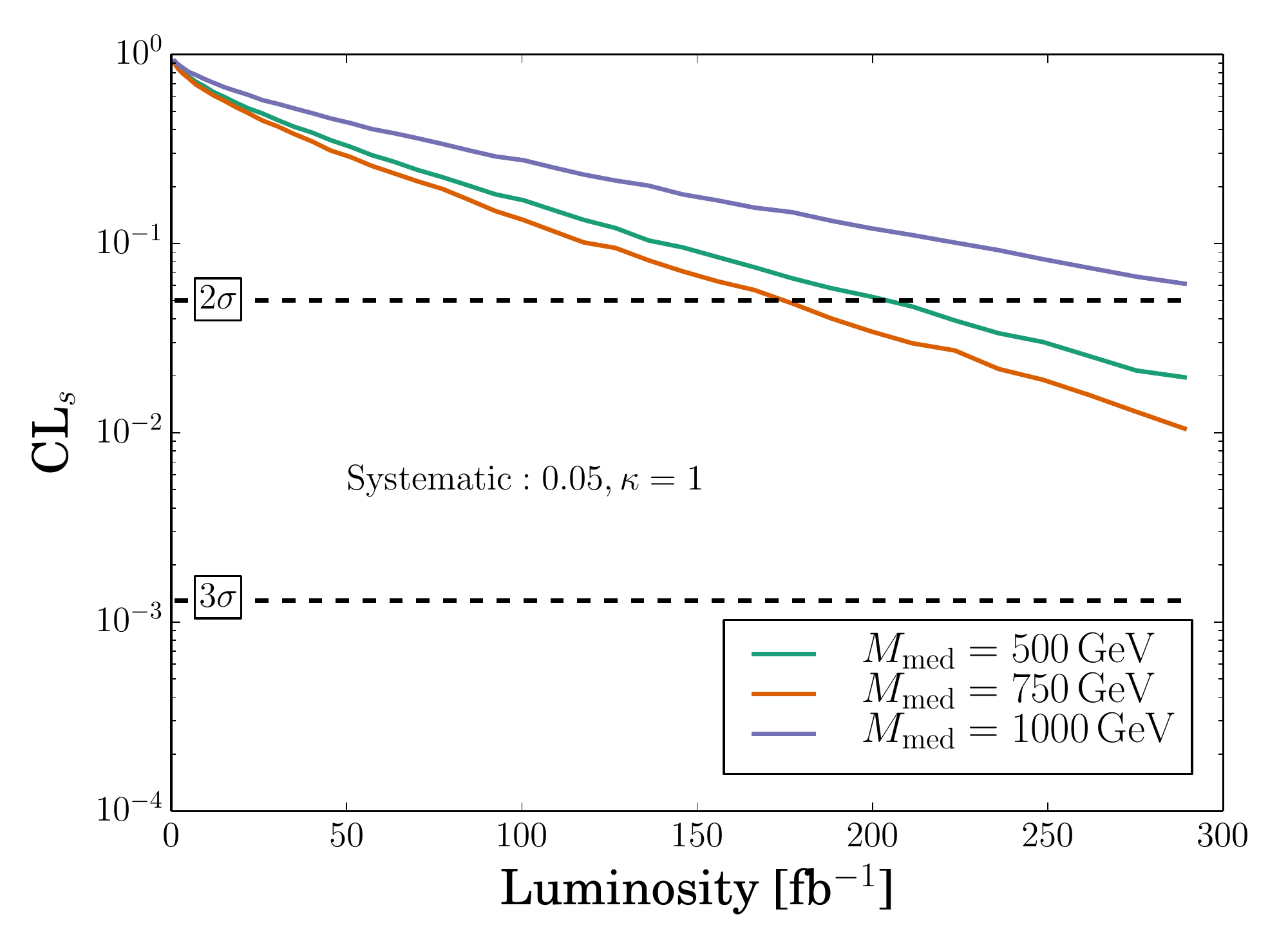}
&
\hspace{.1cm}
\includegraphics[width=0.5\textwidth]{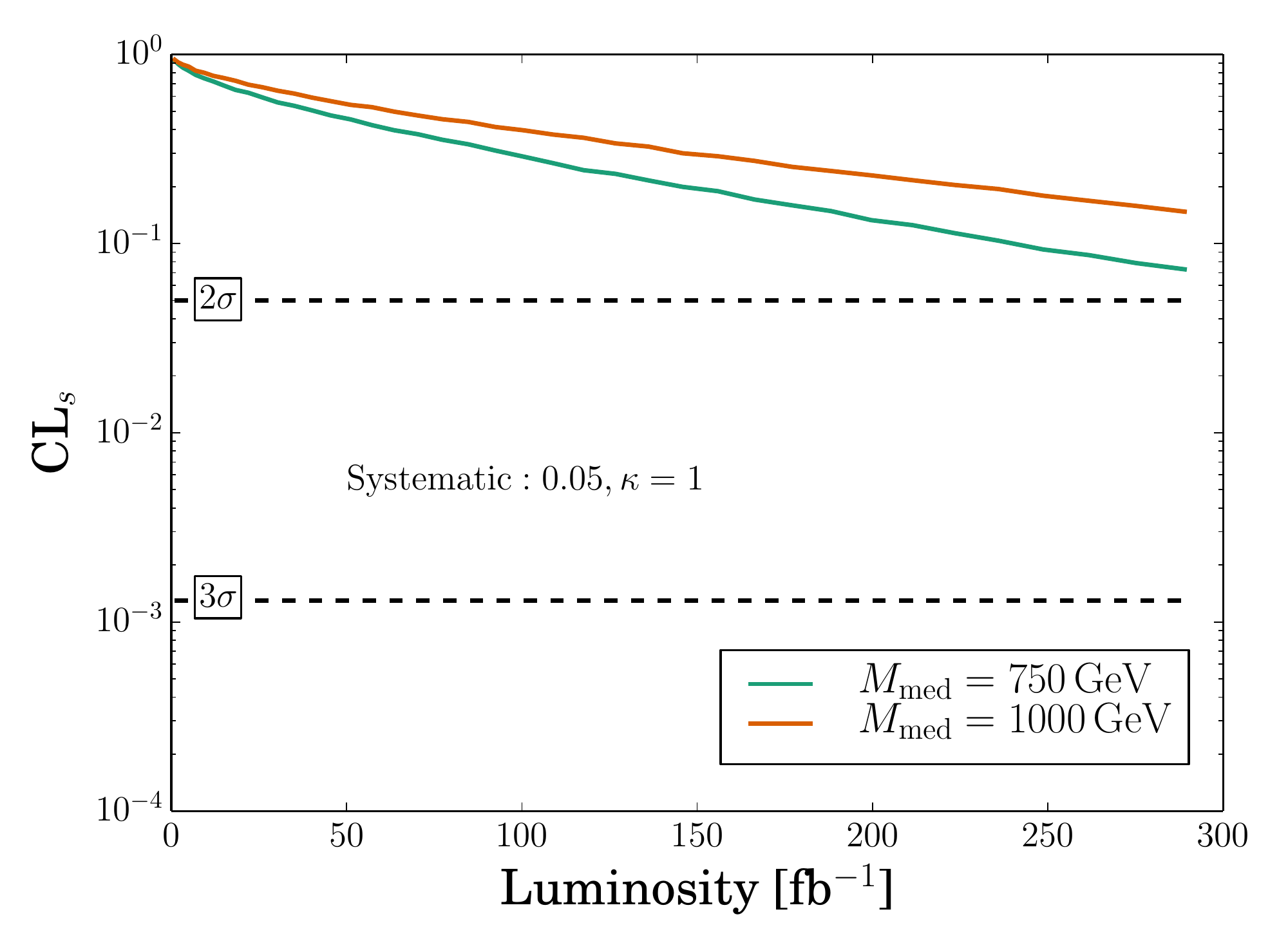}
\\
\end{tabular}
\end{center}
\vskip-0.4cm
\caption{
Differentiating the $\kappa=1$ models  at the LHC. In the left panel we compare to the reference model with $M_{\rm med}=250$ GeV,
and on the right the reference model is 500 GeV.
We assume a systematic uncertainty of 5\% .
}
 \label{fig:lhc_dif_masses}
 \end{figure}

In Figure~\ref{fig:lhc_dif1} we can see how well one can differentiate the models at the LHC with $\kappa=1$.  
Specifically, all the models with $M_{\rm med}= 250,$ 500 and 750 GeV can be distinguished from the 125 GeV mediator.
Within our approach this conclusion is valid even with a relatively high systematic error of 10\%.
We also note that for 
$\kappa=0.15$ it is no longer possible to differentiate any of the models since the cross-section becomes too small. 

So far in Fig.~\ref{fig:lhc_dif1}  we have characterised the simplified model signals by fixing the scaling parameter $\kappa$ to either 1 or 0.15.
Alternatively we can set the signal cross-section to a fixed value corresponding to a 30\% invisible branching ratio of the SM Higgs.
This is shown in  
Fig.~\ref{fig:lhc_dif2} which leads one to conclude that the models with heavier and heavier mediator masses are easier and easier to distinguish 
from the reference model. When comparing models with different mediator masses, there are in general two competing effects: 
 the increased difference 
in the shape of differential distributions,
 and the decrease in the cross-section with the increase of the mediator mass. By fixing the cross-sections in Fig.~\ref{fig:lhc_dif2} 
 the differences between the models are only due to the shapes of differential distributions, while in Fig.~\ref{fig:lhc_dif1} both effects
 are important. This explains why for example the model with $M_{\rm med} =500$ GeV is easier to distinguish than the 250 and 750 GeV models.

We also compare models where the reference model is not the 125 GeV Higgs. The results for using 250 GeV and 500 GeV as reference models 
are shown in Fig.~\ref{fig:lhc_dif_masses}. 
In the same way as for the 125 GeV Higgs, the cross-section for the reference model is set equal to that
of the model we compare it with. We see that the 500 and 750 GeV models can be distinguished from the 250 GeV model at the LHC.
At the same time, the 750 GeV model (and above) cannot be distinguished from the 500 GeV reference point.

\newpage

\section{Scalar Mediator Models at 100 TeV}
\label{sec:100TeV}

We use a similar approach to investigate the models reach and the ability to distinguish between different models at a future 100 TeV circular proton-proton
 collider. The signal and background are simulated in the same way as for the LHC analysis and we use the same binned-log likelihood analysis for exclusion 
 and differentiation of the various models. The main difference is that we use the cuts 
 \beq
 \misspT>100 \,\text{Gev}\,, \quad  M_{jj}>1200\, \text{GeV}\,, \quad \Delta \phi_{jj}<0.5\,, \quad \Delta \eta > 5.5 \,, \quad p_{T,j} >110 \, \text{GeV}\,,
\label{eq:cut100}
\eeq
 instead of the normal VBF cuts in \eqref{eq:cut1}
 as we need to reduce the background more. We also allow for larger jets by using the anti-kt jet algorithm with $R=0.8$. 

 \begin{figure}
\begin{center}
\begin{tabular}{cccc}
\hspace{-0.5cm}
\includegraphics[width=0.5\textwidth]{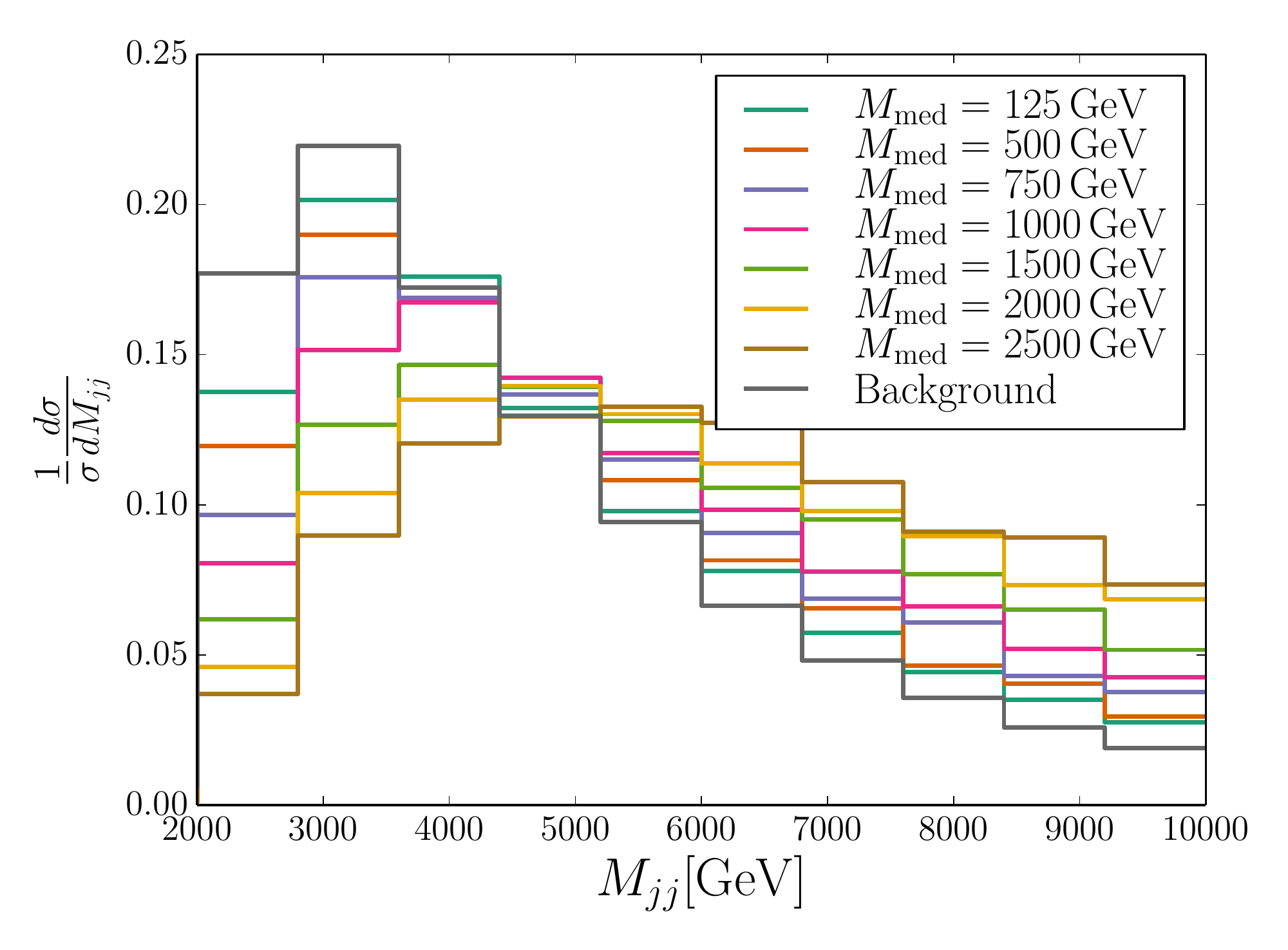}
&
\hspace{.1cm}
\includegraphics[width=0.5\textwidth]{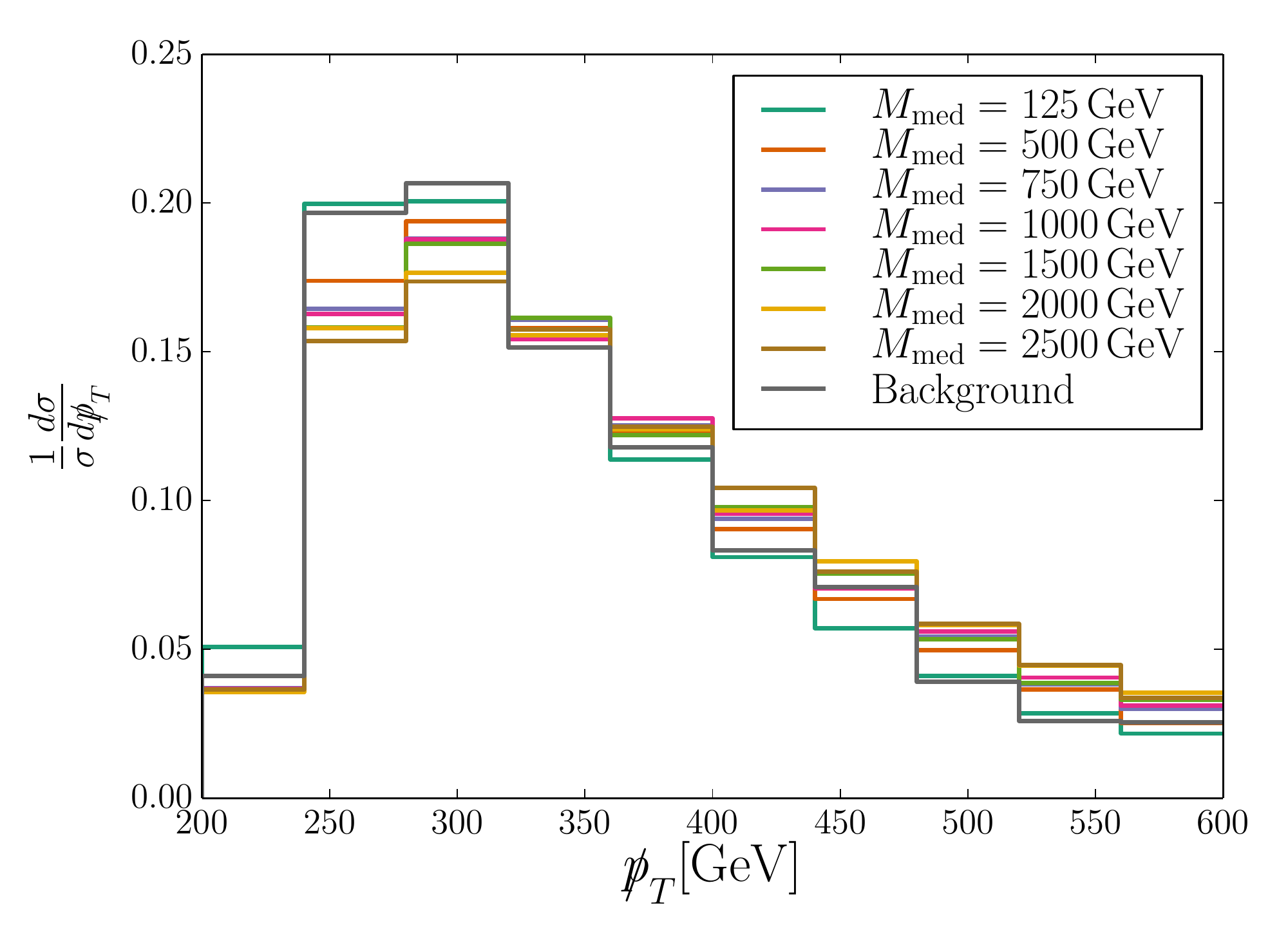}
\\
\hspace{-0.5cm}
\includegraphics[width=0.5\textwidth]{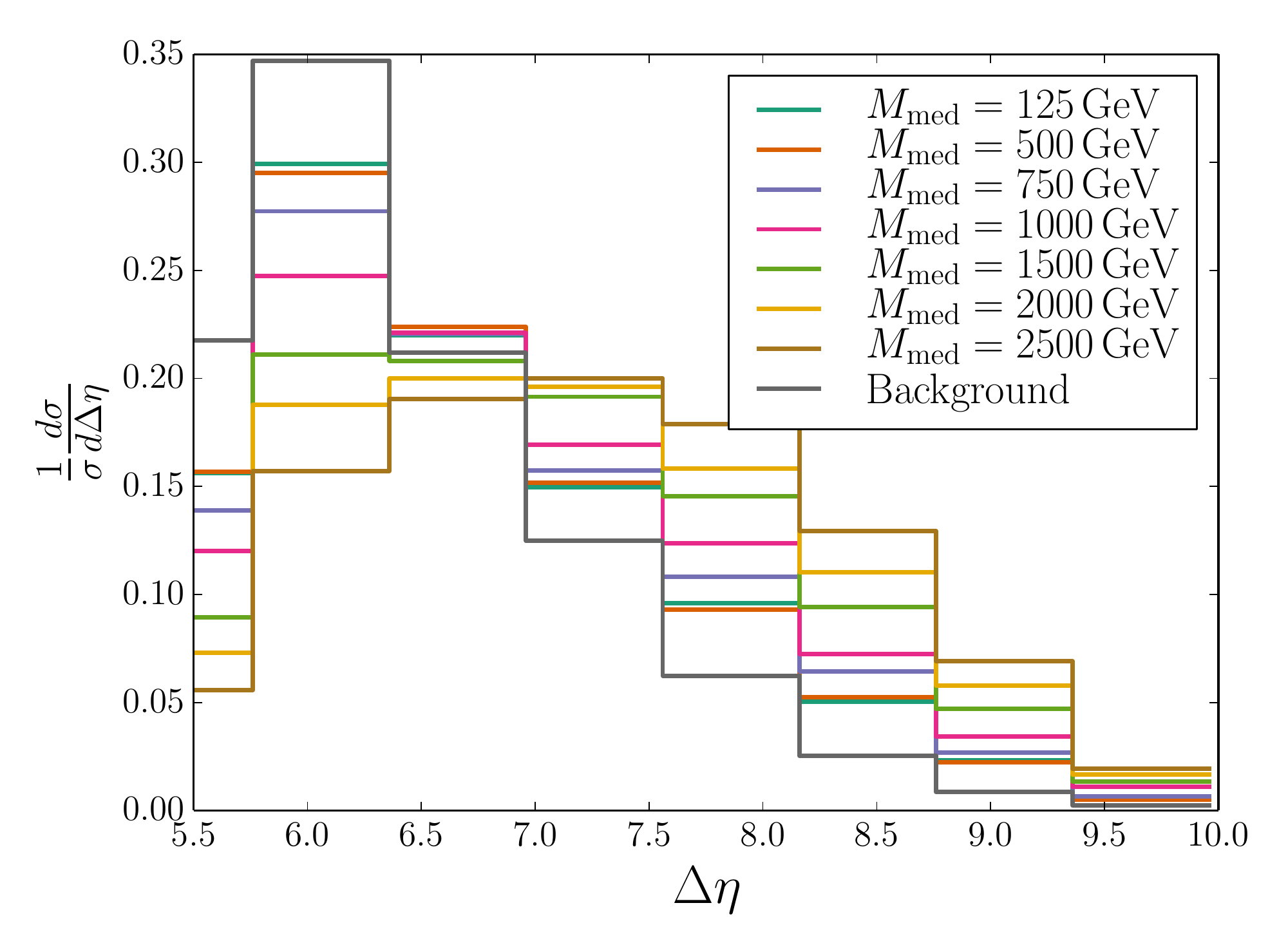}
&
\hspace{.1cm}
\includegraphics[width=0.5\textwidth]{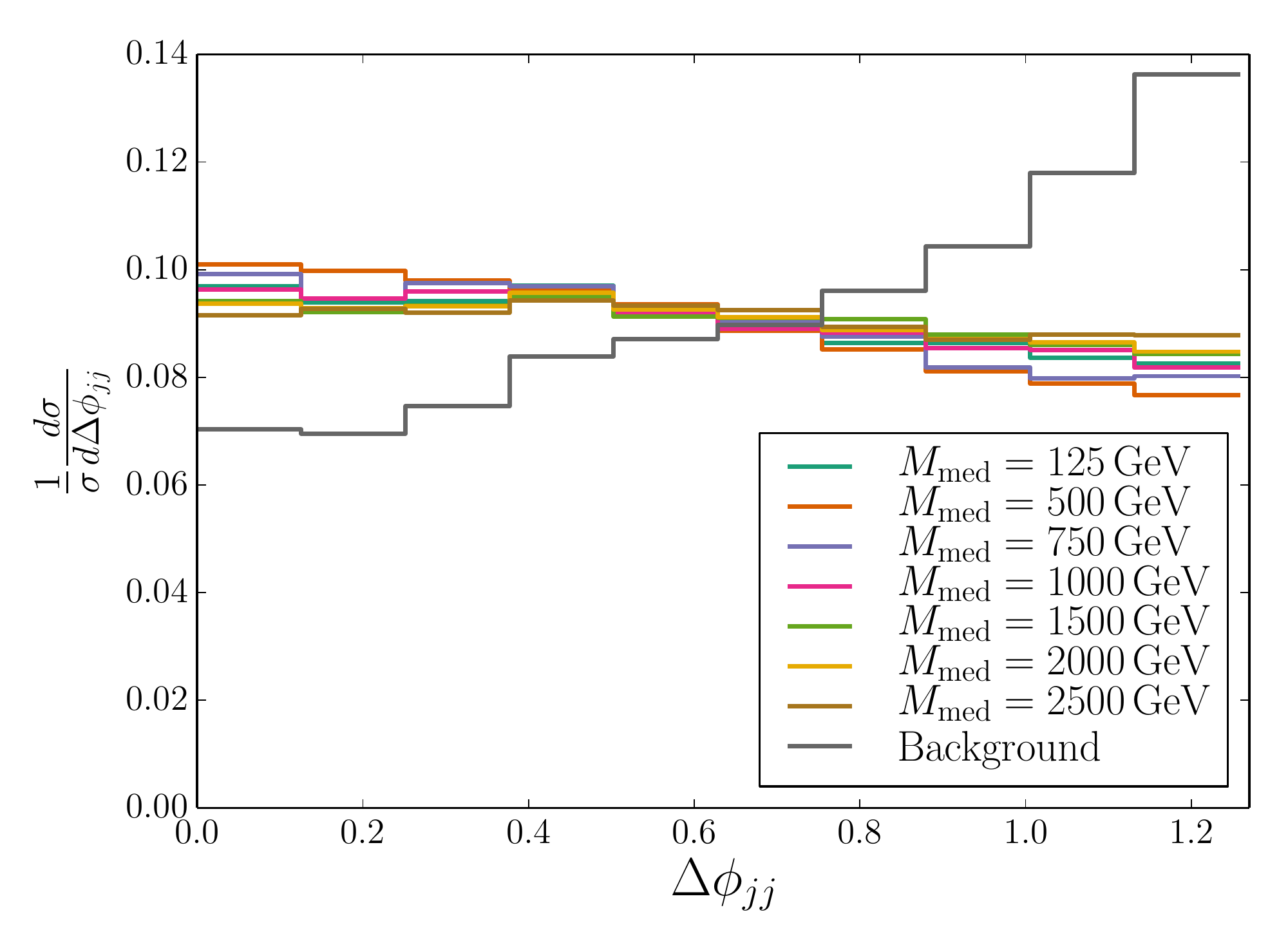}
\\
\end{tabular}
\end{center}
\vskip-0.4cm
\caption{
Kinematic distributions for different values of the mediator mass for the signal, and for the background at a 100 TeV collider. $M_{jj}$ distributions
are shown on the top left panel, $\misspT$ is on top right right, $\Delta \eta$ and $\phi_{jj}$ distributions are on the bottom left and right panels respectively.
}
 \label{fig:Mjj_100}
\end{figure}
 \begin{figure}
\begin{center}
\begin{tabular}{cc}
\hspace{-0.5cm}
\includegraphics[width=0.5\textwidth]{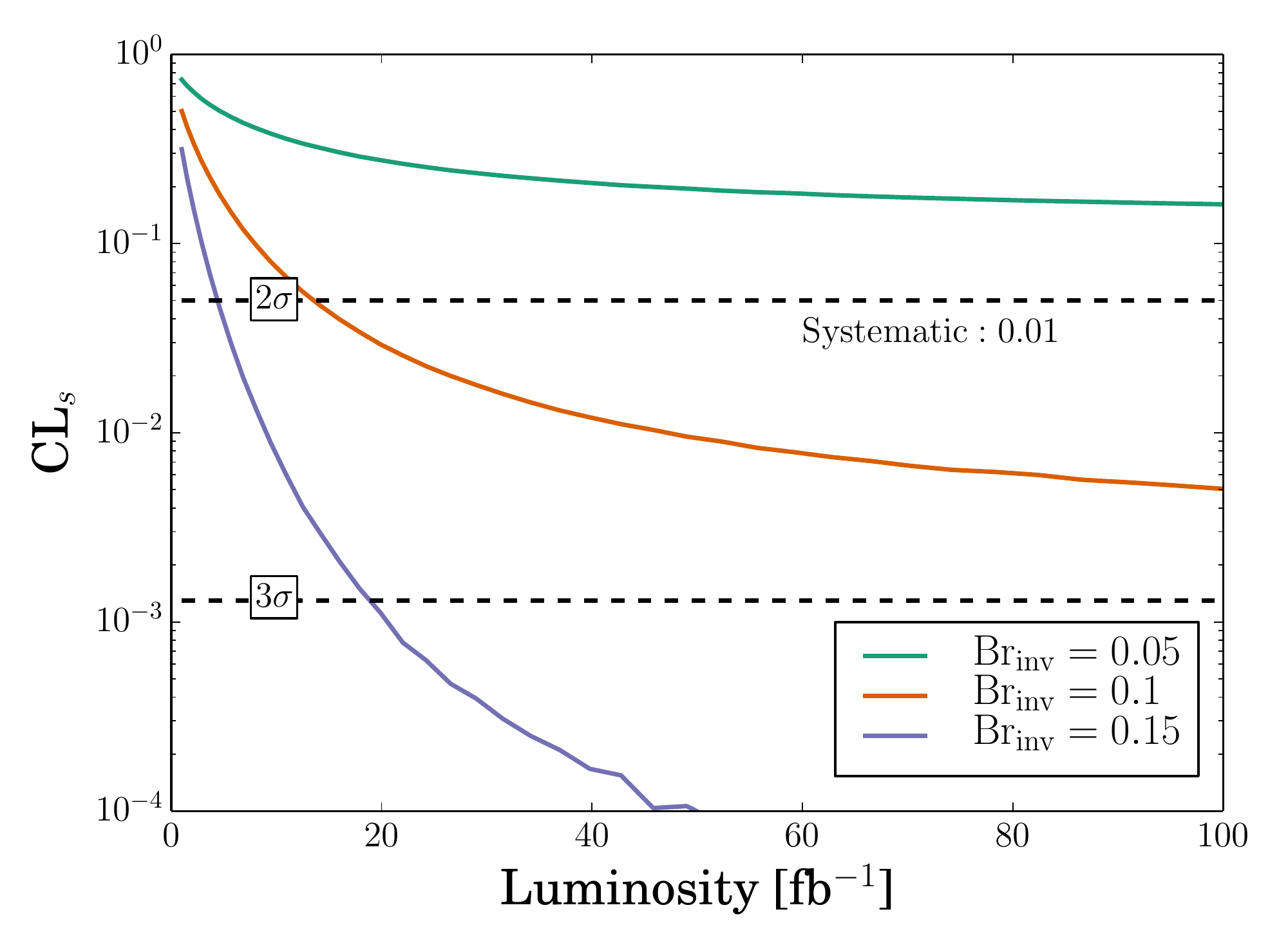}
&
\hspace{.1cm}
\includegraphics[width=0.5\textwidth]{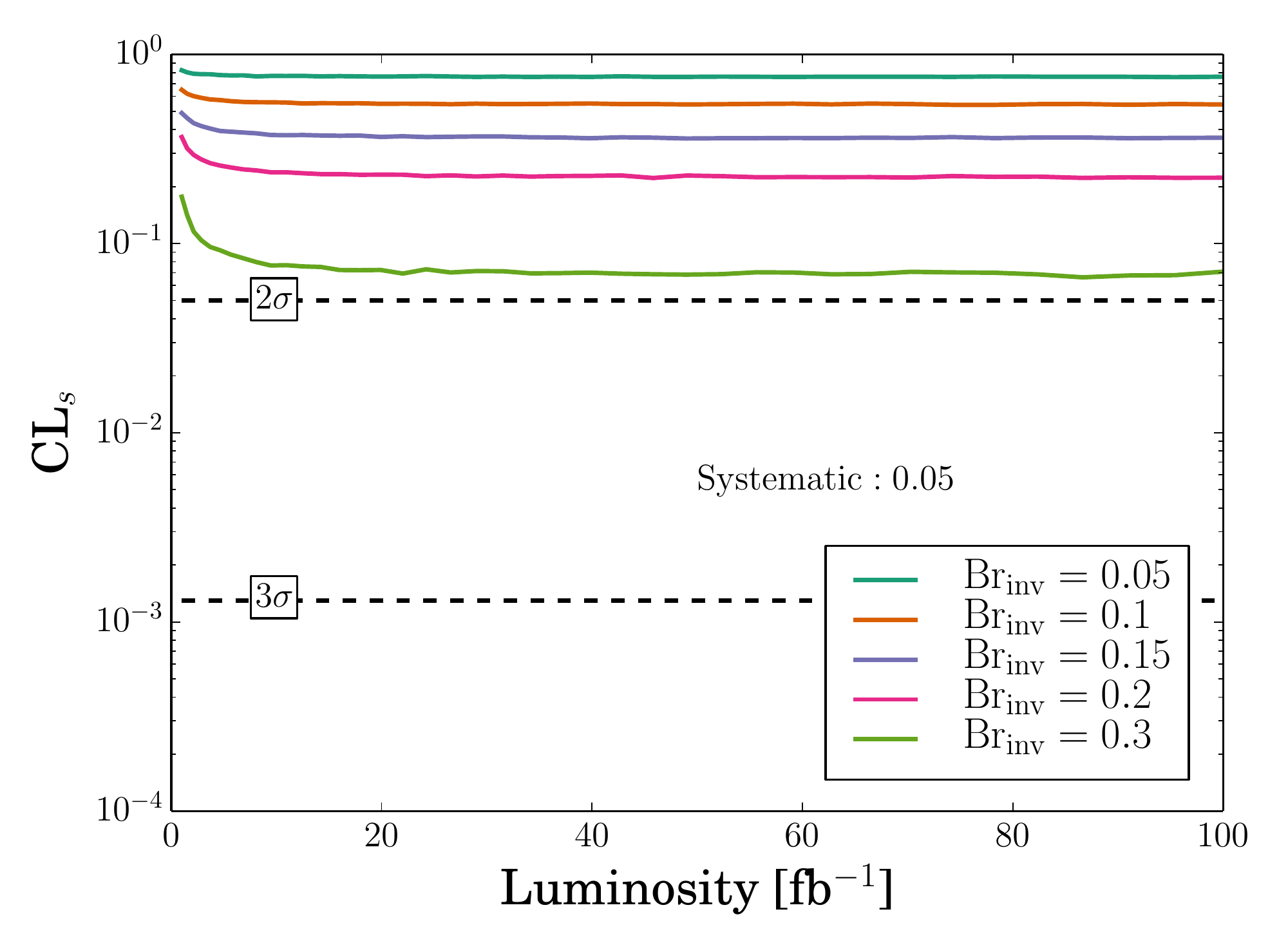}
\\
\end{tabular}
\end{center}
\vskip-0.4cm
\caption{
100 TeV reach for excluding invisible decays of the 125 GeV Higgs boson. 
On the left panel we use a systematic uncertainty of 1\%, and the panel on the right corresponds to 5\% systematic uncertainty  }
 \label{fig:100_inv_br}
 \end{figure}

 \begin{figure}
\begin{center}
\begin{tabular}{cc}
\hspace{-0.5cm}
\includegraphics[width=0.5\textwidth]{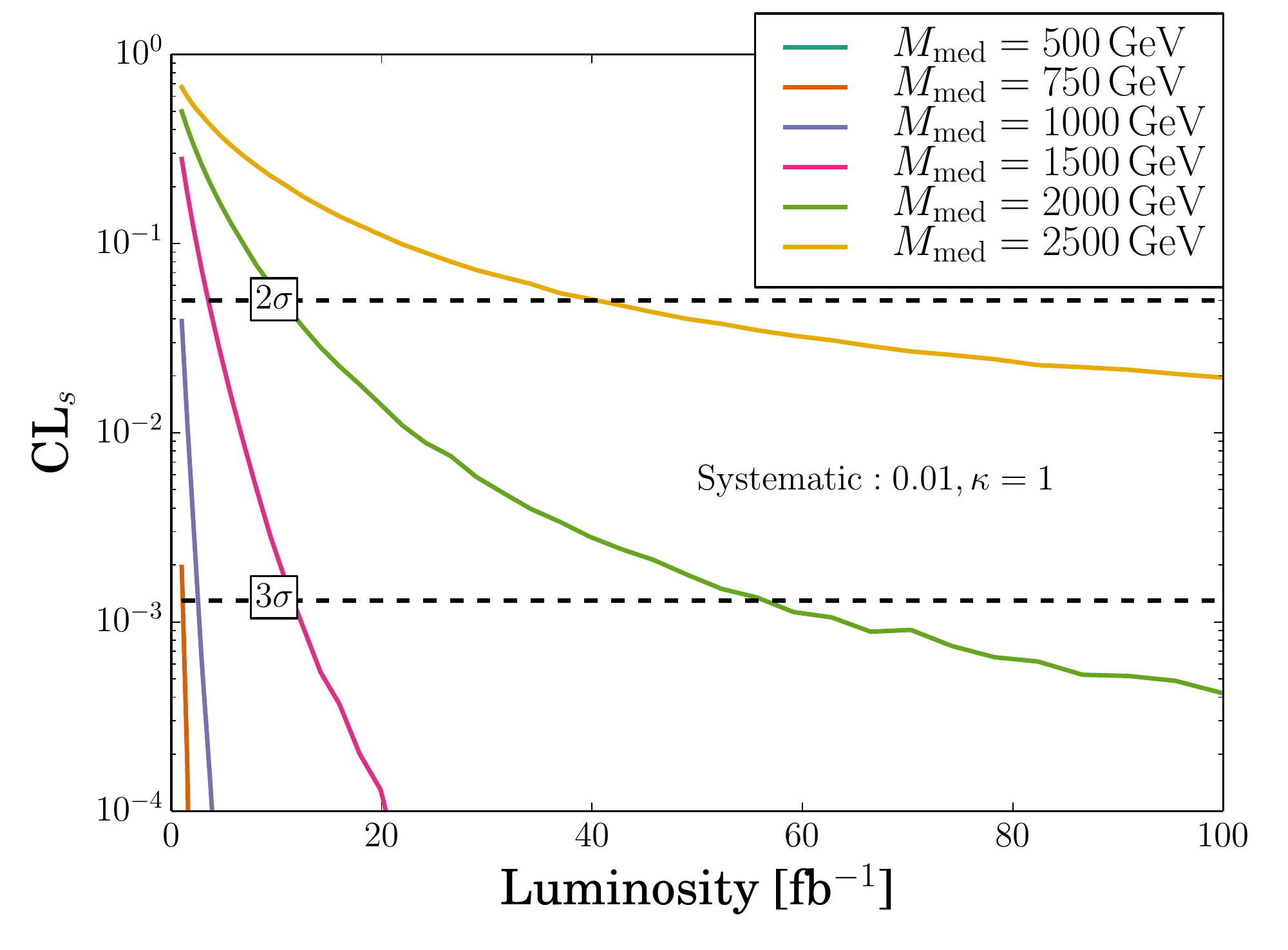}
&
\hspace{.1cm}
\includegraphics[width=0.5\textwidth]{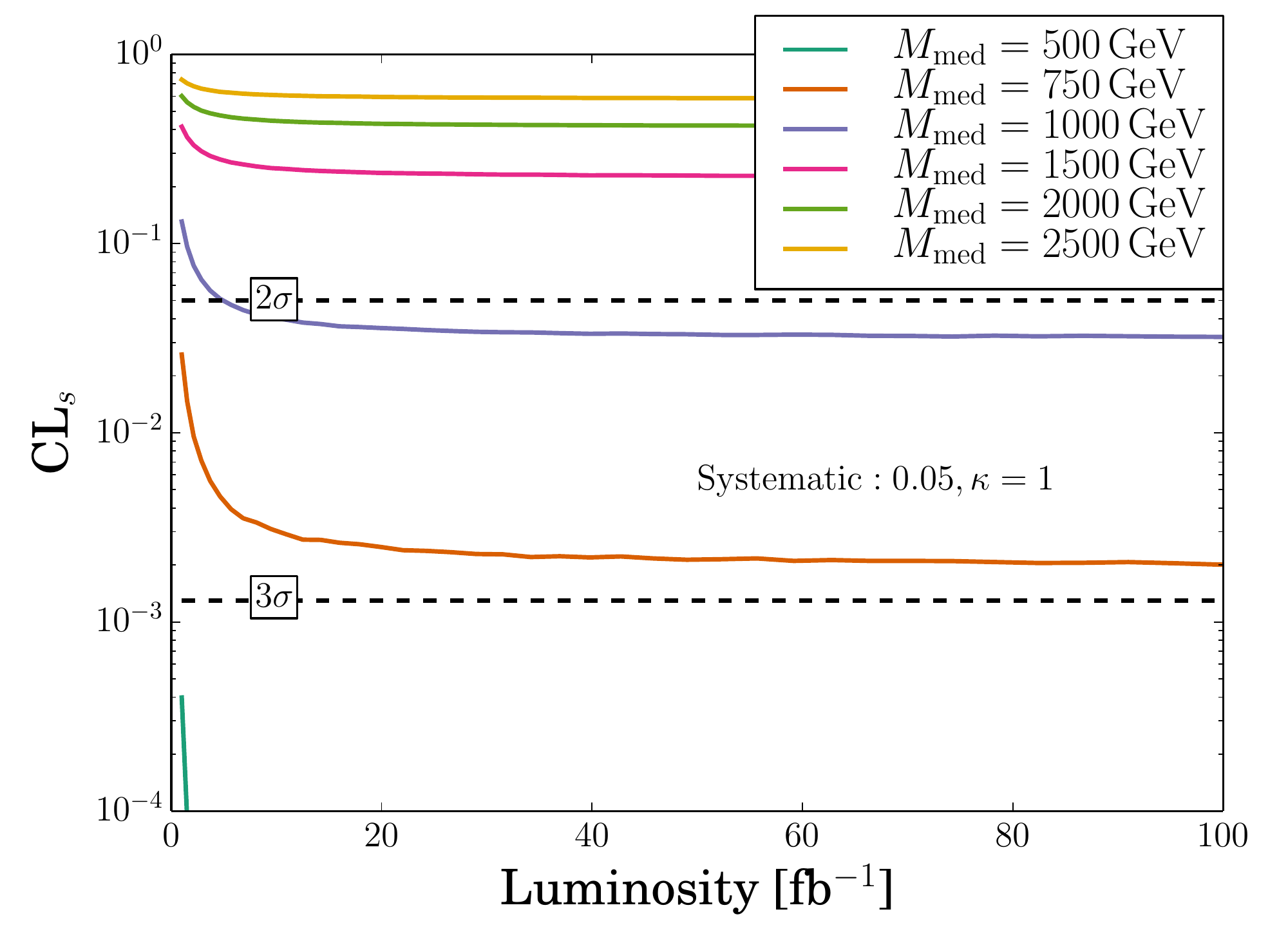}
\\
\end{tabular}
\end{center}
\vskip-0.4cm
\caption{
100 TeV reach for different $M_{\rm med}$ models with $\kappa=1$. On the left panel we use a systematic uncertainty of 1\%, and the panel on the right 
corresponds to 5\% systematic uncertainty. }
 \label{fig:reach100_1}
 \end{figure}
 \begin{figure}
\begin{center}
\begin{tabular}{cc}
\hspace{-0.5cm}
\includegraphics[width=0.5\textwidth]{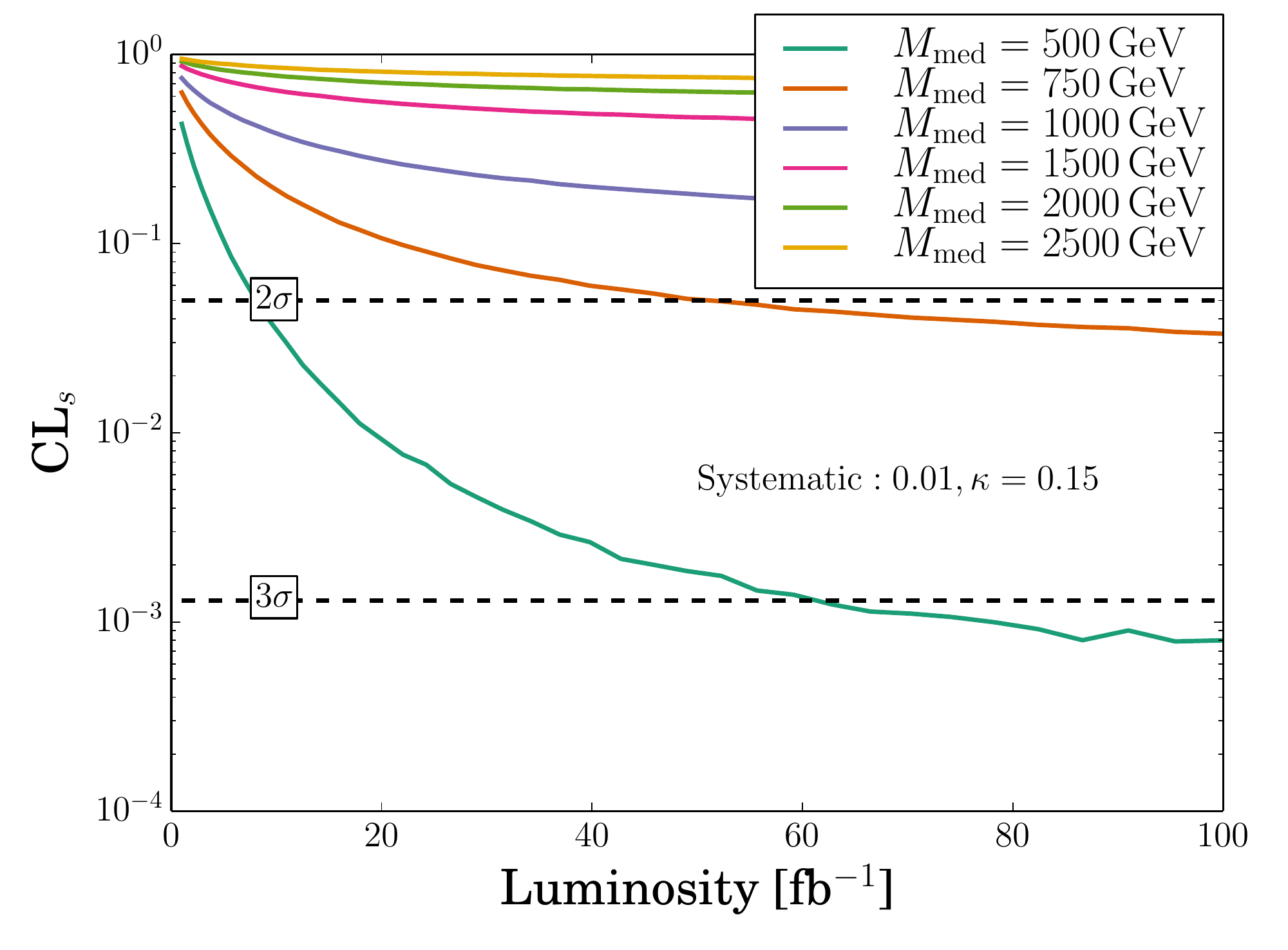}
&
\hspace{.1cm}
\includegraphics[width=0.5\textwidth]{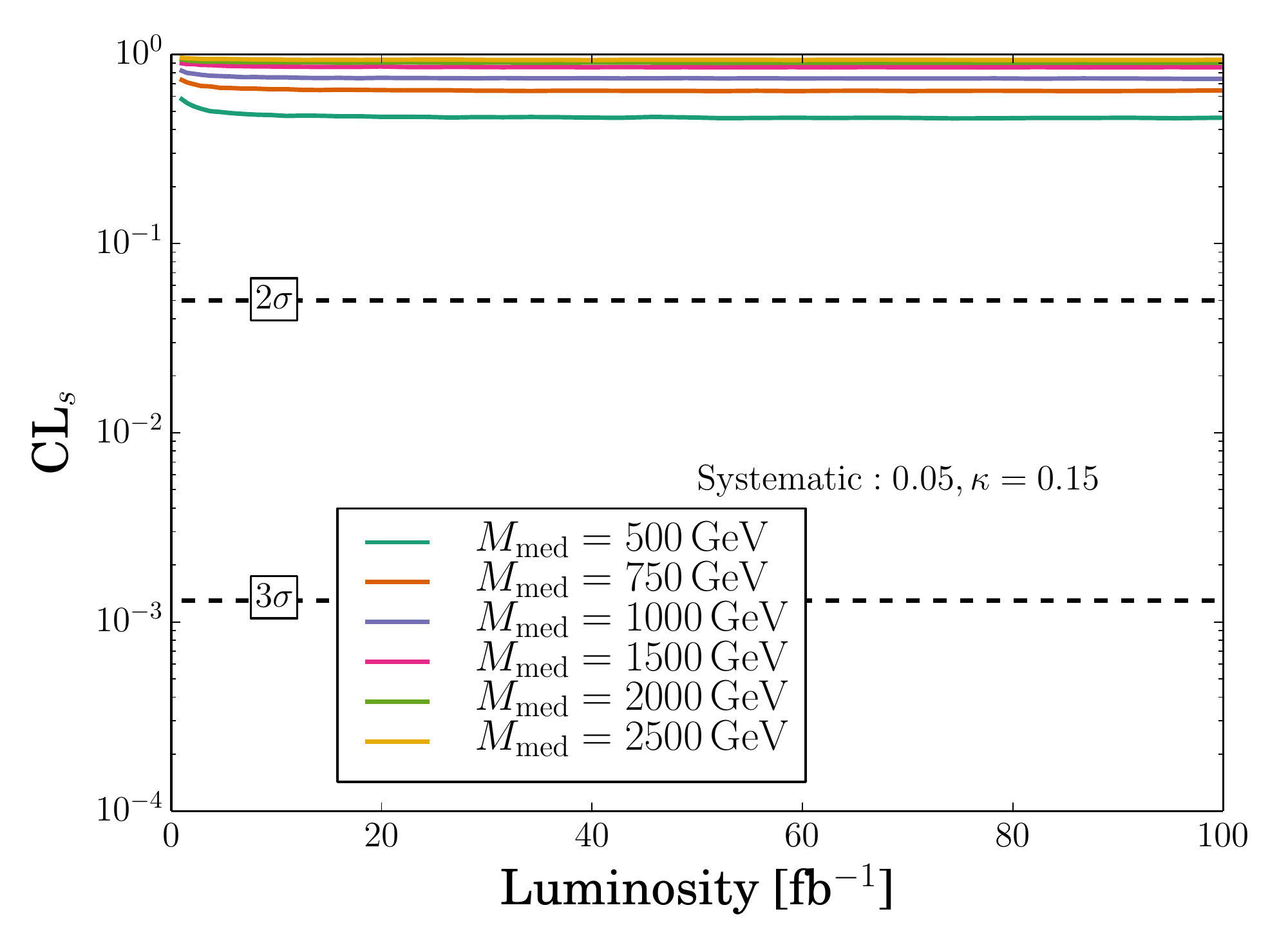}
\\
\end{tabular}
\end{center}
\vskip-0.4cm
\caption{
100 TeV reach for different $M_{\rm med}$ models with $\kappa=0.15$. On the left panel we use a systematic uncertainty of 1\%, and the panel on the right 
corresponds to 5\% systematic uncertainty. }
 \label{fig:reach100_015}
 \end{figure}

The left panel in Table~2 
shows the cross-sections for the signal at a future 100 TeV collider, assuming a 100\% branching 
ratio of the scalar mediators to dark matter.  

\medskip
\begin{center}
\begin{tabular}{cc}
\begin{tabular}{ l | c | c | c}
  \hline                       
  $M_{\rm med}$ & VBF & GGF &Total  \\
  125 GeV &120&100&220  \\
  500 GeV &76&110&185  \\
  750 GeV &53&46&99  \\
  1000 GeV &40&20&60  \\
  1500 GeV &22&5&27  \\
  2000 GeV &14&2&16  \\
  2500 GeV &8.7&0.6&9.4  \\
  \hline 
\end{tabular}
\hspace{.5cm}
&
\hspace{.5cm}
\begin{tabular}{ l |  l}
  \hline                       
  Background & Cross-section(fb)  \\
   Zjj&239  \\
   W$^+$jj&298 \\
   W$^-$jj&157  \\
  \hline 
\end{tabular}
\end{tabular}
\end{center}
\centerline{Table 2. Cross-sections (fb) at partonic level after the cuts \eqref{eq:cut100} at 100 TeV. }
\medskip
We have set $\kappa=1$ and assumed 100\% branching ratio of the mediator to $\bar{\chi}\chi$ DM.
The cross-sections for SM backgrounds are shown in the table on the right.

\medskip
Figure \ref{fig:Mjj_100} plots the kinematic distributions for $M_{jj},\, \misspT,\,\Delta \eta, \,\Delta \phi_{jj}$  
for models with different values of $M_{\rm med}$ and the SM background.
We first investigate the exclusion limits at 100 TeV for invisible decays of the 125 GeV Higgs.
In Figure~\ref{fig:100_inv_br} we plot exclusion limits on the branching rations to invisibles and conclude 
that for small systematic errors one can exclude ${\rm Br}_{\rm inv} \gtrsim 10$\%.

We can now consider simplified models with different values of the mediator mass.
In Figure~\ref{fig:reach100_1} we show 
the expected reach for a 100 TeV collider for these models with $\kappa=1$ for 1\% and 5\% level of systematic uncertainty. 
Figure~\ref{fig:reach100_015} gives the expected exclusion limits for models with $\kappa=0.15$. 
From these Figures we conclude that in the case of small systematic uncertainties the 100 TeV collider would provide a very significant increase
in the exclusion reach for new physics models with mediator masses up to 2.5 TeV. For a larger systematic uncertainty at the level of 5\%
the reach in $M_{\rm med}$ is 1 TeV.
For the more restricted case of $\kappa = 0.15$ Higgs-mixing models, the FCC could exclude models with up to 750 GeV.

\begin{figure}
\begin{center}
\begin{tabular}{cc}
\hspace{-0.5cm}
\includegraphics[width=0.5\textwidth]{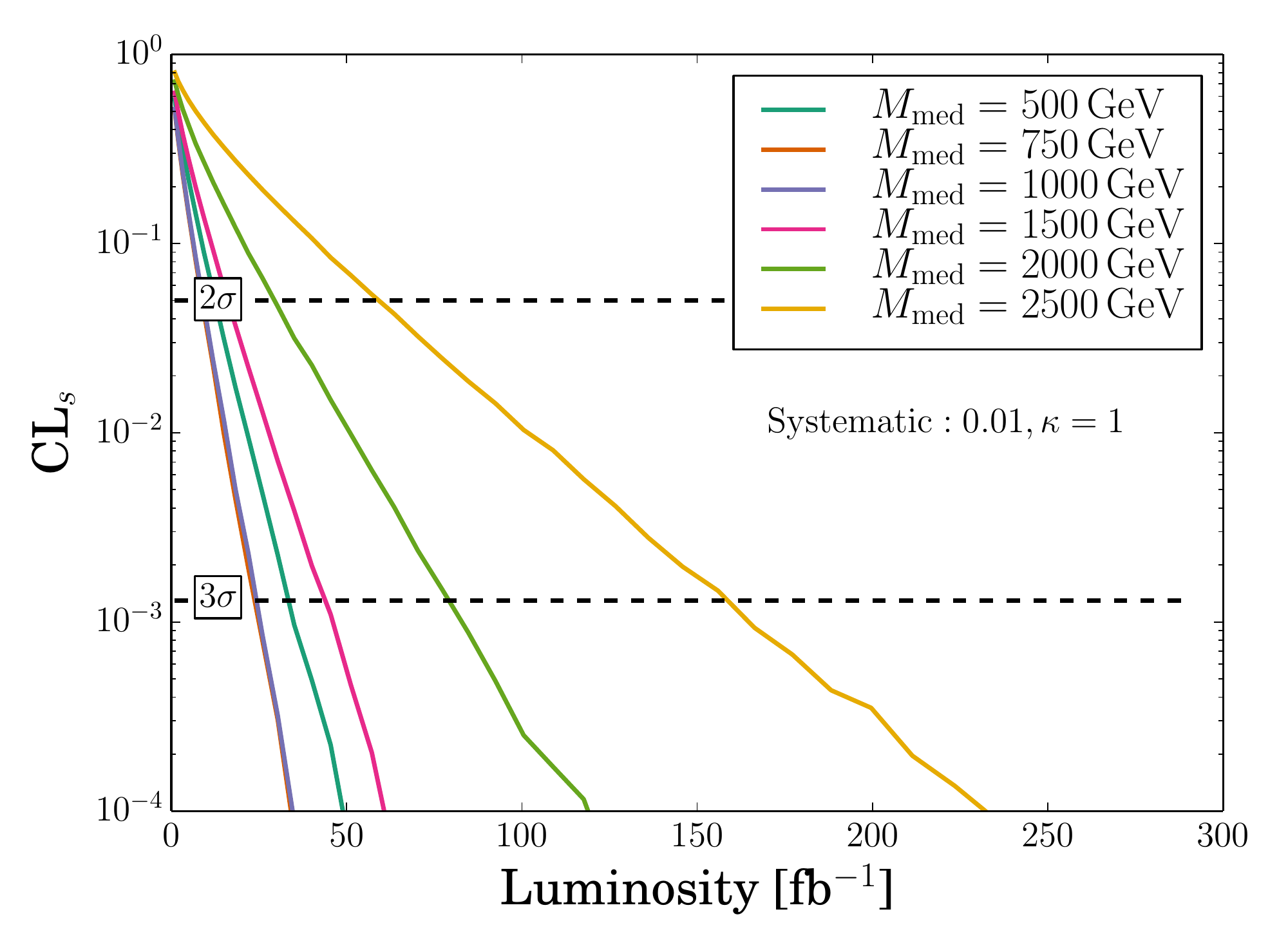}
&
\hspace{.1cm}
\includegraphics[width=0.5\textwidth]{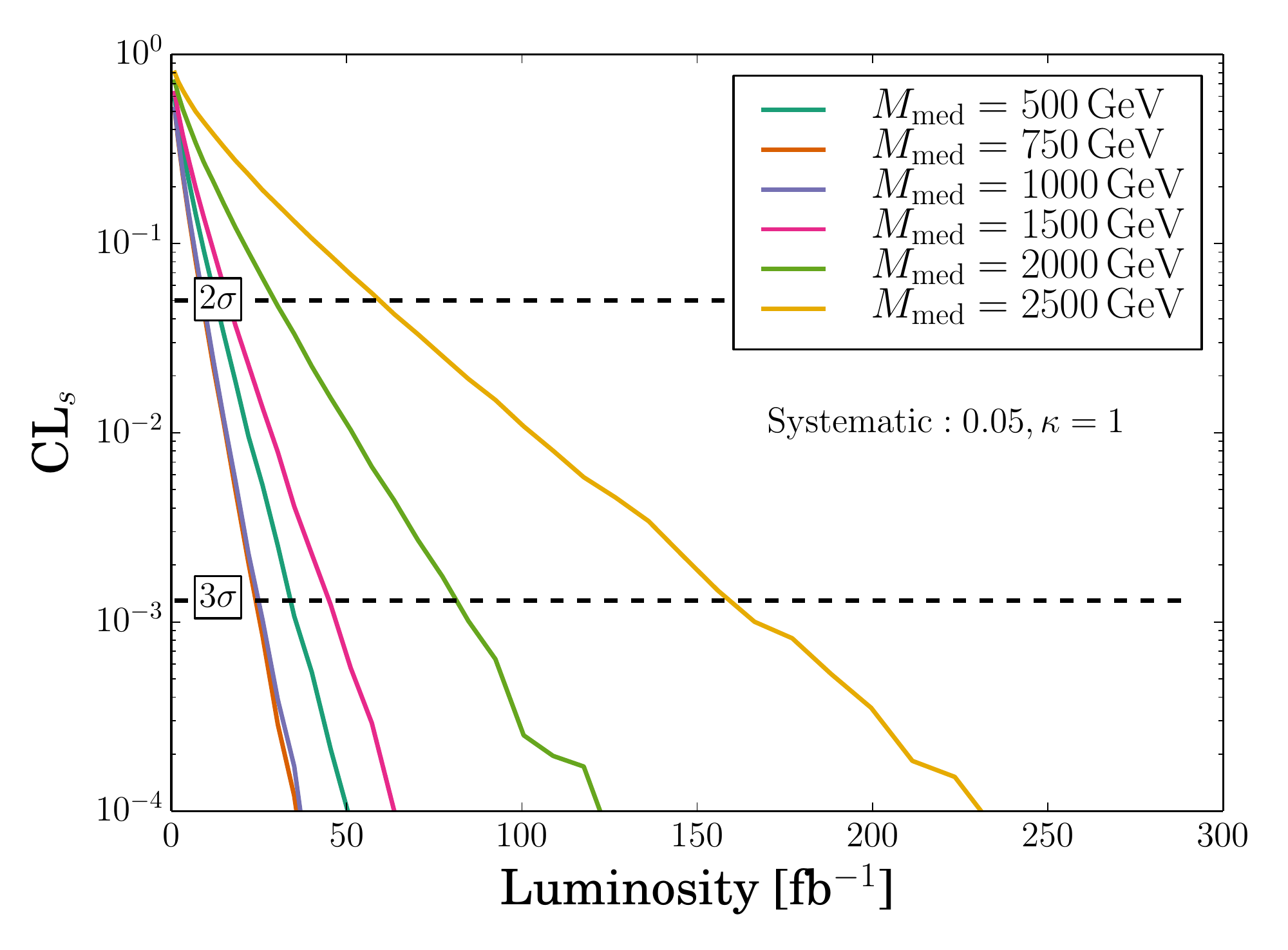}
\\
\end{tabular}
\end{center}
\vskip-0.4cm
\caption{
Differentiating the models at $\kappa=1$ at 100 TeV. For each value of $M_{\rm med}$ between $500$ and $2500$ GeV the models are compared to the reference model with the 125 GeV mediator.  We assume a systematic uncertainty of 1\% (left panel) and 5\% (right panel).
}
 \label{fig:100_dif1}
 \end{figure}

\begin{figure}
\begin{center}
\begin{tabular}{cc}
\hspace{-0.5cm}
\includegraphics[width=0.5\textwidth]{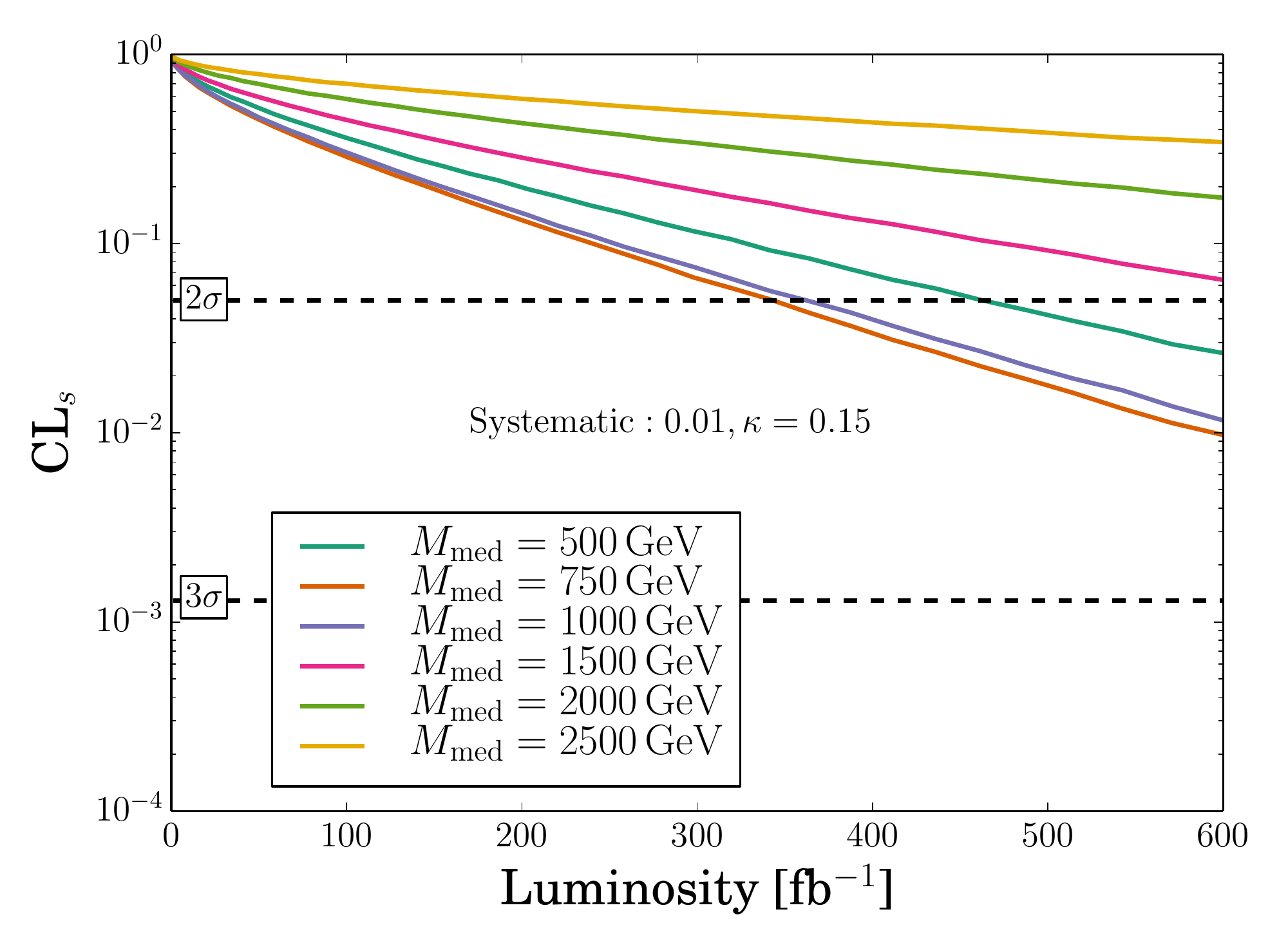}
&
\hspace{.1cm}
\includegraphics[width=0.5\textwidth]{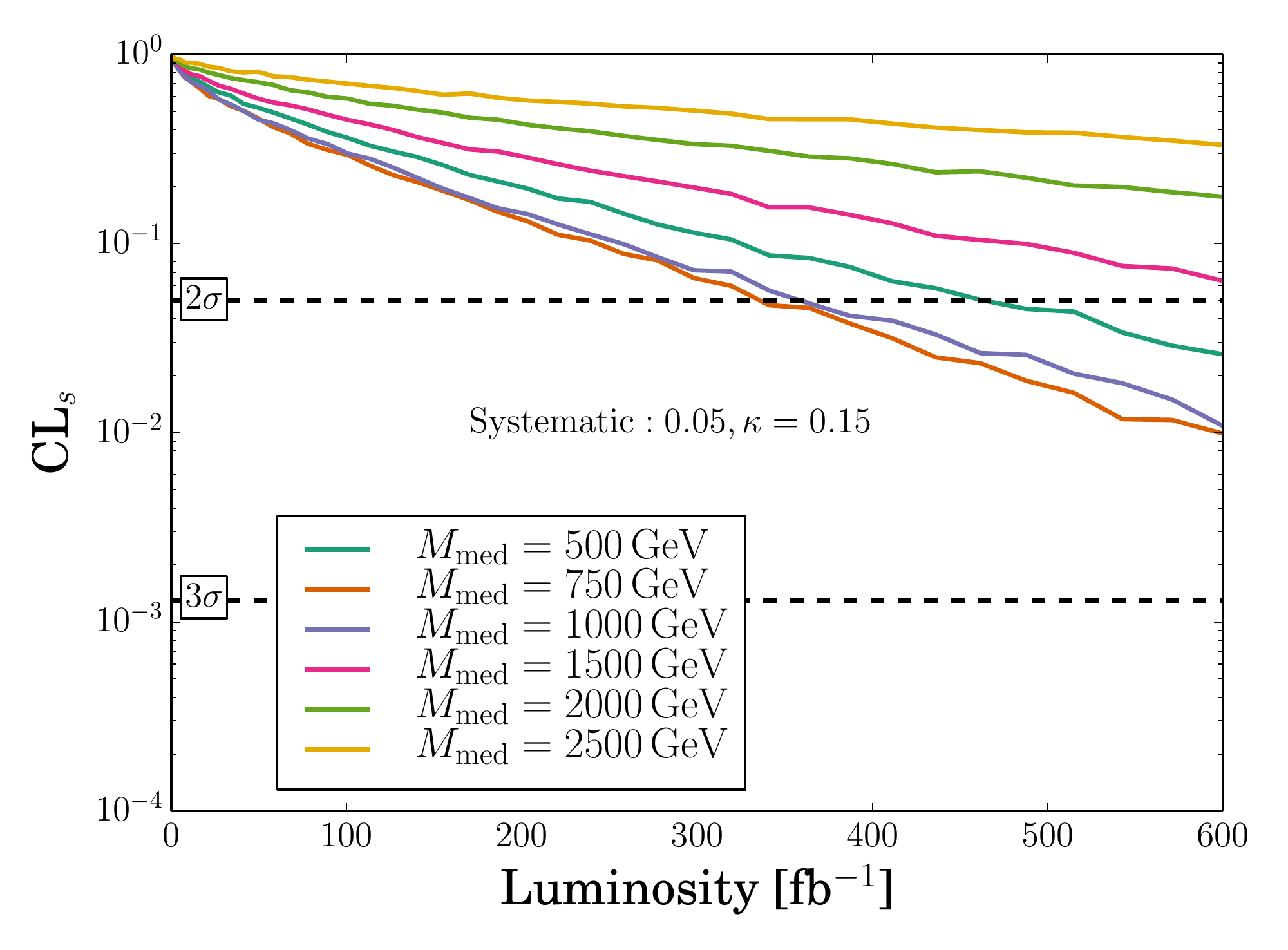}
\\
\end{tabular}
\end{center}
\vskip-0.4cm
\caption{
Differentiating the models at $\kappa=0.15$ at 100 TeV. For each value of $M_{\rm med}$ between $500$ and $2500$ GeV the models are compared to the reference model with the 125 GeV mediator.  We assume a systematic uncertainty of 1\% (left panel) and 5\% (right panel).
}

 \label{fig:100_dif2}
 \end{figure}

Finally, as we have done at the LHC energies before, we can use the
kinematic distributions at 100 TeV to differentiate  pairwise between models with different mediator masses. We first compare each model
with a reference model with a 125 GeV mediator.
The results for this analysis are presented
in Fig.~\ref{fig:100_dif1} for $\kappa=1$ and  Fig.~\ref{fig:100_dif2} for the models with $\kappa=0.15$. 
At $\kappa=1$ we can distinguish all the reachable models with mediators up to 2.5 TeV from the 125 GeV reference model.
For the small-$\kappa$ models in Fig.~\ref{fig:100_dif2}  we can distinguish between the models up to 1 TeV (this requires higher luminosities
up to 600 ${\rm fb}^{-1}$).

\begin{figure}[h!]
\begin{center}
\begin{tabular}{cc}
\hspace{-0.5cm}
\includegraphics[width=0.5\textwidth]{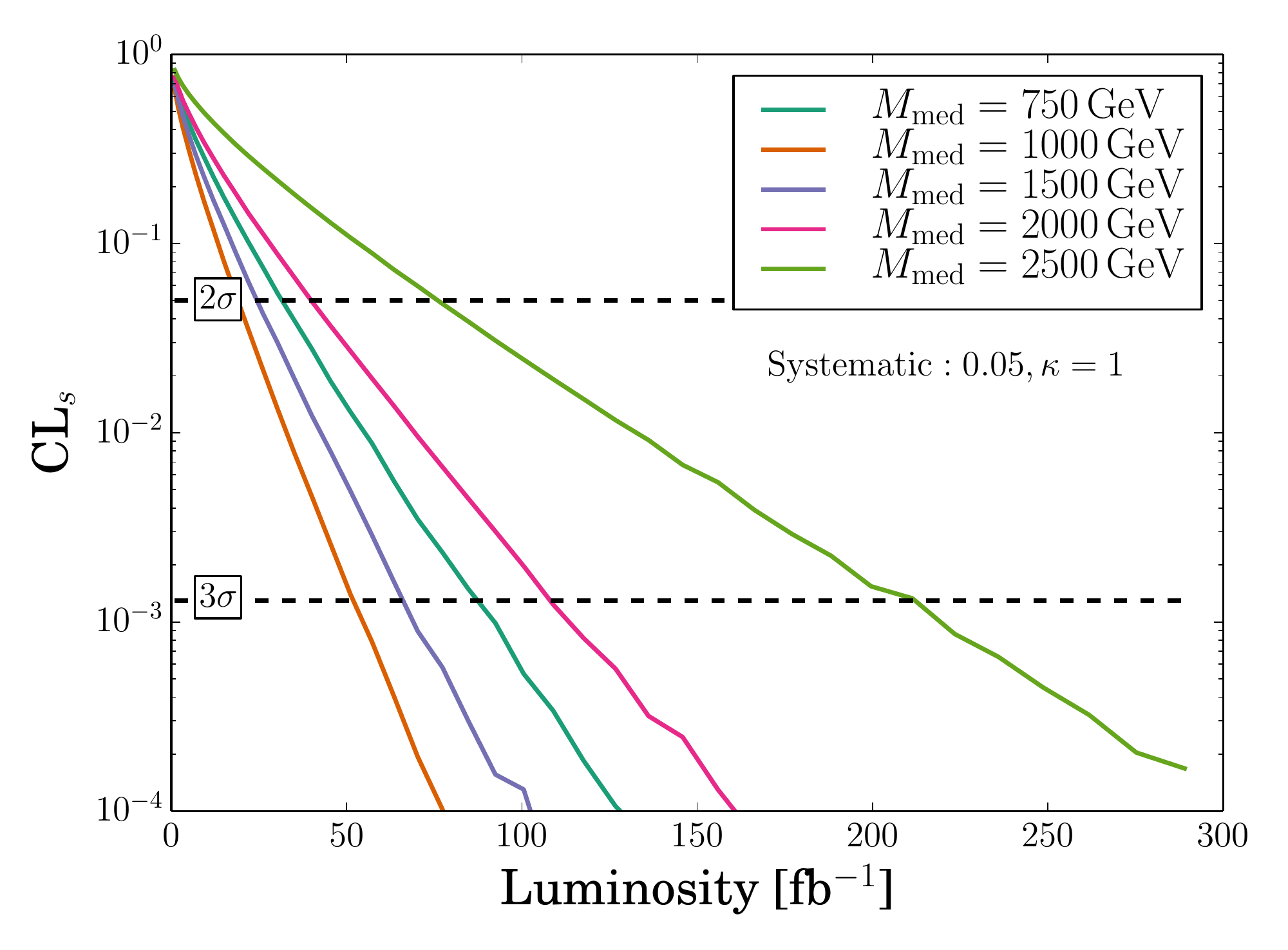}
&
\hspace{.1cm}
\includegraphics[width=0.5\textwidth]{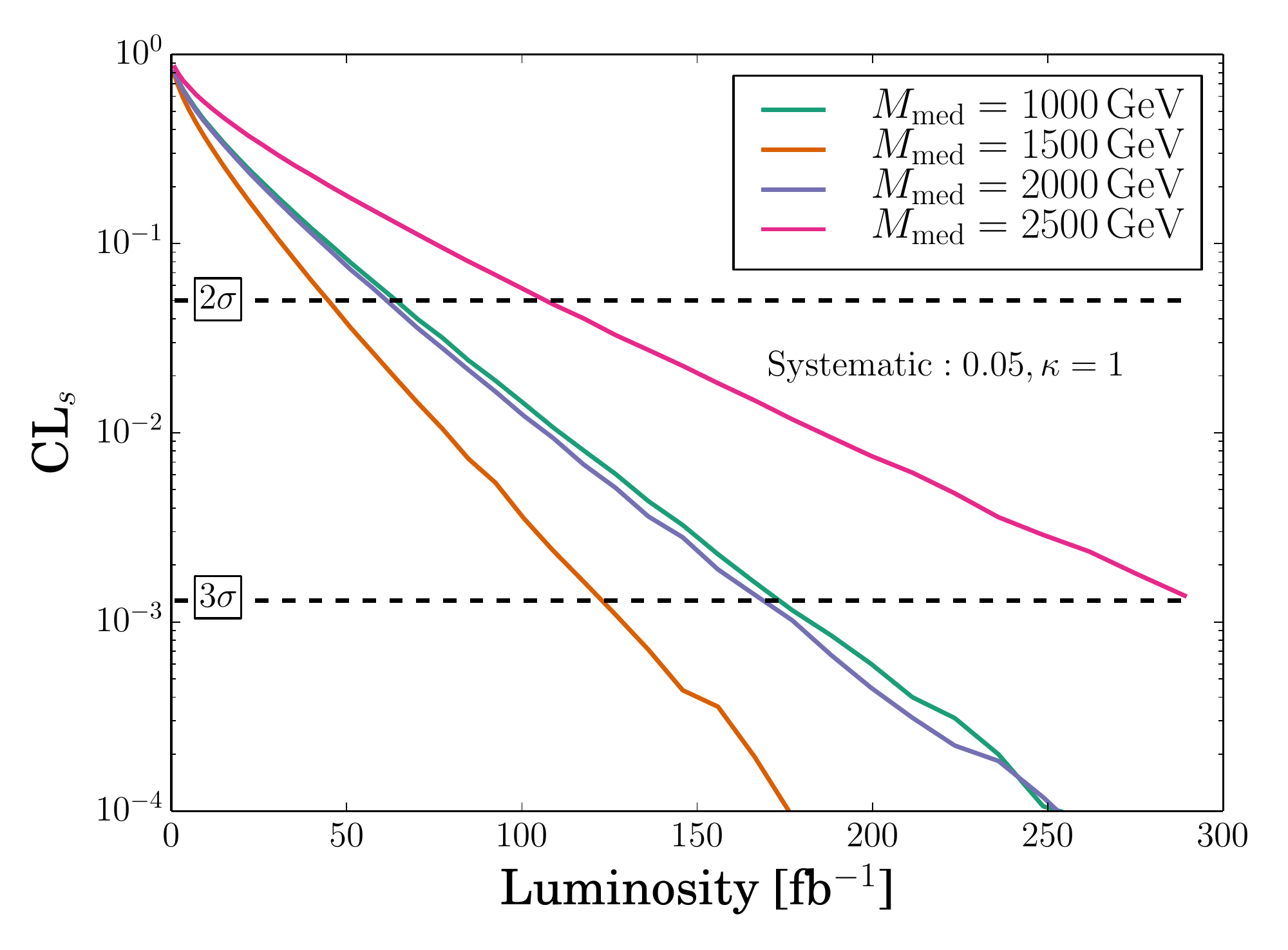}
\\
\end{tabular}
\end{center}
\vskip-0.4cm
\caption{
Differentiating the models at $\kappa=1$ at a future 100 TeV Collider. For each value of $M_{\rm med}$ between the reference model mass and $2500$ GeV the models are compared to the reference model with the 500 GeV mediator in the left panel and 750 GeV in the right panel.  We assume a systematic uncertainty of 5\% .
}
 \label{fig:100_dif_masses}
 \end{figure}

\begin{figure}[h!]
\begin{center}
\begin{tabular}{cc}
\hspace{-0.5cm}
\includegraphics[width=0.5\textwidth]{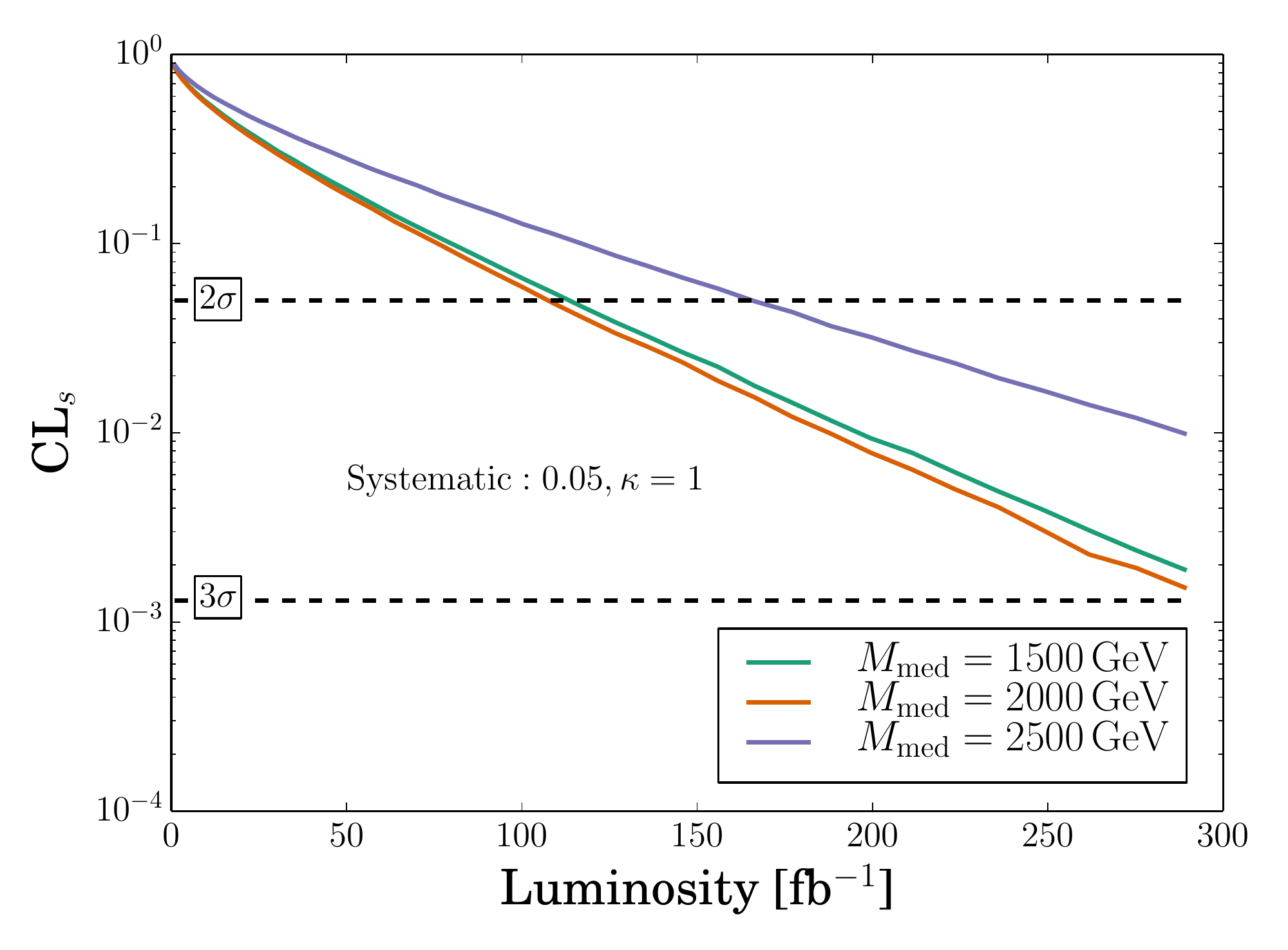}
&
\hspace{.1cm}
\includegraphics[width=0.5\textwidth]{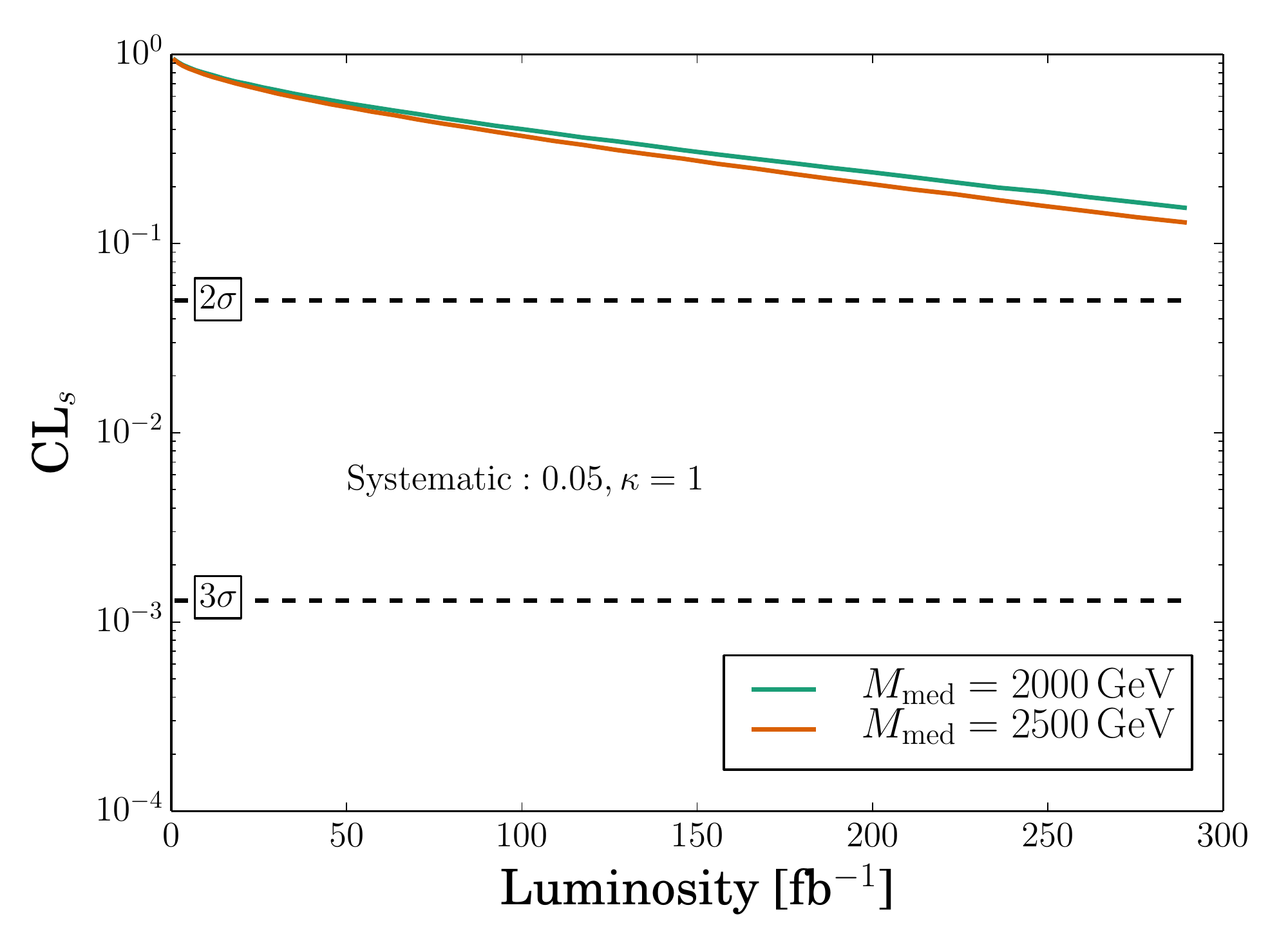}
\\
\end{tabular}
\end{center}
\vskip-0.4cm
\caption{
Differentiating the models at $\kappa=1$ at a future 100 TeV Collider. For each value of $M_{\rm med}$ between the reference model mass and $2500$ GeV the models are compared to the reference model with the 1000 GeV mediator in the left panel and 1500 GeV in the right panel.  We assume a systematic uncertainty of 5\% .
}
 \label{fig:100_dif_masses_2}
 \end{figure}

As before we can also compare to different choices of the reference model. 
In Figures~\ref{fig:100_dif_masses}-\ref{fig:100_dif_masses_2} we plot the results with reference models corresponding to mediator masses
of 500, 750, 1000 and 1500 GeV.
We conclude that for all the reference models with up to 1 TeV mediator masses we can successfully distinguish all higher mass reachable models.


\section{Summary and Conclusions}
\label{sec:conc}

We have studied collider limits for simplified models of dark matter. In the case we consider, dark matter is coupled to the Standard Model
by a scalar field mediator which interacts with both, the SM fermions and the vector bosons. Our main goal was two-fold, first to determine the
projected reach of the Run 2 LHC and the future circular hadron collider for excluding such models. Second, was to use the kinematic distributions to
be able to extract information on the values of the mediator masses. To achieve this we focussed on the two jets plus missing transverse energy 
final states. In our case both production mechanisms are important: the gluon fusion process which effectively probes the coupling of mediators to
quarks, and the weak vector boson fusion which is due to the interactions with vector bosons.

\medskip

We found that at the 13 TeV LHC one can probe dark matter simplified models with mediator masses up to 750 GeV. At a 100 TeV collider
the reach is increased to 2.5 TeV mediators. These measurements are fairly dominated by systematic uncertainties which we have kept 
relatively small and have varied between the 1\% and 10\% level.

\medskip

In order to `measure' the mediator mass -- which is the most relevant new physics scale of these models at colliders -- we have compared the models with
different values of $M_{\rm med}$ pairwise. In both cases, at 13 and at 100 TeV, we found that we can differentiate between essentially all the discoverable 
benchmark models we have considered. In particular, we can distinguish a 125 GeV reference model from models with mediator masses of
250, 500 and 750 GeV at the LHC. For the 100 TeV case the list of distinguishable models extends to 2.5 TeV.

\medskip  

For the more constrained case of the Higgs portal models, where scalar mediators mix with the SM Higgs and the mixing angle is small,
$\sin^2 \theta = 0.15$, the collider reach is reduced due to the overall decrease in the signal cross-section. We found that the LHC reach is
500 GeV for an optimistic 1\% systematic uncertainty, and this is increased to 750 GeV at a 100 TeV collider.

\newpage

\subsubsection*{Acknowledgements} 
We thank Simon Platzer and Alix Wilcock for useful discussions.
This work is supported by the STFC through the IPPP grant, 
and for VVK in part by the  Royal Society Wolfson Research Merit Award.
GR acknowledges the receipt of a Durham Doctoral Studentship.

\newpage



\begin{thebibliography}{99}

\bibitem{Khachatryan:2014rra}
  V.~Khachatryan {\it et al.}  [CMS Collaboration],
  ``Search for dark matter, extra dimensions, and unparticles in monojet events in proton-proton collisions at $\sqrt{s}$ = 8 TeV,''
  arXiv:1408.3583 [hep-ex].
  
\bibitem{Khachatryan:2014rwa}
  V.~Khachatryan {\it et al.}  [CMS Collaboration],
  ``Search for new phenomena in monophoton final states in proton-proton collisions at $\sqrt{s}$ = 8 TeV,''
  arXiv:1410.8812 [hep-ex].
  
\bibitem{Aad:2015zva}
  G.~Aad {\it et al.}  [ATLAS Collaboration],
  ``Search for new phenomena in final states with an energetic jet and large missing transverse momentum in pp collisions at $\sqrt{s}=8$ TeV with the ATLAS detector,''
  arXiv:1502.01518 [hep-ex].
    
\bibitem{Birkedal:2004xn} 
  A.~Birkedal, K.~Matchev and M.~Perelstein,
  ``Dark matter at colliders: A Model independent approach,''
  Phys.\ Rev.\ D {\bf 70}, 077701 (2004)
  [hep-ph/0403004].

\bibitem{Goodman:2010yf} 
  J.~Goodman, M.~Ibe, A.~Rajaraman, W.~Shepherd, T.~M.~P.~Tait and H.~B.~Yu,
  ``Constraints on Light Majorana dark Matter from Colliders,''
  Phys.\ Lett.\ B {\bf 695}, 185 (2011)
  [arXiv:1005.1286 [hep-ph]].

\bibitem{Goodman:2010ku} 
  J.~Goodman, M.~Ibe, A.~Rajaraman, W.~Shepherd, T.~M.~P.~Tait and H.~B.~Yu,
  ``Constraints on Dark Matter from Colliders,''
  Phys.\ Rev.\ D {\bf 82}, 116010 (2010)
  [arXiv:1008.1783 [hep-ph]].

\bibitem{Fox:2012ee} 
  P.~J.~Fox, R.~Harnik, R.~Primulando and C.~T.~Yu,
  ``Taking a Razor to Dark Matter Parameter Space at the LHC,''
  Phys.\ Rev.\ D {\bf 86}, 015010 (2012)
  [arXiv:1203.1662 [hep-ph]].
  
\bibitem{Abdallah:2014hon}
  J.~Abdallah, A.~Ashkenazi, A.~Boveia, G.~Busoni, A.~De Simone, C.~Doglioni, A.~Efrati and E.~Etzion {\it et al.},
  ``Simplified Models for Dark Matter and Missing Energy Searches at the LHC,''
  arXiv:1409.2893 
  
\bibitem{Malik:2014ggr}
  S.~Malik, C.~McCabe, H.~Araujo, A.~Belyaev, C.~Boehm, J.~Brooke, O.~Buchmueller and G.~Davies {\it et al.},
  ``Interplay and Characterization of Dark Matter Searches at Colliders and in Direct Detection Experiments,''
  arXiv:1409.4075 

\bibitem{Buckley:2014fba}
  M.~R.~Buckley, D.~Feld and D.~Goncalves,
  ``Scalar Simplified Models for Dark Matter,''
  Phys.\ Rev.\ D {\bf 91} (2015) 1,  015017
  [arXiv:1410.6497 [hep-ph]].

\bibitem{Harris:2014hga}
  P.~Harris, V.~V.~Khoze, M.~Spannowsky and C.~Williams,
  ``Constraining Dark Sectors at Colliders: Beyond the Effective Theory Approach,''
  Phys.\ Rev.\ D {\bf 91} (2015) 5,  055009
  [arXiv:1411.0535 [hep-ph]].

\bibitem{Chala:2015ama} 
  M.~Chala, F.~Kahlhoefer, M.~McCullough, G.~Nardini and K.~Schmidt-Hoberg,
  ``Constraining Dark Sectors with Monojets and Dijets,''
  arXiv:1503.05916 [hep-ph].
  
\bibitem{Haisch:2013fla}
  U.~Haisch, A.~Hibbs and E.~Re,
  ``Determining the structure of dark-matter couplings at the LHC,''
  Phys.\ Rev.\ D {\bf 89} (2014) 3,  034009
  [arXiv:1311.7131 [hep-ph]].
  
  \bibitem{Robens:2015gla}
  T.~Robens and T.~Stefaniak,
  ``Status of the Higgs Singlet Extension of the Standard Model after LHC Run 1,''
  Eur.\ Phys.\ J.\ C {\bf 75} (2015) 3,  104
  [arXiv:1501.02234 [hep-ph]].
 

 
\bibitem{Falkowski:2015iwa}
  A.~Falkowski, C.~Gross and O.~Lebedev,
  ``A second Higgs from the Higgs portal,''
  arXiv:1502.01361 [hep-ph].
  
\bibitem{Englert:2011yb} 
  C.~Englert, T.~Plehn, D.~Zerwas and P.~M.~Zerwas,
  ``Exploring the Higgs portal,''
  Phys.\ Lett.\ B {\bf 703}, 298 (2011)
  [arXiv:1106.3097 [hep-ph]].
  
\bibitem{Englert:2013gz}
  C.~Englert, J.~Jaeckel, V.~V.~Khoze and M.~Spannowsky,
  ``Emergence of the Electroweak Scale through the Higgs Portal,''
  JHEP {\bf 1304} (2013) 060
  [arXiv:1301.4224 [hep-ph]].

\bibitem{Lopez-Val:2014jva} 
  D.~López-Val and T.~Robens,
  ``$\Delta r$ and the W-boson mass in the singlet extension of the standard model,''
  Phys.\ Rev.\ D {\bf 90}, no. 11, 114018 (2014)
  [arXiv:1406.1043 [hep-ph]].
  
\bibitem{Hambye:2013dgv}
  T.~Hambye and A.~Strumia,
  ``Dynamical generation of the weak and Dark Matter scale,''
  Phys.\ Rev.\ D {\bf 88} (2013) 055022
  [arXiv:1306.2329 [hep-ph]].
  
\bibitem{Carone:2013wla}
  C.~D.~Carone and R.~Ramos,
  ``Classical scale-invariance, the electroweak scale and vector dark matter,''
  Phys.\ Rev.\ D {\bf 88} (2013) 055020
  [arXiv:1307.8428 [hep-ph]].
  
\bibitem{Khoze:2014xha}
  V.~V.~Khoze, C.~McCabe and G.~Ro,
  ``Higgs vacuum stability from the dark matter portal,''
  JHEP {\bf 1408} (2014) 026
  [arXiv:1403.4953 [hep-ph], arXiv:1403.4953].
  
\bibitem{Khoze:2014woa}
  V.~V.~Khoze and G.~Ro,
  ``Dark matter monopoles, vectors and photons,''
  JHEP {\bf 1410} (2014) 61
  [arXiv:1406.2291 [hep-ph]].
    
\bibitem{Altmannshofer:2014vra}
  W.~Altmannshofer, W.~A.~Bardeen, M.~Bauer, M.~Carena and J.~D.~Lykken,
  ``Light Dark Matter, Naturalness, and the Radiative Origin of the Electroweak Scale,''
  JHEP {\bf 1501} (2015) 032
  [arXiv:1408.3429 [hep-ph]].
  
\bibitem{Heikinheimo:2013fta}
  M.~Heikinheimo, A.~Racioppi, M.~Raidal, C.~Spethmann and K.~Tuominen,
  ``Physical Naturalness and Dynamical Breaking of Classical Scale Invariance,''
  Mod.\ Phys.\ Lett.\ A {\bf 29} (2014) 1450077
  [arXiv:1304.7006 [hep-ph]].

\bibitem{Backovic:2013dpa}
  M.~Backovic, K.~Kong and M.~McCaskey,
  Physics of the Dark Universe {\bf 5-6} (2014) 18
  [arXiv:1308.4955 [hep-ph]].

\bibitem{Backovic:2015cra}
  M.~Backovic, K.~Kong, A.~Martini, O.~Mattelaer and G.~Mohlabeng,
  arXiv:1505.04190 [hep-ph].
  
\bibitem{Ade:2013zuv}
  P.~A.~R.~Ade {\it et al.} [Planck Collaboration],
  Astron.\ Astrophys.\  {\bf 571} (2014) A16
  [arXiv:1303.5076 [astro-ph.CO]].

\bibitem{Akerib:2013tjd}
  D.~S.~Akerib {\it et al.} [LUX Collaboration],
  Phys.\ Rev.\ Lett.\  {\bf 112} (2014) 091303
  [arXiv:1310.8214 [astro-ph.CO]].
    
\bibitem{Eboli:2000ze}
  O.~J.~P.~Eboli and D.~Zeppenfeld,
  ``Observing an invisible Higgs boson,''
  Phys.\ Lett.\ B {\bf 495} (2000) 147
  [hep-ph/0009158].
  
\bibitem{Bernaciak:2014pna}
  C.~Bernaciak, T.~Plehn, P.~Schichtel and J.~Tattersall,
  ``Spying an invisible Higgs boson,''
  Phys.\ Rev.\ D {\bf 91} (2015) 3,  035024
  [arXiv:1411.7699 [hep-ph]].
 
  \bibitem{fastjet}
 M.~Cacciari and G.~P.~Salam,
  ``Dispelling the $N^{3}$ myth for the $k_t$ jet-finder,''
  Phys.\ Lett.\  B {\bf 641}, 57 (2006);
M.~Cacciari, G.~P.~Salam and G.~Soyez,
  ``FastJet user manual,''
  Eur.\ Phys.\ J.\ C {\bf 72}, 1896 (2012);
 \url{http://fastjet.fr}
 
\bibitem{Haisch:2012kf}
  U.~Haisch, F.~Kahlhoefer and J.~Unwin,
  ``The impact of heavy-quark loops on LHC dark matter searches,''
  JHEP {\bf 1307} (2013) 125
  [arXiv:1208.4605 [hep-ph]].
 
\bibitem{Baglio:2014uba}
  J.~Baglio, J.~Bellm, F.~Campanario, B.~Feigl, J.~Frank, T.~Figy, M.~Kerner and L.~D.~Ninh {\it et al.},
  ``Release Note - VBFNLO 2.7.0,''
  arXiv:1404.3940 [hep-ph].
  
\bibitem{Arnold:2011wj}
  K.~Arnold, J.~Bellm, G.~Bozzi, M.~Brieg, F.~Campanario, C.~Englert, B.~Feigl and J.~Frank {\it et al.},
  ``VBFNLO: A Parton Level Monte Carlo for Processes with Electroweak Bosons -- Manual for Version 2.5.0,''
  arXiv:1107.4038 [hep-ph].

\bibitem{Arnold:2008rz}
  K.~Arnold, M.~Bahr, G.~Bozzi, F.~Campanario, C.~Englert, T.~Figy, N.~Greiner and C.~Hackstein {\it et al.},
  ``VBFNLO: A Parton level Monte Carlo for processes with electroweak bosons,''
  Comput.\ Phys.\ Commun.\  {\bf 180} (2009) 1661
  [arXiv:0811.4559 [hep-ph]].
  
\bibitem{Alwall:2011uj}
  J.~Alwall, M.~Herquet, F.~Maltoni, O.~Mattelaer and T.~Stelzer,
  ``MadGraph 5 : Going Beyond,''
  JHEP {\bf 1106} (2011) 128
  [arXiv:1106.0522 [hep-ph]].

\bibitem{Bellm:2013lba}
  J.~Bellm, S.~Gieseke, D.~Grellscheid, A.~Papaefstathiou, S.~Platzer, P.~Richardson, C.~Rohr and T.~Schuh {\it et al.},
  ``Herwig++ 2.7 Release Note,''
  arXiv:1310.6877 [hep-ph].
  
\bibitem{Bai:2011wz} 
  Y.~Bai, P.~Draper and J.~Shelton,
  ``Measuring the Invisible Higgs Width at the 7 and 8 TeV LHC,''
  JHEP {\bf 1207}, 192 (2012)
  [arXiv:1112.4496 [hep-ph]].
 
\bibitem{Ghosh:2012ep} 
  D.~Ghosh, R.~Godbole, M.~Guchait, K.~Mohan and D.~Sengupta,
  ``Looking for an Invisible Higgs Signal at the LHC,''
  Phys.\ Lett.\ B {\bf 725}, 344 (2013)
  [arXiv:1211.7015 [hep-ph]].
    
\bibitem{Junk:1999kv}
  T.~Junk,
  ``Confidence level computation for combining searches with small statistics,''
  Nucl.\ Instrum.\ Meth.\ A {\bf 434} (1999) 435
  [hep-ex/9902006].
  
\bibitem{Read:2002hq}
  A.~L.~Read,
  ``Presentation of search results: The CL(s) technique,''
  J.\ Phys.\ G {\bf 28} (2002) 2693.
 
\bibitem{Andersen:2013rda} 
  J.~R.~Andersen, M.~Rauch and M.~Spannowsky,
  ``Dark Sector spectroscopy at the ILC,''
  Eur.\ Phys.\ J.\ C {\bf 74}, 2908 (2014)
  [arXiv:1308.4588 [hep-ph]].
 
\bibitem{Chacko:2013lna} 
  Z.~Chacko, Y.~Cui and S.~Hong,
  ``Exploring a Dark Sector Through the Higgs Portal at a Lepton Collider,''
  Phys.\ Lett.\ B {\bf 732}, 75 (2014)
  [arXiv:1311.3306 [hep-ph]].
 
\bibitem{Plehn:2001nj} 
  T.~Plehn, D.~L.~Rainwater and D.~Zeppenfeld,
  ``Determining the structure of Higgs couplings at the LHC,''
  Phys.\ Rev.\ Lett.\  {\bf 88}, 051801 (2002)
  [hep-ph/0105325].

\bibitem{Hankele:2006ja} 
  V.~Hankele, G.~Klamke and D.~Zeppenfeld,
  ``Higgs + 2 jets as a probe for CP properties,''
  hep-ph/0605117.
  
\bibitem{Dolan:2014upa} 
  M.~J.~Dolan, P.~Harris, M.~Jankowiak and M.~Spannowsky,
  ``Constraining $CP$-violating Higgs Sectors at the LHC using gluon fusion,''
  Phys.\ Rev.\ D {\bf 90}, no. 7, 073008 (2014)
  [arXiv:1406.3322 [hep-ph]].

  
\end{thebibliography}
\end{document}